\documentclass[a4paper,11pt]{article}
\pdfoutput=1 

\usepackage{jheppub}
\usepackage{color} 
\usepackage{float}
\usepackage{ulem}
\usepackage{slashed}
\usepackage{braket}
\usepackage{dsfont}
\usepackage{amsmath,graphicx,amssymb,subfigure,bbm}
\usepackage[all]{xy}

\begin{document}

\title{Nucleons and vector mesons  in a confining holographic QCD model}

\author[a]{Alfonso Ballon-Bayona,}
\author[a]{Ad\~ao S. da Silva Junior}
\affiliation[a]{Instituto de F\'{i}sica, Universidade
Federal do Rio de Janeiro, \\
Caixa Postal 68528, RJ 21941-972, Brazil.}       
\emailAdd{aballonb@if.ufrj.br}
\emailAdd{adaossj@pos.if.ufrj.br}

\abstract{We present a simple holographic QCD model that provides a unified description of vector mesons and nucleons in a confining background based on Einstein-dilaton gravity. For the confining background we consider analytical solutions of the Einstein-dilaton equations where the dilaton is a quadratic function of the radial coordinate far from the boundary. We build actions for the 5d gauge field and the 5d Dirac field dual to the 4d flavor current and the 4d nucleon interpolator respectively.  In order to obtain asymptotically linear Regge trajectories we impose for each sector the condition that the effective Schr\"odinger equation has a potential that grows quadratically in the radial coordinate far from the boundary. For the vector mesons we show that this condition is automatically satisfied by a 5d Yang-Mills action minimally coupled to the metric and the dilaton. For the nucleons we find that the mass term of the 5d Dirac action needs to be generalised to include couplings to the metric and the dilaton. Using Sturm-Liouville theory we obtain a spectral decomposition for the hadronic correlators consistent with large $N_c$ QCD. Our setup contains only three parameters: the mass scale associated with confinement, the 5d gauge coupling and the 5d Dirac coupling. The last two are completely fixed by matching the correlators at high energies to  perturbative QCD. 
We calculate masses and decay constants and compare our results against available experimental data.  Our model can be thought of as a consistent embedding of soft wall models in Einstein-dilaton gravity. }

\maketitle

\flushbottom
\section{Introduction}
The origin of hadron masses and its relation to confinement is one of the challenging problems in quantum chromodynamics (QCD) due to the necessity of non-perturbative techniques. An important mechanism for mass generation in QCD is related to the spontaneous breaking of chiral symmetry, see for example \cite{Miransky:1994vk}. The order parameter of spontaneous chiral symmetry breaking is the quark condensate defined as the VEV (vacuum expectation value) of the quark mass operator, i.e. $\langle \bar{q}q \rangle$. Another important quantity associated with mass generation, confinement and the QCD vacuum is the so-called gluon condensate defined as the  VEV  of the Yang-Mills operator, i.e. $\langle {\rm Tr} F^2 \rangle$. The gluon and quark condensates are both related to the QCD trace anomaly and the vacuum energy density of QCD, see for instance \cite{Shuryak:2004pry}.  

 
Several approaches have emerged in an attempt to describe QCD in the non-perturbative regime, such as the Nambu-Jona-Lasinio model \cite{Nambu:1961tp,Nambu:1961fr}, the linear sigma model \cite{Gell-Mann:1960mvl},  chiral perturbation theory (for a review see \cite{Scherer:2002tk}), lattice QCD \cite{Wilson:1974sk} and QCD sum rules  \cite{Shifman:1978bx} (for a review see \cite{Colangelo:2000dp}). The Nambu-Jona-Lasinio and the linear sigma models describe spontaneous chiral symmetry breaking, generating a mass scale. Chiral perturbation theory is a systematic approach that exploits the approximate chiral symmetry of QCD at low energies and allows describing some hadronic properties. Lattice QCD  is a numerical approach based on the discretisation of spacetime, enabling the calculation of correlation functions of QCD operators through numerical simulations on a lattice. This approach allows for the investigation of non-perturbative properties of hadrons, such as masses. Lastly, the QCD sum rules approach considers correlation functions of composite operators built from quark fields and then uses  operator product expansions (OPE) and spectral functions to estimate hadronic properties such as masses and decay constants. The composite operators are usually known as interpolating fields (this terminology is also used in lattice QCD).

 The AdS/CFT correspondence is an alternative approach for investigating QCD in the strong coupling regime. This conjecture establishes a duality between string theories on $AdS_{d+1}\times M$ ($AdS$ is Anti-de Sitter spacetime and $M$ is a compact space) and conformal field theories (CFT) in $d$ dimensions \cite{Maldacena:1997re,Witten:1998qj, Gubser:1998bc}. In this work, we will restrict ourselves to the particular example of the duality that relates IIB string theory on $AdS_{5}\times S^{5}$ to the $\mathcal{N} = 4$ Super Yang-Mills theory in four dimensions. After the conjecture was proposed, some models emerged that became known as AdS/QCD, which aim to capture low-energy aspects of QCD by breaking the conformal symmetry. These models incorporate QCD properties such as confinement and chiral symmetry breaking. 

There are two main approaches in AdS/QCD: the bottom-up and top-down. The bottom-up approach aims to capture QCD properties mapping the deformation of the $CFT _{4}$ to deformations of the $AdS_{5}$ space. In this approach, a minimal set of 5d fields is introduced to describe the dynamics of 4d operators similar to those appearing in real QCD. The actions are usually minimal models for the 5d fields that reproduce the symmetries of the dual 4d operators. Examples of models within this approach include the hard wall model \cite{Polchinski:2001tt,Boschi-Filho:2002xih,Erlich:2005qh,DaRold:2005mxj}, the soft wall model \cite{Karch:2006pv}, and the Einstein-dilaton models \cite{Gursoy:2007cb,Gursoy:2007er,Gubser:2008ny,Cai:2012xh,Li:2013oda,Ballon-Bayona:2017sxa}. In the hard wall model, specific boundary conditions are imposed on the AdS space, while the soft wall model introduces a scalar field, known as the dilaton, in the action. The Einstein-dilaton model, distinct from the other two, is consistent with Einstein's equations and allows for a description of a non-trivial vacuum in the 4d dual theory. A more rigorous approach in AdS/QCD is the top-down approach, which aligns with string theory principles and introduces the breaking of conformal symmetry and supersymmetry through a setup of D-branes. Models within this approach include the D3/D7  model \cite{Karch:2002sh} and the D4/D8 model \cite{Witten:1998qj,Sakai:2004cn}. 

An important test for AdS/QCD is the description of hadronic masses and decay constants and its relation to confinement. In this work we will focus on the description of light vector mesons and nucleons using the bottom-up approach. Light vector mesons have been investigated previously in the bottom-approach using hard wall models \cite{Erlich:2005qh,Grigoryan:2007vg}, soft wall models \cite{Karch:2006pv,Grigoryan:2007my} and metric deformations in AdS 
\cite{Forkel:2007cm,dePaula:2008fp, 
 FolcoCapossoli:2019imm} considering a 5d Yang-Mills action. There has also been some progress on the description of light vector mesons in holographic models inspired by string theory \cite{Gursoy:2007er,Iatrakis:2010jb,Arean:2013tja} and models based on Einstein-dilaton gravity \cite{He:2013qq,Ballon-Bayona:2023zal}.  Nucleons have been investigated previously in holographic QCD following two approaches. In the first approach one builds a 5d Dirac action for the 5d Dirac field dual to a 4d nucleon interpolator, see  \cite{deTeramond:2005su,Hong:2006ta} for the hard wall model,  \cite{Brodsky:2008pg,Abidin:2009hr,Gutsche:2011vb} for the soft wall model and \cite{Forkel:2007cm, FolcoCapossoli:2019imm,Nascimento:2023dzx} for AdS deformations. The second approach consists of mapping 5d solitons to 4d skyrmions; see, for example \cite{Hata:2007mb, Hashimoto:2008zw,Pomarol:2008aa,Cherman:2009gb,Sutcliffe:2015sta,Hayashi:2020ipd,Jarvinen:2022gcc}. We also noticed some recent progress on the description of fermionic states qualitatively similar to baryons considering a fermionic action for $Dp/Dq$ brane models in string theory \cite{Abt:2019tas,Nakas:2020hyo}.

We present in this paper a simple 5d holographic QCD model that provides a unified description of light vector mesons and nucleons in a confined background, the latter arising from Einstein-dilaton gravity. Our model contains only three parameters; two of them are 5d coupling constants that are fixed matching the result for the two-point correlators at high energies to perturbative QCD, the third parameter is the mass scale associated with confinement which can be fixed matching for instance the mass of the $\rho(770)$ meson to the mass of the fundamental vector meson state, i.e. $m_{\rho^0}$. We calculate the spectrum of light vector mesons  and nucleons  as well as their decay constants. In order to provide a clean comparison to previous models and experimental data we will present our results dividing all the observables by the appropriate power of $m_{\rho^0}$. We impose for vector mesons and nucleons a condition that guarantees an asympotically linear spectrum, namely that the Schr\"odinger effective potential grows quadratically in the radial coordinate far from the boundary. We find for vector mesons that this condition is automatically satisfied a 5d Yang-Mills action minimally coupled to the metric and dilaton. For the nucleons we find that the mass term of the  5d Dirac action needs to be extended to include non-minimal couplings to the metric and dilaton. We use Sturm-Liouville theory to obtain  spectral decompositions for the two-point correlation functions associated with the 4d flavour current and the 4d nucleon interpolator (Ioffe current). We show that the spectral decompositions for the hadronic correlators are consistent with QCD in the large $N_c$ limit. This in turn allows us to obtain a holographic dictionary for the decay constants of vector mesons and nucleons valid for a general class of holographic models based on Einstein-dilaton gravity. 

From the theoretical point of view, our model improves previous bottom-up approaches allowing to make predictions from a consistent five-dimensional background that satisfies the confinement criterion. From the phenomenological point of view, our model leads to results for the vector meson and nucleon masses that are very close to experimental data. Moreover, the model leads to very clean results for the vector meson and nucleon decay constants. In the latter case we will compare for the very first time against lattice QCD results and provide predictions for the excited states that could be a useful guide for future phenomenological studies.

The organisation of this paper is as follows: In section \ref{Sec:models} we review the action and field equations of Einstein-dilaton gravity and present two analytical solutions that satisfy the confinement criterion. These two concrete backgrounds will be used in the rest of the paper. In section \ref{Sec:VectorMesons}, we investigate the light vector mesons. We describe the 5d action, the field equations, the holographic dictionary for the VEV of the 4d flavour current and  the bulk to boundary propagator. Using Sturm-Liouville theory we obtain a spectral decomposition for the current correlator and finally solving the Schr\"odinger equation we obtain the vector meson spectrum and decay constants. In section \ref{nucleonseinsteindilaton}, we investigate the nucleons. We present the Dirac action, the field equations, the VEV of the 4d nucleon operator and the bulk to boundary propagator. Using  Sturm-Liouville theory we find the spectral decomposition for the nucleon correlator and finally  solving the Schrödinger equations we obtain the spectrum of nucleons and the nucleon decay constants. Our conclusions are presented in section \ref{Sec:Conclusions} and additional material is described in four appendices. Appendix \ref{App:sturmlitheory} briefly reviews the Sturm-Liouville theory and the spectral decomposition. The Proca field propagator associated with vector mesons is described in appendix \ref{App:procapropagator}. The vector mesons and nucleons in the soft wall model and hard wall model are described in appendices \ref{App:reviewsoftwall} and \ref{App:reviewhardwall} respectively.

\section{Confining holographic QCD models from Einstein-dilaton gravity}
\label{Sec:models}

\subsection{The action}

Holographic QCD models based on Einstein-dilaton gravity are  described by the following action in the string frame \cite{Gursoy:2007er,Li:2013oda,Ballon-Bayona:2017sxa}
\begin{eqnarray}
    S_{s} =\sigma \int_{\mathcal{M}} d^{5}x \sqrt{-g_{s}}\,e^{-2\Phi_{s}}\left[R_{s} + \mathcal{L}_{\Phi}^{s} \right] \, . \label{action1}
\end{eqnarray}
In this expression, $\sigma = 1/16\pi G_5$, where $G_5$ represents the five-dimensional Newton's constant, $R_{s}$ is the Ricci scalar, $\Phi$ is the dilaton and $\mathcal{L}_{\Phi}^{s}$ is the dilaton Lagrangian express by
\begin{eqnarray}
    \mathcal{L}_{\Phi}^{s} = 4\partial_{\mu}\Phi\partial^{\mu}\Phi + \ell^{-2} V_{s}(\Phi)
\end{eqnarray}
The subscript "s" in the previous equations indicates that they are written in the string frame. The parameter $\ell$ is the AdS radius.  We omit here an additional surface term that is required from the variational principle \footnote{ This term, usually called the Gibbons - Hawking - York boundary term, is also important in the study of vacuum energy in holographic QCD.}.

In Einstein-dilaton gravity, there are two interesting frames: the string frame and the Einstein frame. We can write the action \eqref{action1} in the Einstein frame using the following transformations
\begin{eqnarray}
    g_{mn} &=& g_{mn}^{s}\, e^{-\frac43\Phi}\, , \label{gsge}\\
         V(\Phi) &=& V_{s} \, e^{\frac{4}{3}\Phi}\, .\label{vevs}
\end{eqnarray}
Plugging \eqref{gsge} and \eqref{vevs} into \eqref{action1}, the action in Einstein frame becomes
\begin{eqnarray}
    S_{E} =\sigma\int d^{5}x\, \sqrt{-g}\left[R + \mathcal{L}_{\Phi} \right]\, ,\label{action2}
\end{eqnarray}
where
\begin{align}
\mathcal{L}_{\Phi} = - \dfrac{4}{3}g^{mn}\partial_{m}\Phi\partial_{n}\Phi + \ell^{-2}  V(\Phi) \,.    
\end{align}

\subsection{The field equations}

By varying the action \eqref{action2} with respect to the metric, we obtain 
\begin{eqnarray}
    R_{mn} - \dfrac{R}{2}g_{mn} &=&\dfrac{1}{2\sigma}T_{mn}\, ,\label{eqr}\\
   \nabla^{2}\Phi + \dfrac{3}{8 \ell^2}  \dfrac{dV}{d\Phi} &=& 0\, ,\label{KGeq}
\end{eqnarray}
where the tensorial equation in \eqref{eqr} corresponds to the Einstein equations in the presence of scalar matter and \eqref{KGeq} is the generalization of the Klein-Gordon equation in curved space. The energy-momentum tensor $T_{mn}$, is given by
\begin{align}
    T_{mn} = \sigma \left[\dfrac{8}{3}\partial_{m}\Phi\partial_{n}\Phi + g_{mn}\mathcal{L}_{\Phi}\right]\, 
\end{align}
and the Einstein equations can also be written in the Ricci form
\begin{align}
    R_{mn} = \dfrac{4}{3}\partial_{m}\Phi\partial_{n}\Phi - \dfrac{1}{3 \ell^2}g_{mn}V\, .
\end{align}
We now consider the following ansatz for the 5d metric:
\begin{eqnarray}
    ds^{2} =\dfrac{1}{\zeta(z)^{2}}\left[ - dt^{2} + dx_{i}^{2} + dz^{2}\right]\label{poincarecoordinates}
\end{eqnarray}
This metric preserves Poincaré symmetry. Plugging this ansatz into the Einstein-dilaton equations we find the following field equations:
\begin{align}
& \zeta'' - \frac49 \zeta \Phi'^2 = 0 \, , \label{zetaPhieq} \\ 
&  \ell^{-2} V -\zeta^5 (\zeta^{-3} )''=0    \, ,  \label{Veq}  \\
& \frac83 \zeta^2 \Big [  \Phi'' - 3 \frac{\zeta'}{\zeta} \Phi' \Big ] + \ell^{-2} \frac{d V}{d \Phi} = 0 \, . \label{dVeq}
\end{align}
where $'=d/dz$.  The equation \eqref{dVeq}  comes from the scalar differential equation \eqref{KGeq} or from the Bianchi identity $\nabla_n T^{mn}=0$.  This equation is not independent because it can be obtained from equations \eqref{zetaPhieq} and \eqref{Veq}. The inverse scale factor $\zeta(z)$ is usually written in terms of the warp factor $A(z)$ using the relation
\begin{equation}
\zeta(z) = \exp(-A(z)).     
\end{equation}

The warp factor in the string frame takes the form
\begin{eqnarray}
  A_s(z) &=& A(z) + \frac{2}{3}\Phi  = -\ln \zeta + \dfrac{2}{3}\Phi  \, . \label{AeAs}  
\end{eqnarray}

\subsection{Confining holographic QCD models}

In this work we will consider holographic QCD models where the dilaton is quadratic far from the boundary (infrared regime), i.e. 
\begin{equation}
\Phi( z \to \infty) = k z^2 \,.   \label{IRcond}  
\end{equation}
Near the boundary we only impose that the metric is asymptotically AdS, i.e. 
\begin{equation}
\zeta(z \to 0) = \frac{z}{\ell} \,.    \label{UVcond} 
\end{equation}
The IR asymptotic behaviour \eqref{IRcond} was originally proposed by Karch, Katz, Son and Stephanov \cite{Karch:2006pv} as a condition that guarantees approximate linear Regge trajectories for mesons. It was later proven by Gursoy, Kiritsis and Nitti \cite{Gursoy:2007er} that this asymptotic behaviour is compatible with the confinement criterion and also leads to a linear spectrum for glueballs, see also \cite{Ballon-Bayona:2017sxa}. Later in this section we will use the warp factor in the string frame $A_s(z)$ to show that models satisfying this asymptotic behaviour satisfy the confinement criterion developed in \cite{Kinar:1998vq}.

In order to build concrete models we consider two simple analytical solutions of the Einstein-dilaton equations that satisfy the conditions \eqref{IRcond} and \eqref{UVcond}. 

The first model is given by
\begin{eqnarray}
    \Phi_I(z) = k z^{2} \quad , \quad 
    \zeta_I(z) = \Gamma (5/4) \left (\frac{3}{k} \right )^{1/4} \frac{\sqrt{z}}{\ell} \, I_{1/4} \left (\frac23 k z^2 \right )   \, .\label{model 1}
\end{eqnarray}
In this case we considered a simple ansatz for the dilaton field $\Phi(z)$ and found the inverse scale factor $\zeta(z)$ using the Einstein-dilaton equation \eqref{zetaPhieq}. This model was investigated by Huang and Li in \cite{Li:2013oda}. 

The second model is given by
\begin{eqnarray}
\Phi_{II}(z) = \dfrac{1}{2}\sqrt{k}z\sqrt{9 + 4 k z^{2}} + \dfrac{9}{4}\sinh^{-1}\left(\dfrac{2}{3}\sqrt{k}z\right)   \quad , \quad 
\zeta_{II}(z) = \frac{z}{\ell} \exp\left(\dfrac{2}{3}kz^{2}\right)\,. \label{model 2}
\end{eqnarray}    
In this case we took a simple ansatz for the inverse scale factor $\zeta(z)$ and use the Einstein-dilaton equation \eqref{zetaPhieq} to find the dilaton $\Phi(z)$. This model was proposed by Gursoy, Kiritsis, and Nitti in \cite{Gursoy:2007er} as a simple analytical model for describing confinement \footnote{A similar analytical model  was proposed earlier in the string frame as a phenomenological approach for the quark-antiquark potential without actually solving the Einstein-dilaton equations \cite{Andreev:2006ct}.}. 

\begin{figure}[htp!]
    \centering
    {{\includegraphics[width=7cm]{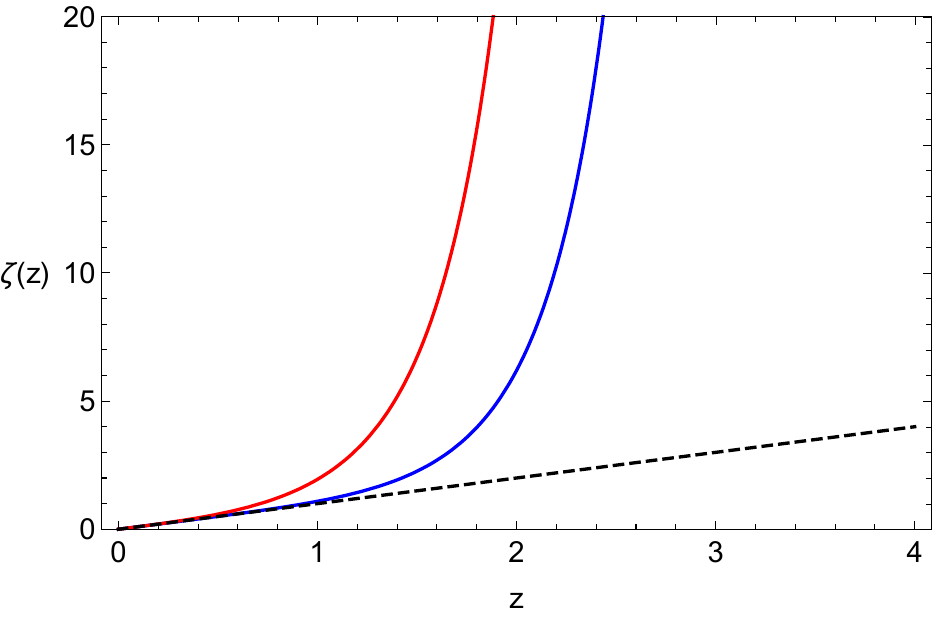} }}    
    {{\includegraphics[width=7cm]{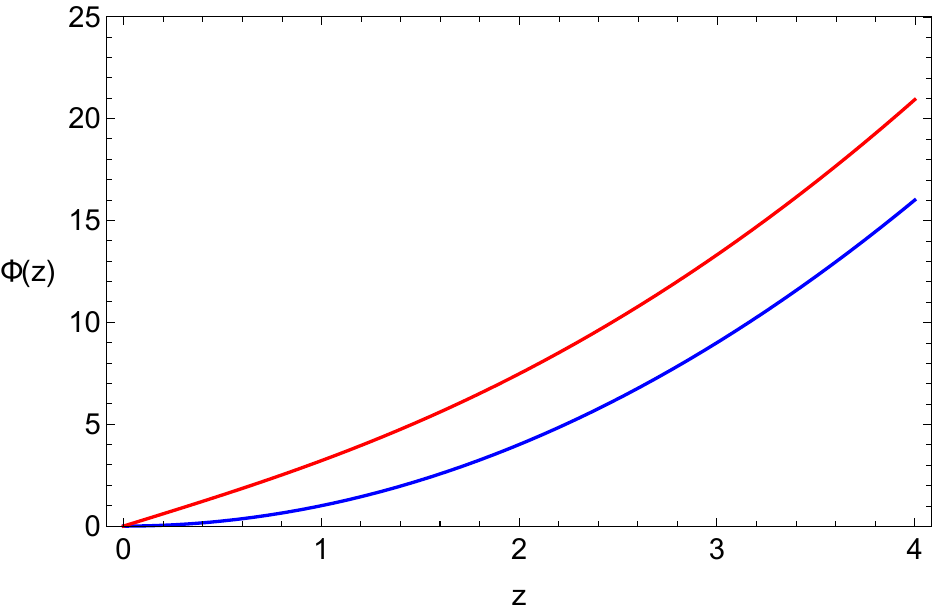} }} 
    \caption{Left panel: Inverse scale factor $\zeta(z)$ in the Einstein frame for model I (blue) and model II (red). The black dashed line represents the AdS limit $\zeta(z) = z$. Right panel:  Dilaton field $\Phi(z)$ for model I (blue) and model II (red). The plots were obtained in units of $k=1$ and $\ell=1$.}
    \label{warpfactors}%
\end{figure}

 The behaviour of the inverse  scale factor, in the Einstein frame, for models I and  II is shown on the left panel of figure \ref{warpfactors}. In both cases the inverse scale factor behaves as $\zeta(z) = z/\ell + \dots$ at small $z$ (AdS asymptotics) and becomes $\exp( \frac23 k z^2 + \dots)$ at very large $z$. The dilaton field $\Phi(z)$ is displayed on the right panel of figure \ref{warpfactors}. In model I the dilaton field is always $\Phi(z) = kz^2$ whilst in model II it evolves from  $\Phi(z) = 3 \sqrt{k} z + \dots$ at small z to $\Phi(z) = kz^2$ at large z.

\begin{figure}[htp!]
    \centering
    {{\includegraphics[width=7cm]{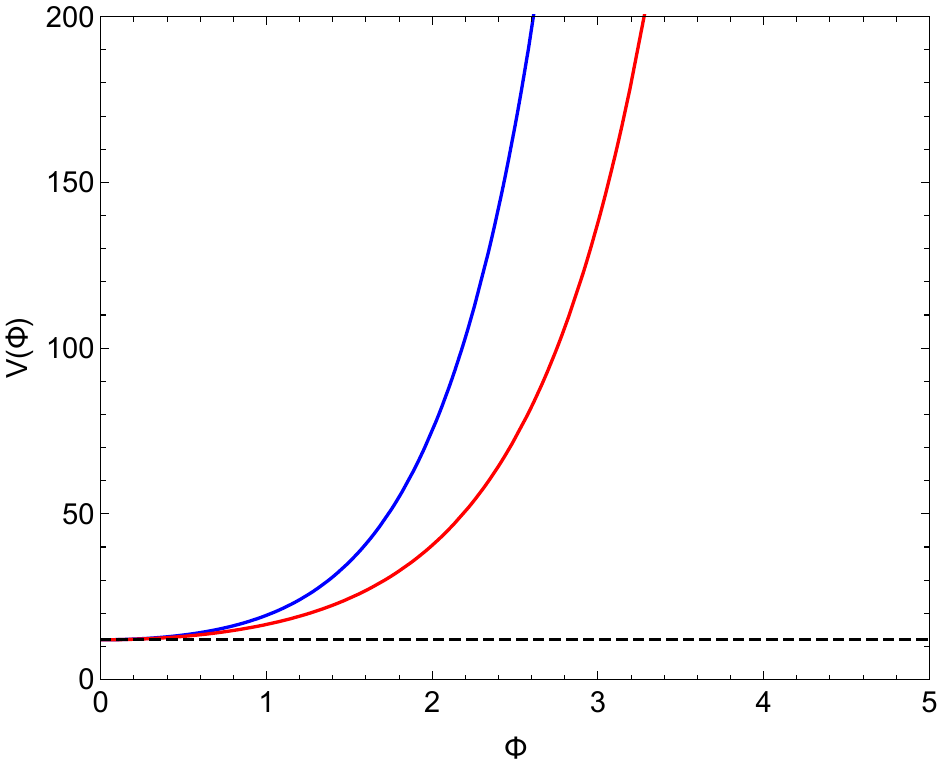} }}%
        \caption{Dilaton potential in the Einstein frame for model I (blue) and model II (red). The black dashed line represents the AdS limit $V=12$ when $\Phi$ goes to zero.}%
    \label{Potentials Einstein Dilaton Model}%
\end{figure}

Using the Einstein-dilaton equation \eqref{Veq} we can reconstruct the dilaton potential $V(\Phi)$ for models I and II. This is shown in figure \ref{Potentials Einstein Dilaton Model} where we also show the limit $V(\Phi \to 0)=12$ that corresponds to the negative cosmological constant for AdS space. 

\subsection{Conformal symmetry breaking and confinement}
\begin{figure}[htp!]
    \centering
    {{\includegraphics[width=7cm]{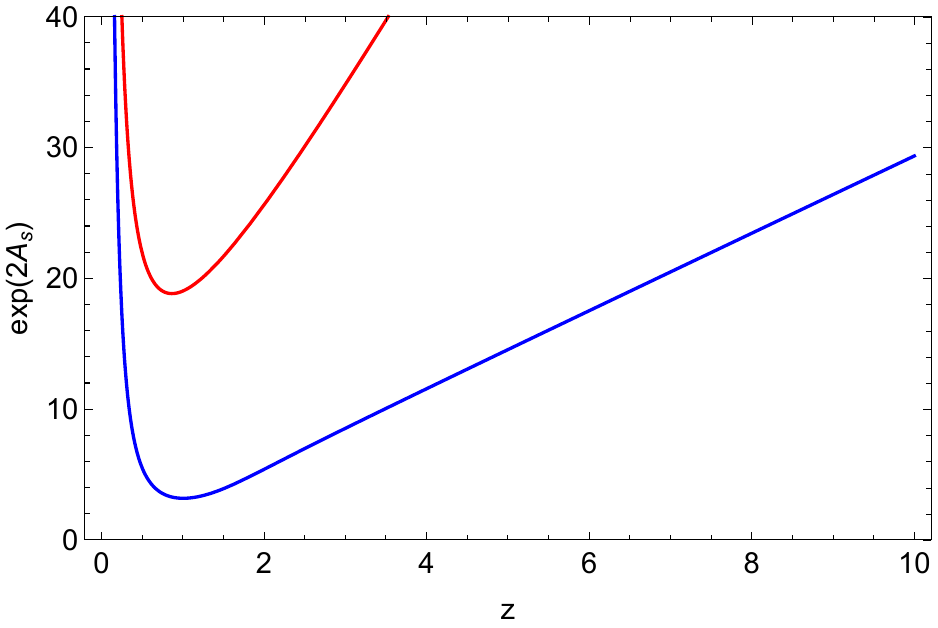} }}%
        \caption{The function $f=\exp(2A_s)$ as a function of $z$ for model I (blue) and model II (red). The plot was done in units of $k=1$ and $\ell=1$.}%
    \label{Plotf}%
\end{figure}

The models presented in the previous section describe an explicit breaking of conformal symmetry and guarantee confinement. In this section we briefly describe the confinement criterion discussed in \cite{Gursoy:2007er} for Einstein-dilaton models based on the general criterion found in \cite{Kinar:1998vq}. The behaviour of the potential energy of a heavy quark-antiquark pair, described by a rectangular Wilson loop, for a review see \cite{Ramallo:2013bua}, when the distance between them is large is given by
\begin{eqnarray}
    E(L) =\mu \, f(z^{\star}) L \label{potentialqq}
\end{eqnarray}
where $E(L)$ is the potential energy of the quark-antiquark pair as a function of the distance $L$, $\mu$ is the fundamental tension of the string, and $f =\exp(2A_{s})$ is a function of the string warp factor $A_s$. A non-zero minimum for $f$, located at  $z=z^{\star}$, guarantees a non-zero string tension for the quark-antiquark potential. Note that confinement in Einstein-dilaton models involves the string frame warp factor 
\begin{eqnarray}
    A_{s}(z) = A(z) + \dfrac{2}{3}\Phi(z) \,. 
\end{eqnarray}
As explained previously in this section, in order to guarantee a linear spectrum for mesons and glueballs, the dilaton  field must behave at large $z$ as 
\begin{equation}
   \Phi( z \to \infty) = k z^2 + \dots \, ,   \label{PhiLargez}
\end{equation}
and this in turn implies that the Einstein frame warp factor should behave as
\begin{equation}
A (z \to \infty) = - \frac23 k z^2 + \dots  \,.
\end{equation}
The dots in the equations above represent subleading terms for $\Phi(z)$ and $A(z)$. As described in \cite{Gursoy:2007er}, at large $z$ the dilaton and warp factor should satisfy the condition
\begin{eqnarray}
    \Phi(z) + \frac32 A(z) =  \frac34 \ln |A'(z)| + ...  \quad (z \to \infty) \, .
\end{eqnarray}
This in turn implies that the string frame warp factor behaves at large z as
\begin{eqnarray}
A_s(z \to \infty) = \frac12 \ln |A'(z)| + \dots = \frac12 \ln (\sqrt{k} z) + \dots   \,. \label{AsLargez}
\end{eqnarray}
On the other hand AdS asymptotics at small z implies that 
\begin{eqnarray}
A_s(z \to 0) = - \ln (z/\ell) + \dots \,.     
\end{eqnarray}
These results when applied to the function $f = \exp (2 A_s)$ imply that this function behaves at large z as $f(z \to \infty) = \sqrt{k} z+ \dots$ and at small z at $f(z \to 0) = (\ell/z)^2 + \dots $ \,. Then the function $f(z)$ is non-monotonic in $z$ and possesses a minimum at some $z=z^{*}$. This is shown in figure \ref{Plotf} where we plot the function  $f = \exp (2 A_s)$ for models I and II. 

We finish this section with an important remark. In the holographic QCD models presented here conformal symmetry breaking and confinement are driven by a single (infrared) mass scale $\sqrt{k}$. If we set $k$ to zero the dilaton vanishes and we recover the AdS space and conformal symmetry.  The analog of this situation in QCD is the presence of a gluon condensate associated with a non-zero trace for the stress energy tensor and conformal symmetry breaking (the QCD trace anomaly). In fact, the problem of generation of hadron masses is expected to be understood in terms of a non-trivial stress energy tensor. 

In the following sections we will incorporate vector mesons and nucleons in models I and II. We will obtain a unified description of vector meson and nucleon masses in terms of the single mass scale $\sqrt{k}$. When investigating the spectrum of vector mesons and nucleons we will focus on mass ratios, since those are independent of the choice of $k$.  We will also find that the two point correlation functions for the vector meson and nucleon interpolating fields satisfy a spectral decomposition consistent with QCD in the large N limit. We will use this decomposition to extract the  decay constants of vector mesons and nucleons.  

\section{Vector mesons in confining holographic QCD }\label{Sec:VectorMesons}

In this section, we will describe vector mesons in confining holographic QCD models based on Einstein-dilaton gravity. Firstly, we will present the 5d action and the equations of motion in both coordinate and momentum space. The VEVs and their connections with 4d currents are discussed. Subsequently, we will study the on-shell action and the bulk to boundary propagator, allowing us to obtain the two-point function. The spectral decomposition for the bulk to boundary propagator is described using Sturm-Lioiuville theory in order to obtain a spectral decomposition for the two-point function consistent with QCD in the large $N_c$ limit. Lastly, we will obtain the spectrum and decay constants of vector mesons for the Einstein-dilaton models I and II described in section \ref{Sec:models}, comparing with previous models and available experimental data. 

\subsection{The 4d flavour currents}

Consider vector mesons in large $N_c$ QCD with $N_f=2$ flavors. The flavour (isospin) currents responsible for creation of vector meson states can be written as 
\begin{equation}
 J^{\mu,a}(x)  =  \bar q (x) \gamma^{\mu} T^a q (x)  \, ,     
\end{equation}
where $q(x)$ is the quark doublet with components $u(x)$ and $d(x)$ and $T^a$, with $a=1,2,3$, are the generators of the $SU(2)$ group. 
For simplicity we assume flavour (isospin) symmetry $m_u = m_d$ so that the flavour current is conserved. 
The matrix element for this current when applied to the vacuum and a vector meson state can be written as
\begin{align}
\langle 0 \vert J^{\mu,a}(0) \vert V^{n,b}(p,\lambda) \rangle = F_{v^n} \, \epsilon^{\mu}(p,\lambda)   \delta^{ab} \,, 
\end{align}
where $\epsilon^{\mu}(p,\lambda)$ is the polarisation of the vector meson state.  The coupling $F_{v^n}$  is associated with the probability amplitude of creating a particular vector meson state from the vacuum. It can be related directly to the weak decay of vector mesons and for this reason it is known as the vector meson decay constant. Since the flavour current is conserved its conformal dimension is equal to its canonical dimension, so $\Delta =3$ at all RG energy scales. Thus the decay constant $F_{v^n}$ has dimension of mass squared \footnote{The meson states are normalised as $\langle \vec{p} \vert \vec{q} \rangle  = 2 E_{\vec{p}} \,  (2 \pi)^3 \delta^{3}(\vec{p}- \vec{q})$. }. 
In large $N_c$ QCD the correlation function for two flavour currents admits the spectral decomposition \cite{Witten:1979kh,Son:2003et}
\begin{equation}
\langle J^{\mu,a} (q) J^{\nu,b}(q') \rangle 
= \delta^4 (q - q') \delta^{ab} P^{\mu \nu}_{\perp}(q) \sum_n \frac{F_{v^n}^2}{q^2 + m_{v^n}^2} \,, \label{currentcorrel}
\end{equation}
where 
\begin{equation}
P^{\mu \nu}_{\perp}(q) = \eta^{\mu \nu} - \frac{q^{\mu} q^{\nu}}{q^2} \, , \label{TProjector}
\end{equation}
is the transverse projector which appear in propagators of  massive spin 1 states (vector mesons). The result in \eqref{currentcorrel} was obtained previously for some particular holographic QCD models \cite{Erlich:2005qh,Grigoryan:2007vg,Grigoryan:2007my}. In this section we will show that a general class of holographic QCD models based on Einstein-dilaton gravity lead to current correlators that satisfy the spectral decomposition \eqref{currentcorrel}.

\subsection{The 5d action and field equations}
\label{sub:actionvec}

We start with a set of 5d gauge fields $V_m^a(z,x)$ dual to the 4d flavour currents $J_{\mu}^{a}(x)$. In order to describe the spectrum of vector mesons we only need a 5d action quadratic on these fields. Assuming a minimal coupling to the metric and dilaton field, the action can be written as
\begin{equation}
S_V = -  \int d^4 x\, dz \frac{1}{4 g_5^2} \sqrt{-g_s} \, e^{-\Phi}
 {v_{mn}^a}^2 \, ,\label{actionvec1}
\end{equation}
where $v_{mn}^a = \partial_m V_n^a - \partial_n V_m^a$ are the (Abelian) field strengths \footnote{Non-Abelian terms are of cubic  or higher order on the fields $V_m^a$ and are relevant only to describe interactions.},  the 5d metric $g_{mn}^s$ is in the string frame and the index $a=1,2,3$ is implicitly summed. The gauge coupling is fixed as $g_{5}^{2} = 12\pi^{2}/N_{c}$ in order to reproduce the perturbative QCD result for the current correlator at small distances \cite{Erlich:2005qh}. The action in \eqref{actionvec1} can be obtained from the vectorial sector of holographic models of chiral symmetry breaking after expanding at quadratic order the 5d Yang-Mills-Higgs action associated with the breaking of $SU(2)_L \times SU(2)_R$ chiral symmetry; see, for example \cite{Erlich:2005qh,Karch:2006pv,Ballon-Bayona:2020qpq,Ballon-Bayona:2021ibm,Ballon-Bayona:2023zal}.

As described in the previous section, in holographic QCD models based on Einstein-dilaton gravity the string frame metric can be written as 
\begin{equation}
g_{mn}^s = e^{2 A_s(z)} \eta_{\hat m \hat n} \, ,   \label{gmns} 
\end{equation}
where $A_s(z)$ is the string frame warp factor and the indices $(\hat{m}, \hat{n})$ correspond to coordinates in the $5d$ flat metric. Then the action in \eqref{actionvec1} becomes
\begin{eqnarray}
    S_{2} = -  \int d^{4}x\int dz\, \dfrac{1}{4 g_{5}^{2}} e^{A_{s} - \Phi}{v_{\hat{m}\hat{n}}^{a}}^{2}\, . \label{actionvec2}
\end{eqnarray}

Varying the action \eqref{actionvec2}, in order to get the field equations, we will have both the contribution of the bulk action and the boundary   
\begin{eqnarray}
    \delta S_{2} = \delta S_{2}^{\textnormal{Bulk}} + \delta S_{2}^{\textnormal{Bdy}}
\end{eqnarray}
where
\begin{eqnarray}
    \delta S_{2}^{\textnormal{Bulk}} =  \int d^{4}x \int dz\, \delta V_{\hat{n}}^{a}\partial_{\hat{m}} \left( \dfrac{1}{g_{5}^{2}} e^{A_{s} - \Phi}v_{a}^{\hat{m}\hat{n}}\right)\, ,
\end{eqnarray}
and 
\begin{eqnarray}
    \delta S_{2}^{\textnormal{Bdy}} = -\int d^{4}x\, \int dz\, \partial_{\hat{m}}\left(\dfrac{1}{g_{5}^{2}}e^{A_{s} - \Phi}v_{a}^{\hat{m}\hat{n}}\delta V_{\hat{n}}^{a}\right)\, .\label{boundaryaction1}
\end{eqnarray}
Imposing periodic boundary conditions in the $x^{\mu}$ coordinates, the boundary term reduces to
\begin{eqnarray}
    \delta S_{2}^{\textnormal{Bdy}} = -\int d^{4}x\, \left(\dfrac{1}{g_{5}^{2}}e^{A_{s} - \Phi}v_{a}^{\hat{z}\hat{\mu}}\delta V_{\hat{\mu}}^{a}\right)_{z =\epsilon}^{z\to \infty}\, .\label{boundaryaction2}
\end{eqnarray}
As described in the previous section, in this work we consider holographic QCD models where the dilaton is quadratic far from the boundary, cf. \eqref{PhiLargez}. The string frame warp factor in that case becomes logarithmic far from the boundary, cf. \eqref{AsLargez}. Using these results we conclude that the surface term at $z \to \infty$ will vanish due to the presence of $e^{-\Phi}$.  Imposing  Dirichlet boundary condition for the fields $V_{\hat \mu}^c$ at the boundary $z =\epsilon$ one guarantees that $\delta S_{2}^{\textnormal{Bdy}} = 0$.

The vanishing of $\delta S_{2}^{\textnormal{Bulk}}$ leads us to the Euler-Lagrange equations 
\begin{eqnarray}
    \partial_{\hat{m}}\left(e^{A_{s} - \Phi}v_{a}^{\hat{m}\hat{n}}\right) = 0\, .\label{eqvec1}
\end{eqnarray}
These can be understood as a generalization of Maxwell equations for the fields $V_{\hat m}^a$ in a background with metric $g_{mn}^s$, given in \eqref{gmns} and a dilaton $\Phi(z)$. These equations are invariant under the gauge transformation
\begin{eqnarray}
    V_{\hat{m}, a}\rightarrow V_{\hat{m}, a} - \partial_{\hat{m}}\lambda_{V}^{a}\, .\label{gaugesymmetry}
\end{eqnarray}
We can write \eqref{eqvec1} in terms of the coordinates $z$ and $\hat{\mu}$ decomposing the gauge field $V_{\hat{m}}^a = (V_{z}^a, V_{\hat{\mu}}^a)$ and the derivatives $\partial_{\hat{m}} = (\partial_{z}, \partial_{\hat{\mu}})$. In this way, the equation \eqref{eqvec1} written in components is expressed as
\begin{align}
 &  \Big [ \partial_z + A_s' - \Phi' \Big ] \left ( \partial_z V^{\hat \mu,a} - \partial^{\hat \mu} V_z^ a \right )
+ \Box V^{\hat \mu,a} - \partial^{\hat \mu} \left (  \partial_{\hat \nu}  V^{\hat \nu,a} \right ) = 0 \, , \nonumber \\
& \Box V_z^a - \partial_z \left ( \partial_{\hat \mu} V^{\hat \mu , a} \right ) = 0 \, .\label{ELeqsfluctsv3}
\end{align}
The gauge symmetry \eqref{gaugesymmetry} allows us to define $V_{z}^a = 0$. The quadri-dimensional vector $V_{\hat{\mu}}$ admit the Lorentz decomposition
\begin{eqnarray}
    V_{\hat{\mu}, a} = V_{\hat{\mu}, a}^{\perp} + \partial_{\hat{\mu}}\xi^{a}
\end{eqnarray}
where $V_{\hat{\mu}, a}^{\perp}$ is the transverse vector field and $\xi^{a}$ are massless scalar fields not present in QCD. Since it is not possible to find normalisable modes for these fields we can set $\xi^{a}$ to zero.

Using these results the equations \eqref{ELeqsfluctsv3} reduce to
\begin{eqnarray}
    \Big [ \partial_z + A_s' - \Phi' \Big ] \partial_z V^{\hat \mu,a}_{\perp} + \Box  V^{\hat \mu,a}_{\perp}  = 0 \label{ELeqsfluctsv4}
\end{eqnarray}
where $ V^{\hat \mu,a}_{\perp}$ is the physical field that describe the vector mesons. Taking the 4d Fourier transform one obtains 
\begin{eqnarray}
    \Big [ \partial_z + A_s' - \Phi' \Big ] \partial_z V^{\hat \mu,a}_{\perp} -q^2  V^{\hat \mu,a}_{\perp}  = 0 \label{ELeqsfluctsv5}
\end{eqnarray}
\subsection{VEVs of the 4d flavour  currents}

In this subsection the vacuum expectation values (VEVs) of  the 4d flavour currents. We start by writing the boundary term \eqref{boundaryaction2} as
\begin{align}
\delta S_2^{Bdy} &=  \int d^4 x \Big[ \frac{1}{g_5^2} e^{A_s - \Phi} v^{\hat z \hat \mu}_a \delta V_{\hat \mu}^a \Big ]_{z=\epsilon} \,. 
\label{deltaS2bdyv3}
\end{align}
As described in the previous susbsection, the surface term at $z \to \infty$ vanishes due to the dilaton asymptotic behaviour. 
At small $z$ (near the boundary) we can approximate the metric by the AdS metric and solve the equation \eqref{ELeqsfluctsv4}. We find that the vector gauge field can be expanded at small $z$ as
\begin{align}
V_{\hat \mu,c}(x,z) &= V_{\hat \mu,c}^{(0)} (x) + \dots + V_{\hat \mu,c}^{(2)}(x) z^2 + \dots \, , \label{vectorUV}
\end{align}
where  $V_{\hat \mu,c}^{(0)} (x)$ are the 4d external sources and $V_{\hat \mu,c}^{(2)}(x)$ are the VEV coefficients.   The VEV of the flavour currents responsible for the creation of vector mesons, according to the holographic dictionary, is given by
\begin{align}
    \langle J^{\hat \mu,a} (x)  \rangle =  \frac{ \delta S_2^{o-s}}{\delta V_{\hat \mu,a}^{(0)}(x)} =   \frac{ \delta S_2^{Bdy}}{\delta V_{\hat \mu,a}^{(0)}(x)} =  \frac{1}{g_5^2} \Big [ e^{A_s - \Phi} v^{\hat z \hat \mu}_a  \Big ]_{z=\epsilon} 
= \frac{1}{g_5^2} \Big [ e^{A_s - \Phi} \partial_z V^{\hat \mu,a} \Big ]_{z=\epsilon} \, , \label{VEVcurrents}
\end{align}
where the last equality holds for the gauge $V_{z}^a=0$. 
Note that it is possible to write the VEVs in \eqref{VEVcurrents} in terms of the VEV coefficients $V_{\hat \mu,c}^{(2)}(x)$. In this work it will be sufficient to use the result \eqref{VEVcurrents}. Later in this section we will derive a Sturm-Liouville expansion for the vector fields that will lead to a spectral decomposition for the correlator of flavour currents.

\subsection{The on-shell action, the bulk to boundary propagator and the two point function} 

In this subsection we will write the on-shell action in terms of  bulk to boundary propagator and the 4d sources. We will establish the connection between the bulk to boundary propagator, the VEV of the 4d flavour currents and the correlator of flavour currents. 

First we evaluate the action in \eqref{actionvec2} on-shell and find
\begin{align}
    S_2^{o-s} &= S_{2,Bdy}^{o-s} + S_{2,Bulk}^{o-s}   \, ,
\end{align}
where
\begin{align}
    S_{2,Bdy}^{o-s} = - \int d^4 x \int dz \,   \partial_{\hat m} \Big [ \frac{1}{2 g_5^2} e^{A_s - \Phi}  v^{\hat m \hat n}_a V_{\hat n}^a \Big ] \, ,
\end{align}
and
\begin{align}
    S_{2,Bulk}^{o-s} = \int d^4 x \int dz \,  V_{\hat n}^a \partial_{\hat m} \Big ( \frac{1}{2 g_5^2} e^{A_s - \Phi} v^{\hat m \hat n}_a \Big )  = 0 \, .
\end{align}
We remind the reader that the indices $(\hat{m}, \hat{n})$ are raised or lowered using the 5d flat metric $\eta_{\hat m \hat n}$.  As expected, the on-shell action becomes a surface term. Using again periodic boundary conditions for the $x^{\mu}$ coordinates and the condition that the surface term at $z \to \infty$ vanishes, due to the asymptotic behavior of the dilaton, the on-shell action is reduced to
\begin{align}
    S_{2}^{o-s} &=  \int d^4 x \Big [ \frac{1}{2 g_5^2} e^{A_s - \Phi} (\partial_z V^{\hat \mu}_a) V_{\hat \mu}^a \Big ]_{z= \epsilon} \, , \label{onshell}
\end{align}
where we also used the gauge condition $V_z^a=0$.
We can define the bulk to boundary propagator in real space by the relation
\begin{align}
     V_{\hat \mu}^a (z,x) = \int d^4 y\, K_{\hat \mu \hat \nu}^{ab} (z, x; y) V^{\hat \nu,0}_b (y).\label{realvecpropagator}
\end{align}
where $ K_{\hat \mu \hat \nu}^{cd} (z, x; y)$  is  the bulk to boundary propagator (in real space) and $V^{\hat \nu,0}_b (y)$ is the 4d external source. Plugging \eqref{realvecpropagator} into \eqref{onshell} yields
\begin{align}
    S_{2}^{o-s} = \int d^4 x \int d^4 y \Big \{\frac{1}{2 g_5^2} V^{\hat \mu,0}_c (x) \Big [ e^{A_s - \Phi} \partial_z  K_{\hat \mu \hat \nu}^{cd} (z,x;y) \Big ]_{z= \epsilon} V^{\hat \nu,0}_d (y)\Big \} \, , \label{onshellv2}
\end{align}
The VEV of the flavour currents \eqref{VEVcurrents} can also be expressed in terms of the bulk to boundary propagator:
\begin{align}
    \langle J_{\hat \mu,c} (x) \rangle &= 
\frac{1}{g_5^2} \Big [ e^{A_s - \Phi} \partial_z V_{\hat \mu,c} \Big ]_{z=\epsilon} 
= \frac{1}{g_5^2} \int d^4 y \Big [ e^{A_s - \Phi} \partial_z  K_{\hat \mu \hat \nu}^{cd} (z,x;y) \Big ]_{z= \epsilon} V^{\hat \nu,0}_d(y) \, .
\end{align}
Varying the on-shell action in \eqref{onshellv2} we obtain
\begin{align}
    \delta S_2^{o-s} =  \int d^4 x\, \langle J_{\hat \mu , c}(x) \rangle  \delta V^{\hat \mu,0}_c\label{varyingonshell} \, ,
\end{align}
as expected. Note that we used the $x \leftrightarrow y$ symmetry in the bulk to boundary propagators. According to the holographic dictionary the correlator of flavour currents in real space corresponds to
\begin{align}
    G_{\hat \mu \hat \nu}^{cd} (x-y) &= \langle J_{\hat \mu, c} (x) J_{\hat \nu,d} (y)  \rangle = \frac{ \delta S_2^{o-s}}{ \delta V^{\hat \mu,0}_c(x) \delta V^{\hat \nu,0}_d(y)} = \frac{1}{g_5^2} \Big [ e^{A_s - \Phi} \partial_z  K_{\hat \mu \hat \nu}^{cd} (z,x;y) \Big ]_{z= \epsilon} \label{dictionvector}
\end{align}
The relation between the VEV and the source is expressed through the two-point function, given by
\begin{align}
    \langle J_{\hat \mu,c} (x) \rangle = 
 \int d^4 y \, G_{\hat \mu \hat \nu}^{cd} (x-y) V^{\hat \nu,0}_d(y)\, .
\end{align}

\subsection{Spectral decomposition for the bulk to boundary propagator}\label{decompositionvectormesons}

In subsection \ref{sub:actionvec} we saw that the vector mesons are described by a transverse vector field. This implies that the bulk to boundary propagator in momentum space takes the form
\begin{equation}
\tilde K_{\hat \mu \hat \nu}^{ab}(z,q) = P^{\mu \nu}_{\perp}(q)  \delta^{ab} \, V(z,q) \,, \label{Kmunu}
\end{equation}
where $P^{\mu \nu}_{\perp}(q)$ is the transverse projector, defined in \eqref{TProjector}, and $V(z, q)$ a scalar function that carries all the information of the the bulk to boundary propagator in momentum space. 

In momentum space, the 5d gauge fields can be written as 
\begin{equation}
\tilde V_{\hat \mu}^a (z, q) = \tilde K_{\hat \mu \hat \nu}^{ab}(z,q)  \tilde V_b^{\hat \nu,0}(q) \, .   
\end{equation}
Using these relations in the field equation \eqref{ELeqsfluctsv5} we obtain 
\begin{equation}
\Big [ \Big ( \partial_z + A_s' - \Phi' \Big ) \partial_z  -  q^2  \Big ] V(z,q)  =0 \,, \label{VqEq}
\end{equation}
which is an ordinary second order differential equation for the bulk to boundary propagator. 
It is  convenient to rewrite this equation as
\begin{equation}
\Big [ \partial_z \Big ( e^{A_s - \Phi} \partial_z \Big )  -  q^2 e^{A_s - \Phi} \Big ] V(z,q)  =0 \,.\label{VqEqL}
\end{equation}
This equation can be written as a Sturm-Liouville equation
\begin{equation}
\Big [ {\cal L} + \lambda \, r(z) \Big ] y(z) = f(z)   \quad  , \quad 
{\cal L} = \partial_z \Big ( p(z) \partial_z \Big )  - s(z) \, ,
\end{equation}
where we identify 
\begin{equation}
p(z) = e^{A_s - \Phi} \, \, , \, \, s(z) = 0 \, \, , \, \, \lambda = - q^2  \, \, , \, \, r(z) = e^{A_s - \Phi} \, \,  \text{and}  \, \,  f(z) = 0 \, \text{(homogeneous)} \, .
\end{equation}
The Sturm-Liouville theory is briefly described in appendix  \ref{App:sturmlitheory}.  In the non-homogeneous case, i.e. $f(z) \neq 0$,  we can define the Green's function by the equation
\begin{equation}
\Big [ {\cal L} + \lambda \, r(z) \Big ] G(z;z') = \delta (z - z') \,.\label{liouvillegreenvec}
\end{equation}
Now we will define an infinite set of eigenfunctions, $v^n(z)$, that obey the eigenvalue equation
\begin{equation}
\Big [ {\cal L} + \lambda_n r(z) \Big ] v^n(z)  = 0  \, ,
\end{equation}
or
\begin{eqnarray}
    \left[\partial_{z}\left(e^{A_{s} - \Phi}\partial_{z}\right) + m_{v^{n}}^2 e^{A_{s} - \Phi}\right]v^{n}(z) = 0
\end{eqnarray}
where $\lambda_{n} =  m_{v^{n}}^2$ are the eigenvalues. Note that these Sturm-Liouville modes are essentially the normalisable modes in holographic QCD.  Indeed, these modes satisfy the orthonormality condition 
\begin{equation}
 \int dz \, e^{A_{s} - \Phi}  v^m (z) v^n (z) = \delta^{mn} 
 \, , \label{normalvec}
\end{equation}
and the Green's function admits the spectral decomposition
\begin{equation}
G(z;z') =  -\sum_n \frac{ v^n(z) v^n(z')}{q^2 + m_{v^n}^2} \, . \label{SpectralVector}
\end{equation}
For more details, see appendix \ref{App:sturmlitheory}. 

We can find a relation between the bulk to boundary propagator $V(z,q)$, corresponding to the homogeneous solution, can be written in terms of the Green's function, associated with the non-homogeneous solution as follows.  

Multiplying both sides of \eqref{liouvillegreenvec} by  $V(q, z)$, integrating over $z$ and using \eqref{VqEqL} we obtain
\begin{eqnarray}
    V(z^{\prime}, q) = \Big [  e^{A_s(z)-\Phi(z)} ( V(z,q) \partial_z G(z;z') - G(z;z') \partial_z V(z,q)) \Big ]_{z=\epsilon}^{z\to \infty}
\end{eqnarray}
For a dilaton that is quadratic at large $z$, it is possible to show that the surface term at $z \to \infty$ vanishes so we end up with the relation
\begin{equation}
V(z',q) = - \Big [  e^{A_s-\Phi} \partial_z G(z;z') \Big ]_{z=\epsilon}   \, , \label{VqfromG}
\end{equation}
where we also used the boundary condition $V(\epsilon,q)=1$.  Substituting the spectral decomposition \eqref{SpectralVector} in \eqref{VqfromG}, we find
\begin{eqnarray}
    V(z',q) = \sum_n c_n (q^2) v^n(z') \, ,
    \label{propagatorcn}
\end{eqnarray}
where
\begin{eqnarray}
 c_n (q^2) =  \dfrac{\left[e^{A_{s} - \Phi}\partial_{z}v^{n}(z)\right]_{z =\epsilon}}{q^{2} + m_{v^{n}}^{2}}   \,. \label{cnvec}
\end{eqnarray}
Using this result and the orthonormality condition \eqref{normalvec} we obtain
\begin{eqnarray}
    \int dz\, e^{A_{s} - \Phi}v^{m}V(z, q) = c_{m}(q^{2})\label{propagatorcm}
\end{eqnarray}
replacing this result in \eqref{propagatorcn} we obtain the completeness relation for the normalisable (Sturm-Liouville) modes
\begin{eqnarray}
    \sum_{n} e^{A_{s} - \Phi}v^{n}(z)v^{n}(z^{\prime}) =\delta (z - z^{\prime})\, . \label{completenessvec}
\end{eqnarray}
Plugging \eqref{propagatorcn} into \eqref{Kmunu} the tensorial bulk to boundary propagator becomes
\begin{align}
\tilde K_{\hat \mu \hat \nu}^{ab}(z,q) &= 
 P^{\mu \nu}_{\perp}(q) \delta^{ab} \sum_n c_n(q^2) v^n(z) \, .\label{kmunumodes}
\end{align}

\subsection{The 4d current correlator} 

The 2-point current correlator in real space was obtained in \eqref{dictionvector} from the bulk to boundary propagator. In momentum space it takes the form
\begin{align}
    G_{ \hat \mu \hat \nu}^{ab}(q) &= \frac{1}{g_5^2} \Big [ e^{A_s - \Phi} \partial_z  \tilde  K_{\hat \mu \hat \nu}^{ab} (z,q) \Big ]_{z= \epsilon}\label{2vecpointmomenta} 
\end{align}
Using the spectral decomposition \eqref{kmunumodes} with the coefficients \eqref{cnvec}, the 2-point function becomes
\begin{align}
     G_{ \hat \mu \hat \nu}^{ab}(q) = \left ( \eta_{\hat \mu \hat \nu} - \frac{q_{\hat \mu} q_{\hat \nu}}{q^2}  \right ) \delta^{ab} \sum_n \frac{ F_{v^n}^2 }{q^2 + m_{v^n}^2}  \,,\label{2pointvecdecay}
\end{align}
where the coefficients $F_{v^n}$ are defined as
\begin{equation}
F_{v^n}= \frac{1}{g_5} \Big [ e^{A_s - \Phi} \partial_z  v_n(z)  \Big ]_{z= \epsilon} \, . \label{vecdecayconstants}
\end{equation}
The $F_{v^n}$ can be interpreted as probability amplitudes associated with the creation of vector mesons from the vacuum.  They are commonly known as vector meson decay constants because they are relevant for describing the weak decay of vector mesons. 

The result in \eqref{2pointvecdecay} is very general for holographic QCD models based on Einstein-dilaton gravity. It is consistent with \eqref{currentcorrel}, which is the spectral decomposition for a current correlator in large $N_c$ QCD.  Note the appearance in \eqref{2pointvecdecay} of 4d propagators for the vector mesons. The vector meson propagator can be obtained as a particular case of the Proca propagator, as described in appendix \ref{App:procapropagator}.

\subsection{Spectrum of vector mesons}

To obtain the spectrum of vector mesons we need to solve the eigenvalue problem for the normalisable (Sturm-Liouville) modes
\begin{eqnarray}
    \left[\partial_{z}\left(e^{A_{s} - \Phi}\partial_{z}\right) + m_{v^{n}}^2 e^{A_{s} - \Phi}\right]v^{n}(z) = 0 \label{ModesEqVM}
\end{eqnarray}
We can write this equation in the form of a Schrödinger equation considering the Bogoliubov transformation
\begin{eqnarray}
     v^n  = e^{- B_{V}}\psi_{V^n}  \quad \textnormal{where}\quad B_{V} =\dfrac{1}{2}\left(A_{s} - \Phi\right) \,.\label{Bogoliubovvec}
\end{eqnarray}
 Plugging \eqref{Bogoliubovvec} into \eqref{ModesEqVM}, we find the following Schrödinger equation
\begin{eqnarray}
    \Big [ \partial_z^2 + m_{v^{n}}^2 - V_V  \Big ] \psi_{V^n} = 0 \, ,\label{schrodingervec}
\end{eqnarray}
$V_{V}$ is expressed by
\begin{equation}
V_V = B_V'' + B_V'^2 \,. \label{potentialvec}
\end{equation}
From the Schrödinger equation, we can derive the mass spectrum and the wave functions associated with the normalisable (Sturm-Liouville) modes. Notice that the dominant contribution to the Schr\"odinger potential at large $z$ (far from the boundary) is given by the  dilaton which is quadratic in $z$ for large $z$. This in turn guarantees the condition that the Schr\"odinger potential is quadratic in $z$ at large $z$ leading to asymptotically linear Regge trajectories. 

\begin{figure}[htp!]%
    \centering
    {{\includegraphics[width=7cm]{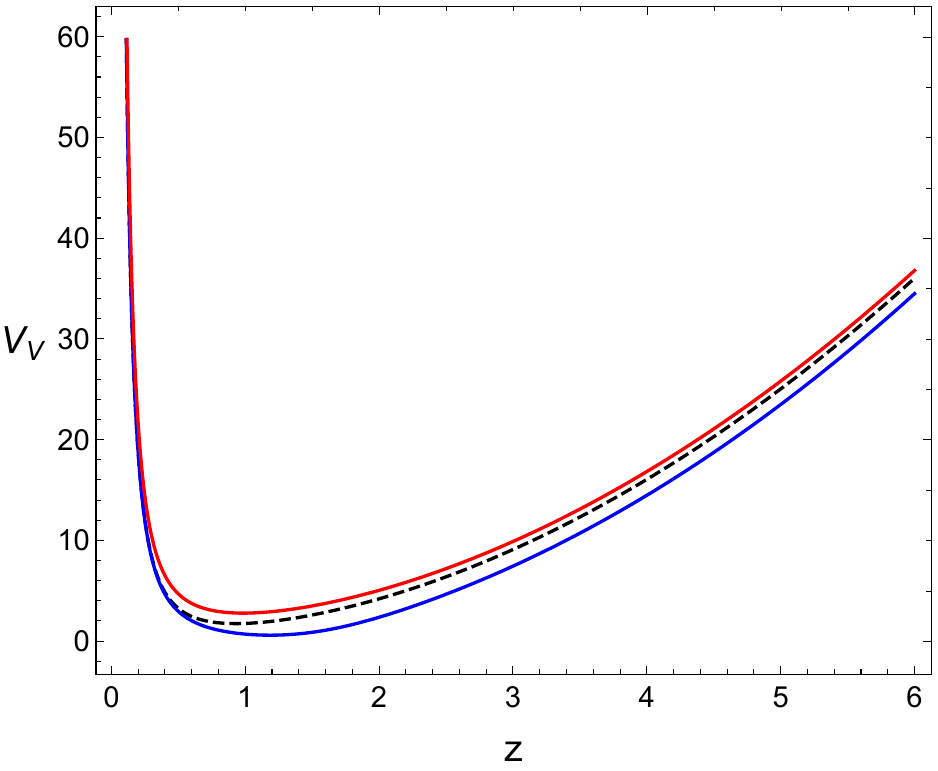}}}%
    
    \caption{Schr\"odinger potentials for vector mesons in Einstein- dilaton models I and II (blue and red solid lines) and soft wall model (black dashed line).}%
    \label{Plot:PotentialVM}%
\end{figure}

Figure \ref{Plot:PotentialVM} displays our numerical results for the Schr\"odinger potential for vector mesons in the Einstein-dilaton models I and II (blue and red lines respectively), compared against the soft wall model (black dashed line). Note that the potentials of the Einstein-dilaton models are very similar to the potential in the soft wall model. The main difference between them is that model I (model II) displays a miminum at lower (higher) energy than the soft wall model.  One may conclude from this analysis that model I (model II) leads to a lower (higher) mass for the fundamental state $\rho_0$. However, the mass of the fundamental state also depends on the infrared parameter $k$ which can be fixed differently for each model. In this work we will consider only dimensionless mass ratios so that we do not need to fix the infrared parameter $k$. 

\medskip 

{\bf Asymptotic solution and numerical integration}

\medskip 

In order to find the spectrum of vector mesons we need to solve the differential equation \eqref{ModesEqVM} or equivalently the Schr\"odinger equation \eqref{schrodingervec}. We first find the asymptotic solution at small $z$
\begin{equation}
v^n(z) = N_{V^n} z^2 + \dots \quad , \quad \text{or} \quad \psi_{V^n}(z) = N_{V^n} z^{3/2} + \dots \, ,  \label{vnsmallz}
\end{equation}
where $N_{V^n}$ is a constant necessary for the normalisation condition
\begin{equation}
 \int dz \, {\psi_{V^n}(z) }^2 = 1 \,. \label{normalvn}
\end{equation}

The eigenvalues of the problem can be obtained integrating numerically either \eqref{ModesEqVM} or \eqref{schrodingervec} and imposing the following behaviour at large $z$
\begin{equation}
\lim_{z \to \infty} \sqrt{z} \, \psi_{V^n} (z) = 0 \, ,    \label{largezcondVM}
\end{equation}
which guarantees that the solution is normalisable. The numerical procedure, commonly known as the shooting method, consists of shooting the value of $m_{V^n}$ until one finds a solution that satisfies the condition \eqref{largezcondVM}. In this way one finds a discrete set of eigenvalues corresponding to the vector meson masses.

\medskip 

{\bf Spectrum}

\medskip

We present in table \ref{Table:VectorMesonMasses} our results for the spectrum of vector mesons in the Einstein-dilaton models I and II described in section \ref{Sec:models}. As described above, we consider only dimensionless mass ratios so that we can compare different models without fixing the infrared parameter $k$. We show in table \ref{Table:VectorMesonMasses} our results for 
the mass ratios $m_{\rho^n}/m_{\rho^0}$ for the first five excited states, i.e.  $n=1,..,5$.  The mass of the fundamental state $m_{\rho^0}$ can later be fixed to the corresponding experimental value fixing the infrared parameter $k$. We compare our results for models I and II with previous results obtained using the soft wall  and hard wall models and also against experimental results.

\begin{table}[ht]
\centering
\begin{tabular}{l |c|c|c|c|c}
\hline 
\hline
Ratio & Model I & Model II & Soft wall  & Hard wall & Experimental \\
\hline 
 $m_{\rho^1}/m_{\rho^0}$ & 1.591  & 1.34   & 1.414  & 2.295 &  $1.652 \pm 0.048$  \\
 $m_{\rho^2}/m_{\rho^0}$ & 2.015  & 1.611  & 1.732  & 3.598   & $1.888 \pm 0.032$  \\
 $m_{\rho^3}/m_{\rho^0}$ & 2.365  &1.843  & 2 & 4.903  & $2.216\pm 0.026$ \\
 $m_{\rho^4}/m_{\rho^0}$ & 2.67  & 2.049 &  2.236 & 6.209   & $2.443 \pm 0.072$  \\
 $m_{\rho^5}/m_{\rho^0}$ & 2.944  &  2.236 &  2.45 & 7.514   & $2.727 \pm 0.265$  \\
\hline\hline
\end{tabular}
\caption{
Ratio of vector meson masses $m_{\rho^n}/m_{\rho^0}$ for the first excited states $n=1,..,5$ in the Einstein-dilaton models I and II, the soft wall model and the hard wall model, compared against experimental results.  The experimental result for $m_{\rho^1}$ was taken from \cite{OBELIX:1997zla} and the experimental results for the other states were obtained from PDG \cite{Workman:2022ynf}, including the mass of the fundamental state $m_{\rho^0} = 0.776 \pm 0.001 \, {\rm GeV}$. The numerical error in our computations of mass ratios in Einstein-dilaton models I and II was of the order of $10^{-6}$.
}
\label{Table:VectorMesonMasses}
\end{table}

\begin{figure}[htp!]%
    \centering
    {{\includegraphics[width=7cm]{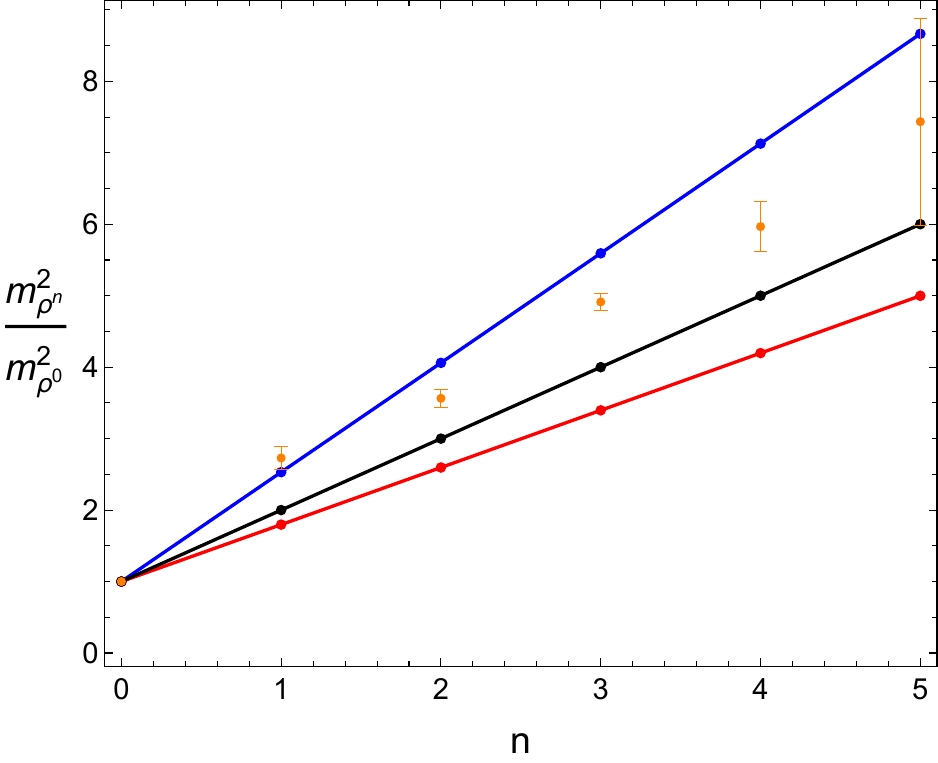}}}%
    
    \caption{Dimensionless squared mass ratios $m_{\rho^n}^2/m_{\rho^0}^2$ for vector mesons as a function of the radial excitation number $n$ in the Einstein-dilaton models I and II (blue and red solid lines and dots) and the soft wall model (black solid line and dots), compared against experimental data (orange dots and error bars).}%
    \label{Plot:MassRatioVM}%
\end{figure}

Figure \ref{Plot:MassRatioVM} shows the behaviour of the squared masses of vector mesons as a function of the radial excitation number in the Einstein-dilaton models I and II (blue and red solid lines) and the soft wall model (black dashed line), compared against experimental data (orange dots and error bars). As expected, the Einstein-dilaton models I and II lead to approximately linear Regge trajectories whilst the Regge trajectory in the soft wall model is exactly linear. The main difference between the Einstein-dilaton models and the soft wall model is that the masses of excited states grow faster (slower) in model I (model II) than in the soft wall model leading to a higher (smaller) slope. Note that the Einstein-dilaton model I and the soft wall model provide results that are closer to the experimental data.

\subsection{Wave functions and vector meson decay constants}

Besides calculating the spectrum, it is important to investigate the vector meson wave functions. This allows us to identify the emergence of the fundamental state $V^0$ and the excited states $V^{n}$ with $n=1,2, \dots$ by a comparison with normal modes in wave mechanics. From the small $z$ behaviour of the vector meson wave functions we will also be able to extract the vector meson decay constants $F_{V^n}$.  

\begin{figure}[htp!]%
    \centering
    {{\includegraphics[width=7cm]{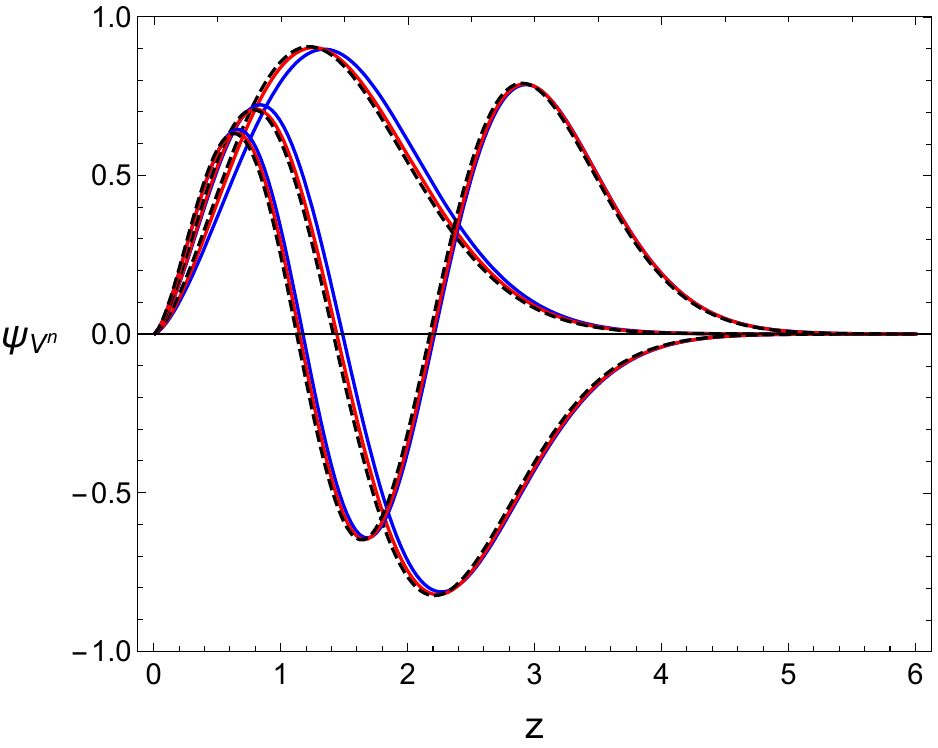} }}%
    
    \caption{Wave functions of vector mesons in the Einstein-dilaton models I and II (blue and red solid curves) and the soft wall model (black dashed curve).}
    \label{Plot:WaveFunctionsVM}
\end{figure}

Figure \ref{Plot:WaveFunctionsVM} illustrates the behavior of the vector meson wave functions in Einstein-dilaton models I and II (blue and red solid curves) and the soft wall model (black dashed curve). Note that the wave functions in models I and II are not very different to the wave functions in the soft wall model. The discrepancy occurs at small and intermediate values of $z$. This is expected because the Einstein-dilaton models affect the differential equation \eqref{ModesEqVM} through the dilaton and the AdS space deformation while in the soft wall model the AdS space is not deformed. At large $z$ the quadratic dependence of the dilaton field is expected to be the dominant contribution to the differential equation which is the same as in the soft wall model. 

We finally evaluate the vector meson decay constant as follows 
\begin{equation}
F_{v^n}= \frac{1}{g_5} \Big [ e^{A_s - \Phi} \partial_z  v_n(z)  \Big ]_{z= \epsilon} = \frac{2}{g_5} N_{V^n} \, , 
\end{equation}
where we used the small $z$ behaviour of the normalisable mode \eqref{vnsmallz}, the AdS asymptotic behaviour for $A_s$ and the property that the dilaton vanishes at the AdS boundary. The normalisation contants $N_{V^n}$ are calculated numerically using the normalisation condition \eqref{normalvn}. In table \ref{Table:VectorMesonDecayConstants} we present our results for the dimensionless ratios $\sqrt{F_{\rho^n}}/m_{\rho^0}$ for Einstein-dilaton models I and II, compared against the soft wall model and  the hard wall model. For the fundamental state we also compare against the experimental result. We conclude that, although the Einstein-dilaton model I provides a better result than the soft wall model, the hard wall model still provides the best result.  We would like to remark that the results for the vector meson decay constants in the case of excited states are theoretical predictions from holographic QCD. In particular, we note that all holographic QCD models predict that the vector meson decay constants grow with the radial excitation number. We hope that these predictions will be tested in the near future.

\begin{table}[ht]
\centering
\begin{tabular}{l |c|c|c|c|c}
\hline 
\hline
Ratio & Model I & Model II & Soft wall  & Hard wall &  Experimental \\
\hline 
$\sqrt{F_{\rho^0}}/m_{\rho^0}$  &  0.3719 & 0.283  & 0.3355  & 0.4246 &  $0.446 \pm 0.0019$  \\
$\sqrt{F_{\rho^1}}/m_{\rho^0}$ & 0.4704  & 0.3407 & 0.3989  & 0.7946    & -  \\
$\sqrt{F_{\rho^2}}/m_{\rho^0}$ & 0.5298  & 0.3798  & 0.4415  & 1.114    & - \\
$\sqrt{F_{\rho^3}}/m_{\rho^0}$ & 0.5741  & 0.41   & 0.4744  & 1.405    & - \\
$\sqrt{F_{\rho^4}}/m_{\rho^0}$ & 0.61  & 0.4351  & 0.5017  & 1.677     & - \\
\hline\hline
\end{tabular}
\caption{
Dimensionless ratios $\sqrt{F_{\rho^n}}/m_{\rho^0}$  for vector meson decay constants in the Einstein-dilaton models I and II, the soft wall model and the hard wall model, compared against the experimental result. The experimental result was obtained using $\sqrt{F_{\rho^0}} = 0.3462 \pm 0.0014 \, {\rm GeV}$  \cite{Donoghue:1992dd} and $m_{\rho^0} = 0.776 \pm 0.001 \, {\rm GeV}$ \cite{Workman:2022ynf}. The numerical error in our computations of $\sqrt{F_{\rho^n}}/m_{\rho^0}$ in Einstein-dilaton models I and II was of the order of $10^{-3}$.
}
\label{Table:VectorMesonDecayConstants}
\end{table}

\section{Nucleons in confining holographic QCD}\label{nucleonseinsteindilaton}

In this section, we describe $1/2$ spin baryons, more specifically the nucleons (proton and neutron). We first present the so-called Ioffe currents which are spinorial operators associated with the creation of nucleons and describe the spectral decomposition for the nucleon correlator in large $N_c$ QCD. Next, we present the 5d action for the Dirac spinor dual to the nucleon operator and derive the equations of motion. Subsequently, we study the VEV of nucleon operators and from the on-shell action we obtain the two-point nucleon correlation function. Next, we investigate the spectral decomposition for the bulk to boundary propagator using the Sturm-Liouville theory and we find a spectral decomposition for the nucleon correlator consistent with large $N_c$ QCD.  Finally, we obtain the spectrum and decay constants for nucleons for the Einstein dilaton models I and II described in section \ref{Sec:models} and compare against previous models and available experimental data.

\subsection{The 4d nucleon operator}
\label{Subsec:nucleonop}

Consider nucleons in large $N_c$ QCD with $N_f=2$ flavours. For simplicity we consider isospin symmetry, i.e. $m_u = m_d$. The creation of nucleon states can be described by nucleon operators built from the quark fields. For the case of the proton, the nucleon operator takes the form of the Ioffe current \cite{Ioffe:1981kw,Cohen:1994wm}
\begin{align}
    {\cal O}(x) &= \epsilon_{abc} \left ( u_a^T(x) C \gamma_{\mu} u_b(x) \right )\gamma_{5}\gamma^{\mu}d_c(x) \, , \label{nucleonop}
\end{align}
where $u$ and $d$ are the quark fields, $a$, $b$, and $c$ are color indices, and $C$ is the charge conjugation operator. The operator \eqref{nucleonop} has $I_z = 1/2$ corresponding to proton states. A similar operator can be constructed for the neutron states ($I_z = -1/2$) replacing the $u u d$ structure by a $d d u$ structure. 
The matrix element for the nucleon operator when applied to the vacuum and a nucleon state can be written as 
\begin{equation}
    \bra{0} {\cal O}(0) \ket{N^n(p)} =\lambda_{N^n}\, u^n(p)
\end{equation}
where $u^n(p)$ is the Dirac spinor corresponding to the nucleon state. The coupling $\lambda_{N^n}$ is associated with the probability amplitude of creating a particular nucleon state from the vacuum. Although there is no direct connection of these couplings to the weak decay of the neutron we will nevertheless call them nucleon decay constants. If the conformal dimension $\Delta$ of the nucleon operator ${\cal O}$  is equal to the canonical dimension we have $\Delta =9/2$ and the nucleon decay constant $\lambda_{N^n}$ has dimension of mass cubed \footnote{The nucleon states are normalised as $\langle \vec{p} \vert \vec{q} \rangle  = 2 E_{\vec{p}} \,  (2 \pi)^3 \delta^{3}(\vec{p}- \vec{q})$ and the Dirac spinors as $\bar u^r(p) u^s(p) = 2 m \, \delta^{rs} $. }. If we take into account the effect of the anomalous dimension one would obtain $\Delta < 9/2$ and $\lambda_n$ would have dimension $M^{\Delta - 3/2}$. In this work we will investigate the spectrum of nucleons for the cases $\Delta =9/2$ and $\Delta = 7/2$ using holographic QCD based on Einstein-dilaton gravity.

In large $N_c$ QCD the nucleon correlator  admits the following spectral decomposition \cite{Witten:1979kh,Leinweber:1995fn}
\begin{eqnarray}
    \langle \mathcal{O}(q)\bar{\mathcal{O}}(q')\rangle = i\delta^{4}(q - q')\sum_{n} \lambda_{N^n}^2 \dfrac{(i\slashed{q} + m_{N^n})}{q^{2} + m_{N^n}^{2}}\, . \label{nucleoncorrel}
\end{eqnarray}
On the right-hand side we identify the Dirac propagators associated with the different nucleon states. In holographic QCD we are interested on the two point correlation function of the right part of the nucleon correlator, namely
\begin{align}
\langle {\cal O}_R(q) \bar {\cal O}_R (q') \rangle
&= P_R  \langle {\cal O}(q) \bar {\cal O}(q') \rangle P_L  \nonumber \\
&= \delta^4 ( q - q') ( - P_R \slashed q ) \sum_n \frac{\lambda_{N^n}^2}{q^2 + m_{N^n}^2} \,,
\label{nucleoncorrelR}
\end{align}
where 
\begin{equation}
P_{R/L} = \dfrac{1}{2} \left(\mathds{1}\pm \gamma^5 \right)  \, ,
\end{equation}
are the right and left chiral projectors. The result in \eqref{nucleoncorrelR} was obtained previously in the soft wall model \cite{Abidin:2009hr}. In this section we will show that a general class of holographic QCD models based on Einstein-dilaton gravity lead to nucleon correlators that satisfy the spectral decomposition \eqref{nucleoncorrelR}.

\subsection{The 5d action and field equations}
\label{sub:actionnucleons}

We start with a 5d Dirac field $\psi(z,x)$ dual to the 4d nucleon operator ${\cal O}$. The dynamics of the 5d Dirac field can be obtained coupling the Dirac spinor to a background given by Einstein-dilaton gravity. The generalised 5d Dirac action action in the string frame can be written as
\begin{eqnarray}
    S_F =  G_F \int d^{5} x \sqrt{-g_s} \, e^{- \Phi}  \Big ( \frac{i}{2} \bar \psi  \slashed D \psi + \text{c.c.} - i \tilde{m} \bar \psi \psi \Big ) + \Delta S \, , \label{DiracAction}
\end{eqnarray}
where $\psi$ and $\bar{\psi}$ are the Dirac spinor and its adjoint respectively, with $\bar{\psi} =\psi^{\dagger} \Gamma^{\hat{0}}$. We have included a surface term $\Delta S$ necessary for the variational principle. The coupling $\tilde{m}$ is a generalisation of the mass term that may include first derivatives of the metric and the dilaton. The 5d  coupling constant $G_F$ will be determined later when comparing the result for the two point nucleon correlator at high energies with the perturbative QCD result. 

The covariant derivative in the Feynman notation is given by
\begin{eqnarray}
    \slashed{D} =\Gamma^{n}D_{n}\label{slashedderivative}
\end{eqnarray}
where the (curved space) gamma matrices, $\Gamma^{n}$, and the covariant derivative, $D_{n}$, explicitly are 
\begin{eqnarray}
    \Gamma^{n} &=& e_{\hat{a}}^{n}\Gamma^{\hat{a}}\,, \label{Gamman} \\
    D_{n} &=&\partial_{n} + \dfrac{1}{8}\omega_{n}^{\hat{a}\hat{b}}\left[\Gamma_{\hat{a}}, \Gamma_{\hat{b}}\right] =\partial_{n} + \dfrac{1}{4}\omega_{n}^{\hat{a}\hat{b}}\Gamma_{\hat{a}\hat{b}}\,. \label{Dn}
\end{eqnarray}
The quantities $e_{\hat{a}}^{n}$ and $\omega_{n}^{\hat{a}\hat{b}}$ are the veilbein and spin connection, respectively, and $\Gamma^{\hat a}$ are the gamma matrices in 5d flat space. For the string frame metric given in \eqref{gmns} they take the form 
\begin{align}
    e_{\hat{a}}^{n} &= e^{-A_{s}(z)}\delta_{\hat{a}}^{n}\,, \label{veilbein} \\
    \omega_{n}^{\hat{a}\hat{b}} &= e^{m\hat{a}}\nabla_{n}e_{m}^{\hat{b}}\, .\label{spinconection}
\end{align}
Note that $m,n$ are  tensorial indices associated with the curved space $g_s^{mn}$ whilst $\hat a,\hat b$ are tensorial indices associated with the tangent flat space $\eta_{\hat a \hat b}$. The gamma matrices in the tangent space satisfy the Clifford algebra
\begin{eqnarray}
    \{\Gamma^{\hat{a}}, \Gamma^{\hat{b}}\} = 2\eta^{\hat{a}\hat{b}}\mathds{1}\,.\label{clifford}
\end{eqnarray}
The coupling $\exp(-\Phi)$ in the Dirac action can be absorbed in the following  redefinition of the Dirac spinor
\begin{eqnarray}
    \psi \to e^{\Phi/2}\psi\label{spinorredef}\, .
\end{eqnarray}
Plugging \eqref{spinorredef} into action \eqref{DiracAction} we obtain
\begin{align}
S_F &=  G_F \int d^{5} x \sqrt{-g_s} \Big ( \frac{i}{2} \bar \psi  \slashed D \psi + \text{c.c.} 
 - i \tilde{m} \bar \psi \psi \Big ) + \Delta S \, . \label{DiracActionv2}
\end{align}
To find the equation of motion we first note that the only non-vanishing components of the spin connection are
\begin{eqnarray}
    \omega_{\mu}^{\hat{z}\hat{\nu}} = - \omega_{\mu}^{\hat{\nu}\hat{z}} = - A_{s}^{\prime}\delta_{\mu}^{\hat{\nu}}\, .\label{componentsspin}
\end{eqnarray}
 Using \eqref{Gamman}, \eqref{Dn}, \eqref{veilbein}  and \eqref{componentsspin}, the Dirac operator acting on the Dirac field takes the form
\begin{align}
\slashed{D} \psi = \Gamma^n D_n \psi &= e^{-A_s} \left ( \Gamma^{\hat a} \partial_{\hat a} 
+ 2  A_s' \Gamma^{\hat z}  \right ) \psi \, ,   \label{DiracOp}
\end{align}
where the indices $\hat a = (\hat z , \hat \mu)$ are contracted using the $5d$ Minkowski metric $\eta_{\hat a \hat b}$. Writing the action \eqref{DiracActionv2} in terms of the operator \eqref{DiracOp} we find
\begin{align}
S_{F} = G_F \int d^{5} x \, e^{4A_s}  \Big ( \frac{i}{2} \bar \psi  \Gamma^{\hat a} \partial_{\hat a} \psi  - \frac{i}{2} (\partial_{\hat a} \bar \psi) \Gamma^{\hat a} \psi  
- i  e^{A_s} \tilde m \bar \psi \psi \Big ) + \Delta S \, . \label{DiracActionv3}
\end{align}
The field equations are found by varying the action \eqref{DiracActionv3} with respect to $\psi$ and $\bar \psi$. We obtain
\begin{align}
& \Big ( \Gamma^{\hat a} \partial_{\hat a} 
+ 2 A_s' \Gamma^{\hat z} - e^{A_s} \tilde m \Big ) \psi  = 0 \, ,   \label{DiracEq}  \\
& \bar \psi \Big ( \overleftarrow{\partial_{\hat a}} \Gamma^{\hat a}    
+ 2 A_s' \Gamma^{\hat z} + e^{A_s} \tilde m \Big )  = 0 \, .   \label{DiracEqAdj}
\end{align}
It is interesting to write the equation \eqref{DiracEq} in terms of left and right chiralities of the Dirac field. Thus, in the decomposition $\psi =\psi_{R} + \psi_{L}$, the left and right components are given by
\begin{eqnarray}
    \psi_{R/L} =\dfrac{1}{2}\left(\mathds{1}\pm \Gamma^{\hat{z}}\right)\psi = P_{R/L}\psi\, \label{projectorpsi}\\
    \bar{\psi}_{R/L} =\bar{\psi}\dfrac{1}{2}\left(\mathds{1}\mp\Gamma^{\hat{z}}\right) =\bar{\psi}P_{L/R}\, \label{projectorpsibar}
\end{eqnarray}
where $P_{R}$ and $P_L$ are the right and left chiral  projectors. The left and right spinors are eigenstates of the chirality operator, $\Gamma^{\hat{z}} =\gamma^{5}$,
\begin{eqnarray}
    \Gamma^{\hat{z}}\psi_{R/L} = \pm \psi_{R/L}\label{chiraloperator}\,.
\end{eqnarray}
Plugging \eqref{projectorpsi}, \eqref{projectorpsibar} and \eqref{chiraloperator} in the Dirac equation, \eqref{DiracEq}, we arrive at the following system of coupled equations
\begin{align}
 \slashed \partial \psi_L=& -\Big ( \partial_z + 2 A_s' - e^{A_s} \tilde m \Big ) \psi_R \, , \label{DiracEqLR} \\
 \slashed \partial  \psi_R =&\Big ( \partial_z + 2 A_s' + e^{A_s} \tilde m \Big )
\psi_L  \, ,\label{DiracEqRL}
\end{align}
and a similar system for their adjoints. Acting on the right with the operator $\slashed{\partial}$ in \eqref{DiracEqLR} and \eqref{DiracEqRL}, we obtain the decoupled second-order differential equations
\begin{eqnarray}
    \Box\psi_{R/L} = -\left(\partial_{z} + 2A^{\prime}_{s} \pm e^{A_{s}}\tilde{m}\right)\left(\partial_{z} + 2A^{\prime}_{s} \mp e^{A_{s}}\tilde{m}\right)\psi_{R/L}\label{DiracEqv4}
\end{eqnarray}
 The general solutions for the left and right Dirac fields can be written as
\begin{eqnarray}
    \psi_{R/L}(x, z) = \int d^{4}q\, e^{iq\cdot x}\, F_{R/L}(q, z)\alpha_{R/L}(q)\label{fouriernucleons}
\end{eqnarray}
where $F_{R/L}(q, z)$ are the bulk to boundary propagators in momentum space for the right and left chiralities  whilst $\alpha_{R/L}(q)$ are left and right spinorial sources in the 4d field theory. Plugging \eqref{fouriernucleons} into \eqref{DiracEqv4}, we obtain the equation for the bulk to boundary propagator 
\begin{align}
&\Big [ \partial_z^2 + 4 A_s' \partial_z + 2 A_s'' + 4 A_s'^2 \mp \partial_z \Big ( e^{A_s}  \tilde m \Big ) - e^{2A_s} \tilde m^2 + Q^2 \Big ] F_{R/L} = 0 \,, \label{fLReq2ndv2}
\end{align}
where $Q =\sqrt{-q^{2}}$. 

Alternatively, we can expand the right and left Dirac fields in terms of 4d modes as follows
\begin{eqnarray}
    \psi_{R/L}(x, z) =\sum_{n} f_{R/L}^{n}(z) \alpha_{R/L}^{n}(x)\, . \label{MucleonsModeExp}
\end{eqnarray}
The 4d modes $\alpha^{n}_{R/L}(x)$ satisfy the coupled equations 
\begin{eqnarray}
   \slashed \partial \alpha_{R/L}^n = m_{N^n} \, \alpha_{L/R}^n \, ,   
\end{eqnarray}
which are equivalent to the Dirac equation
\begin{eqnarray}
(\slashed \partial - m_{N^n} ) \alpha^n = 0 \, ,   
\end{eqnarray}
for the 4d Dirac spinor  modes $\alpha^n(x) = \alpha_R^n(x) + \alpha_L^n(x)$. Using these results in \eqref{DiracEqLR} we find that the normalisable modes $f_{R/L}^{n}(q, z)$ obey the system of coupled equations
\begin{align}
\Big ( \partial_z + 2 A_s' \mp e^{A_s} \tilde m \Big ) f_{R/L}^n  = \mp m_{N^n} \,  f_{L/R}^n \,,\label{fLReq1st}
\end{align}
The second order decoupled equations for these normalisable modes take the form
\begin{align}
&\Big [ \partial_z^2 + 4 A_s' \partial_z + 2 A_s'' + 4 A_s'^2 \mp \partial_z \Big ( e^{A_s}  \tilde m \Big ) - e^{2A_s} \tilde m^2 + m_{N^n}^2 \Big ] f_{R/L}^n = 0 \,. \label{fnlr}
\end{align}
Note that the equations \eqref{fnlr} can be thought as the eigenvalue equations associated with the bulk to boundary propagator satisfying equation \eqref{fLReq2ndv2}.  

In subsection \ref{Subsec:Sdbp}  we will apply the Sturm-Liouville theory to the equations \eqref{fLReq2ndv2} in order to arrive at a spectral decomposition for the bulk to boundary propagator and in subsection \ref{Subsec:fermioniccorrelator} we will obtain the spectral decomposition of the 4d nucleon correlator. In  subsection \ref{spectrumnucleons} we will use the equations \eqref{fnlr} to  find the spectrum of nucleons. But first we will obtain in the following two subsections the VEV of the 4d nucleon operator as well as the dictionary for the nucleon correlator in terms of the bulk to boundary propagator.

\subsection{VEV of the 4d nucleon operator} 

In this subsection we  obtain the holographic dictionary for the VEV of the right projection of the nucleon operator, namely 
\begin{equation}
\langle {\cal O}_R(x) \rangle = P_R \langle {\cal O}(x) \rangle \, ,    
\end{equation}
from the 5d action. The key observation is that this operator couples to a left spinorial source $\alpha_L(x)$ as
\begin{eqnarray}
\int d^4 x  \, \Big ( \bar \alpha_L (x) \langle {\cal O}_R (x) \rangle + c.c. \Big ) \, .
\end{eqnarray}
The 4d spinorial source $\alpha_L(x)$ will appear as the leading term coefficient in the small $z$ (UV) expansion of the 5d left spinor field $\psi_L(x,z)$. 
As described at the beginning of the section, the nucleon operator $\langle {\cal O}_R(x) \rangle$ has conformal dimension $\Delta$. We will consider the cases $\Delta = 9/2$ (canonical dimension) and $\Delta = 7/2$ (including anomalous dimension).  The 4d source $\alpha_L(x)$ have conformal dimension $4 - \Delta$.

Let us start with the variation of the action 

\begin{equation}
    \delta S_F = \delta S^{bulk} + \delta S^{Bdy} \, ,
\end{equation}
where
\begin{align}
\delta S^{Bulk} &=  G_F \int d^5 x\, \Big (i  e^{4 A_s} \delta \bar \psi ( \Gamma^{\hat a} \partial_{\hat a} + 2 A_s' \Gamma^{\hat z} - \tilde m e^{A_s} ) \psi + c. c.  \Big ) = 0 \, , 
\end{align}
and
\begin{align}
\delta S^{Bdy} &= G_F \int d^5 x\, \partial_{\hat a} \left ( - \frac{i}{2} e^{4A_s} \delta \bar \psi \Gamma^{\hat a} \psi 
+ c.c.  \right ) + \delta ( \Delta S ) \,. 
\end{align}
Imposing periodic boundary condition in the $x^{\mu}$ coordinates and using the property that the spinor field solution decays fast enough at $z\to \infty$ vanishes, the on-shell variation reduces to
\begin{align}
\delta S_F &= G_F \int d^4 x  \left (  \frac{i}{2} e^{4A_s} \delta \bar \psi \Gamma^{\hat z} \psi 
   +\, c.c. \right )_{z = \epsilon}\, + \delta ( \Delta S ) \,.\label{varyingsf}  
\end{align}
Decomposing the Dirac spinor in their chiralities we obtain
\begin{align}
\delta S_F = G_F \int d^4 x \left ( \frac{i}{2} e^{4 A_s} \delta \bar \psi_L \psi_R - \frac{i}{2} e^{4 A_s}  \delta \bar \psi_R \psi_L \right )_{z=\epsilon} + c.c.  + \delta (\Delta S) \,.\label{varyingf}
\end{align}
The left and right chiralities of the Dirac field are coupled, which means that it is impossible to fix them simultaneously. As a result, we need to select one of the chiralities. In order to fix the left component, we define the surface term $\Delta S$ as
\begin{align}
\Delta S &= G_F \int d^4 x \left ( \sqrt{- \gamma}  \frac{i}{2} \bar \psi \psi \right )_{z=\epsilon} \nonumber \\
&= G_F \int d^4 x \left ( \frac{i}{2}  e^{4 A_s}  ( \bar \psi_L \psi_R + \bar \psi_R \psi_L ) \right )_{z=\epsilon} \, .\label{ctactionnucleons}
\end{align}
Varying this surface term we obtain
\begin{align}
\delta (\Delta S) &= G_F \int d^4 x \left ( \frac{i}{2} e^{4 A_s} ( \delta \bar \psi_L \psi_R + \delta \bar \psi_R \psi_L ) \right )_{z = \epsilon} + c. c \, .\label{varyingct}
\end{align}
Plugging \eqref{varyingct} into \eqref{varyingf} we obtain the final result for the variation of the action
\begin{align}
\delta S_F &= G_F \int d^4 x \left ( i \, e^{4 A_s} \delta \bar \psi_L \psi_R \right )_{z=\epsilon} + c.c. \nonumber \\
&= \int d^4 x \left ( \delta \bar \psi_L  \Pi_R +  \bar \Pi_R \delta \psi_L \right )_{z=\epsilon} \,,\label{varyingfmomenta}
\end{align}
where we introduced the conjugate momenta
\begin{align}
 \Pi_R = i \, G_F e^{4 A_s} \psi_R  \quad , \quad 
\bar \Pi_R = i \, G_F e^{4 A_s} \bar \psi_R \, .  
\end{align}
Note from \eqref{varyingfmomenta} that fixing the left spinor at the boundary is now consistent with the variational principle. 
Solving at small $z$ (near the AdS boundary) the second order differential equations \eqref{DiracEqv4} for the left and right components we find
\begin{align}
\psi_{L}(x,z) &=   \alpha_L(x) z^{2-m} + \dots + \beta_L (x) z^{3+m} + \dots  \, , \nonumber \\
\psi_{R} (x,z) &= \alpha_R(x) z^{3-m} + \dots + \beta_R (x) z^{2+m} +  \dots \,,  \label{nucleonsUV}
\end{align}
where $m$ is the constant mass which is the asymptotic value of the 5d mass coupling $\tilde m(z)$ in the limit $z \to 0$ (near the AdS boundary). The 4d spinors $\alpha_{L}(x)$ and $\alpha_{R}(x)$ are the source coefficients associated with the non-normalisable sector of the 5d spinors $\psi_{L}(x, z)$ and $\psi_{R}(x, z)$ respectively. The 4d spinors $\beta_{L}(x)$ and $\beta_{R}(x)$ are the VEV coefficients corresponding to the normalisable sector of the 5d spinors $\psi_{L}(x, z)$ and $\psi_{R}(x, z)$ respectively.

As described above, we take $\alpha_L(x)$ as the only independent 4d source. Note that it has conformal dimension $2+ m$ since the 5d spinor has conformal dimension zero near the AdS boundary.  This source couples to the operator $O_R$ of conformal dimension $\Delta = 2 + m$ so we can find the VEV of this operator using the holographic dictionary. From the action variation in \eqref{varyingfmomenta} we obtain
\begin{align}
\langle {\cal O}_R \rangle &=  \frac{ \delta S_F}{\delta \bar \alpha_L}  =    \left ( z^{2-m}  \Pi_R \right )_{z=\epsilon} \nonumber \\
&= i \, G_F \left ( z^{2-m} e^{4 A_s} \psi_R \right )_{z=\epsilon} 
\label{SpinorOs}
\end{align}
Using \eqref{DiracEqRL} we can write the result in \eqref{SpinorOs} in terms of the left spinor:
\begin{align}
\langle {\cal O}_R \rangle = i \, G_F\left ( z^{2-m} e^{4 A_s} \dfrac{\slashed \partial}{\partial^{2}} (\partial_z + 2 A_s' + e^{A_s} \tilde m  ) \psi_L \right )_{z= \epsilon} \,. 
\label{SpinorOsv2}
\end{align}
The VEV, according to the results \eqref{SpinorOs} and \eqref{SpinorOsv2}, is the one-point function in the presence of the 4d source $\bar{\alpha}_{L}$. In the next subsection we will obtain the two-point function of the nucleon operator from the bulk to boundary propagator and will find a relation with the VEV. 

\subsection{The on-shell action, the bulk to boundary propagator and the two point function}
\label{onshellbulkpropagator} 

The on-shell and the bulk to boundary propagator of the Dirac field allow us to obtain the VEV and the two-point function for nucleons. Our starting point is the Dirac action in \eqref{DiracActionv3} with the additional surface term given in \eqref{ctactionnucleons}. Evaluating this action on-shell we obtain
\begin{align}
 S_F^{o-s} &= S_{Bulk}^{o-s} + S_{Bdy}^{o-s}    \, ,
\end{align}
where the bulk action is given by
\begin{align}
S_{Bulk}^{o-s} &= G_F \int d^5 x \, e^{4 A_s}  \Big ( \frac{i}{2} \bar \psi ( -2 A_s' \Gamma^{\hat z}  + e^{A_s} \tilde m ) \psi \nonumber \\
&- \frac{i}{2} \bar \psi ( -2 A_s' \Gamma^{\hat z}  - e^{A_s} \tilde m ) \psi  - i  e^{A_s} \tilde m \bar \psi \psi \Big )\label{nucleonsbulkaction}
= 0 \, ,
\end{align}
and the boundary action is
\begin{align}
 S_{Bdy}^{o-s} &= \Delta S^{o-s} = G_F \int d^4 x \left ( \frac{i}{2} e^{4 A_s} \bar \psi_L \psi_R + c.c. \right )_{z=\epsilon} \nonumber \\
 &=  G_F \int d^4 x \left ( \frac{i}{2} e^{4 A_s} \bar \psi_L \,  \dfrac{\slashed \partial}{\partial^{2}} (\partial_z + 2 A_s' + e^{A_s} \tilde m) \psi_L + c.c. \right )_{z=\epsilon}\label{nucleonsBdyaction}
 \, .
\end{align}
Note that in \eqref{nucleonsbulkaction} we used the equations \eqref{DiracEq}-\eqref{DiracEqAdj} for the Dirac field and in \eqref{nucleonsBdyaction} we used  \eqref{DiracEqRL}. 

The bulk to boundary propagator written in coordinate space can be expressed by the following relation
\begin{equation}
\psi_L(z,x) = \int d^4 y \, F_L (z,x ; y)\, \alpha_L (y) \,, \label{DiracBulkToBdy} 
\end{equation}
where $\psi_{L}(z, x)$ is the left component of the Dirac field in 5d, $F_L (z,x ; y)$ is a real scalar representing bulk to boundary propagator and $\alpha_L (y)$ is the  4d left spinorial source. Substituting \eqref{DiracBulkToBdy} in \eqref{nucleonsBdyaction} and \eqref{nucleonsbulkaction}, the on-shell action becomes
\begin{align}
 S_F^{o-s} 
 &=  G_F \int d^4 x \int d^4 y \Big  ( \frac{i}{2} \bar \alpha_L (x)  \left ( z^{2-m} e^{4 A_s} \,  \dfrac{\slashed \partial_{x-y}}{\partial^{2}} (\partial_z + 2 A_s' + e^{A_s} \tilde m) F_L(z, x; y) \right )_{z=\epsilon} \nonumber \\
 & \alpha_L(y) + c.c. \Big )  \, .\label{actionfnucleons}
\end{align}
where we also used the asymptotic behavior \eqref{nucleonsUV}. 

Note that the VEV in \eqref{SpinorOsv2} can be written in terms of the bulk to boundary propagator as
\begin{align}
\langle {\cal O}_R (x) \rangle &= i \, G_F \int d^4 y \left ( z^{2-m} e^{4 A_s} \dfrac{\slashed \partial_{x-y}}{\partial^{2}} (\partial_z + 2 A_s' + e^{A_s} \tilde m  ) F_L (z, x ; y) \right )_{z= \epsilon} \alpha_L (y)  \,. 
\label{SpinorOsv3}
\end{align}
Varying the on-shell action we obtain
\begin{equation}
\delta S_F^{o-s} = \int d^4 x \left (   \delta \bar \alpha_L (x) \langle  {\cal O}_R (x) \rangle  + c.c. \right ) \, ,
\end{equation}
as expected. Varying once more we obtain the two-point function
\begin{align}
\Gamma_R (x-y) = \langle {\cal O}_R (x) \bar {\cal O}_R (y) \rangle &=\frac{ \delta S_F^{o-s}}{ \delta \bar \alpha_L (x) \delta \alpha_L (y)} 
 =  P_{R}\frac { \delta \langle \bar {\cal O}_R (y) \rangle }{ \delta \bar \alpha_L (x)} \nonumber \\
&=  i G_F  P_{R} \dfrac{\slashed \partial_{x-y}}{\partial^{2}} \left ( z^{2-m} e^{4 A_s}  (\partial_z + 2 A_s' + e^{A_s} \tilde m  ) F_L (z, x ; y) \right )_{z= \epsilon} \,.
\end{align}
The relation between the one-point and two-point functions is
\begin{equation}
\langle {\cal O}_R (x) \rangle = \int d^4 y \, \Gamma_R (x-y)\, \alpha_L (y) \,. 
\end{equation}
 The equation \eqref{DiracBulkToBdy} can be written in momentum space as
\begin{equation}
\psi_L (z,q) = F_L (z,q)\,  \alpha_L (q) \, . 
\end{equation}
The VEV \eqref{SpinorOsv3} in momentum space takes the form
\begin{equation}
\langle {\cal O}_R (q) \rangle =\Gamma_R (q)\, \alpha_L (q) \,,
\end{equation}
where
\begin{equation}
 \Gamma_R (q) =  - G_F  P_{R}\dfrac{\slashed q}{Q^{2}}\left ( z^{2-m} e^{4 A_s} (\partial_z + 2 A_s' + e^{A_s} \tilde m  ) F_L (z, q) \right )_{z= \epsilon}\, . \label{GReq}
\end{equation}
We end this subsection fixing the coupling constant $G_F$ that characterises the 5d Dirac action. In order to do that we evaluate the correlator \eqref{GReq} in the limit $q^2 \to \infty$ (UV). In this limit the 4d theory becomes conformal and  the bulk to boundary propagator can be approximated by the (analytical) solution corresponding to 5d AdS space. For $m$ half-integer we find that
\begin{equation}
 \Gamma_R (q) =  G_F a_m  P_{R} \, \slashed{q} \,  q^{2m-1} \ln q^2  \, , \label{GReqAdS}
\end{equation}
where
\begin{equation}
a_m = \frac{(-1)^{m-1/2}}{2^{2m} {\Gamma (m+1/2)}^2} \,.    
\end{equation}
For $m=5/2$ we have $\Delta =9/2$, which is the canonical dimension of the nucleon operator. In this case we can compare against the perturbative QCD result \cite{Cohen:1994wm}
\begin{equation}
 \Gamma_R (q) = \frac{1}{64 \pi^4}  P_{R} \, {\slashed q} \, q^4 \ln q^2 \quad \text{(perturbative QCD)}\, , \label{GRpQCD}
\end{equation}
and obtain
\begin{equation}
G_F = \frac{2}{\pi^4} \,.   \label{GF} 
\end{equation}
In the following subsections we will obtain a spectral decomposition for the bulk to boundary propagator using Sturm-Liouville theory. From this result we will finally obtain the spectral decomposition of the nucleon correlator. The spectrum of nucleons then will be obtained from the eigenvalue problem and the nucleon decay constants will be extracted from the coefficients of the spectral decomposition. 

\subsection{Spectral decomposition for the bulk to boundary propagator}
\label{Subsec:Sdbp}

In this subsection, we will use Sturm-Liouville theory to find a spectral decomposition for the bulk to boundary propagator. We will proceed in a similar way as in the case of vector mesons, described in subsection \ref{decompositionvectormesons}. 

We start writing the equation in \eqref{fLReq2ndv2} for the left bulk to boundary propagator in the following form: 
\begin{eqnarray}
    \left[\left(\partial_{z} + 4A_{s}^{\prime}\right)\partial_{z} + \theta_{L} + Q^{2}\right]F_{L}(q, z) = 0\label{eq1}
\end{eqnarray}
where
\begin{eqnarray}
    \theta_{L} = 2A_{s}^{\prime\prime} + 4A_{s}^{\prime\, 2} + \partial_{z}(e^{A_{s}}\tilde{m}) - e^{2A_{s}}\tilde{m}^{2}\, .
\end{eqnarray}
Rewriting \eqref{eq1} as
\begin{eqnarray}
\left[\partial_{z}\left(e^{4A_{s}}\partial_{z}\right) + e^{4A_{s}}\theta_{L} + Q^{2}e^{4A_{s}}\right]F_{L}(q, z) = 0,\label{sturmnucleons}
\end{eqnarray}
we identify this equation with the Sturm-Liouville equation
\begin{equation}
\Big [ {\cal L}_L + \lambda \, r(z) \Big ] y(z) = f(z)   \quad  , \quad 
{\cal L}_L = \partial_z \Big ( p(z) \partial_z \Big )  - s_L(z) \, ,
\end{equation}
where 
\begin{eqnarray}
    p(z) = r(z) = e^{4A_{s}},\quad s_L(z) = - e^{4A_{s}}\theta_{L},\quad \lambda = Q^{2} \quad \textnormal{and} \, \, f(z) = 0 \,  \text{(homogeneous)}.
\end{eqnarray}
The Sturm-Liouville theory is briefly described in appendix \ref{App:sturmlitheory}.
The  Green's function $G_{L}(z; z^{\prime})$ corresponding to the non-homogeneous case satisfies the equation
\begin{eqnarray}
    \left[\mathcal{L}_{L} + \lambda \, r(z)\right]G_{L}(z; z^{\prime}) =\delta(z - z^{\prime})\label{liouvillegreennucleons}
\end{eqnarray}
Again, it is convenient to define an infinite set of eigenfunctions $f_L^n(z)$ by the eigenvalue equation
\begin{eqnarray}
    \left[\mathcal{L}_{L} + \lambda_{n} r(z)\right] f_L^n(z) = 0 \, ,
\end{eqnarray}
or 
\begin{eqnarray}
\left[\partial_{z}\left(e^{4A_{s}}\partial_{z}\right) + e^{4A_{s}}\theta_{L} + \lambda_n e^{4A_{s}}\right] f_L^n(z) = 0 \, ,
\end{eqnarray}
where $\lambda_{n} = m_{N^n}^2$ are the eigenvalues. Comparing this equation with \eqref{fnlr} we see that the Sturm-Liouville modes are the normalisable modes in holographic QCD. These modes satisfy the  orthonormality condition
\begin{equation}
\int dz \, e^{4A_{s}} f_L^m(z) f_L^n(z) = \delta^{mn} \,,
\end{equation}
and the Green's function admits the following spectral decomposition
\begin{equation}
G_L(z;z') = \sum_n  \frac{ f_L^n(z)  f_L^n(z')}{q^{2} + m_{N^n}^{2}} \,. \label{GLdecomp}
\end{equation}
For more details, see appendix \ref{App:sturmlitheory}.

Multiplying both sides of \eqref{liouvillegreennucleons} by $F_L(q,z)$, integrating over $z$ and using \eqref{sturmnucleons} we obtain the relation between the bulk to boundary propagator and the Green's function:
\begin{align}
F_L(q,z') = \Big [ e^{4A_s(z)} \Big ( F_L(q,z)  \partial_z G_L(z;z')  - G_L(z;z') \partial_z F_L(q,z) \Big ) \Big ]_{z=\epsilon}^{\infty} \,.\label{wronskiannucleonsprime}
\end{align}
Assuming that $F_L(q,z)$ and $G_L(z;z')$ vanish sufficiently fast in the limit $z \to \infty$  and using the spectral decomposition \eqref{GLdecomp} we obtain
\begin{align}
F_L(q,z') &=  \sum_n \frac{  f_{L,n}(z')}{q^{2} + m_{N^n}^{2}}  \Big [ e^{4A_s(z)} \Big ( F_L(q,z)   f_{L,n}'(z)   - f_{L,n}(z)  F_L'(q,z) \Big ) \Big ]_{z=\epsilon}\label{wronskiannucleons}
\end{align}
From \eqref{fouriernucleons} and \eqref{nucleonsUV} we see that $F_L(q,z)$ behaves as $z^{2-m}$ at small $z$. Using also the asymptotic behaviour for the warp factor the equation \eqref{wronskiannucleons} reduces to
\begin{align}
F_L(q, z') = \sum_n \frac{  f_n m_{N^n} f_{L,n}(z')}{q^2 + m_{N^n}^2 }  \, ,\label{spectralFL} 
\end{align}
where we also used the coupled equations \eqref{fLReq1st} and the coefficients $f_n$ are defined as
\begin{equation}
f_n = \Big [ z^{-2-m}  f_{R,n}(z) \Big ]_{z= \epsilon} \,.  \label{fn}
\end{equation}
In the next subsection we will relate these coefficients correspond to the nucleon decay constants. 
Lastly, it is easy to show that the Sturm-Liouville modes satisfy the completeness relation 
\begin{eqnarray}
    \sum_{n} e^{4A_{s}}f_{L, n}(z)f_{L, n}(z^{\prime}) =\delta (z - z^{\prime})\, .
\end{eqnarray}
For more details see appendix \ref{App:sturmlitheory}.
In the following subsection we will obtain the spectral decomposition forthe 4d nucleon correlator and show the compatibility with the spectral decomposition expected in large $N_c$ QCD.
\subsection{The 4d nucleon correlator}
\label{Subsec:fermioniccorrelator}

In subsection \ref{onshellbulkpropagator} we obtained  the holographic dictionary \eqref{GReq} that relates the 4d nucleon correlator   $\Gamma_R(q)$ to the 5d bulk to boundary propagator $F_L(z,q)$. In subsection \ref{Subsec:Sdbp} we obtained the spectral decomposition \eqref{spectralFL} for the bulk to boundary propagator. Then using \eqref{GReq} and \eqref{spectralFL} we finally obtain the spectral decomposition for the nucleon correlator:
\begin{align}
 \Gamma_R (q) = - P_{R}\dfrac{\slashed{q}}{Q^{2}}\sum_n \frac{\lambda_{N^n}^2 m_{N^n}^2}{ q^2 + m_{N^n}^2}\, ,
\label{GReqv2}
\end{align}
where 
\begin{equation}
\lambda_{N^n} = \sqrt{G_F} f_n = \sqrt{G_F} \Big [ z^{-2-m}  f_{R,n}(z) \Big ]_{z= \epsilon} \, ,    \label{lambdaNn}
\end{equation}
 and we also used the coupled equations \eqref{fLReq1st}. 
It is interesting to write the correlator in \eqref{GReqv2} as
\begin{eqnarray}
    \Gamma_{R}(q) = - P_{R}\slashed{q}\left(\dfrac{1}{Q^{2}}\sum_{n} \lambda_{N^n}^2 + \sum_{n}\dfrac{\lambda_{N^n}^2}{q^{2} + m_{N^n}^{2}}\right)\, . \label{correlatorrewritten}
\end{eqnarray}
The first term in \eqref{correlatorrewritten} diverges. This UV divergence is expected since we have worked with the original on-shell action without introducing holographic renormalisation. Subtracting this UV divergence we obtain the renormalised correlator 
\begin{eqnarray}
    \Gamma_{R}^{\textnormal{ren}}(q) = - P_{R}\slashed{q}\sum_{n}\dfrac{\lambda_{N^n}^{2}}{q^{2} + m_{N^n}^{2}}\, . \label{GRren}
\end{eqnarray}
This final result \eqref{GRren} for the nucleon correlator is valid for a general class of holographic QCD models based on Einstein-dilaton gravity and it is consistent with the spectral decomposition \eqref{nucleoncorrelR} obtained in large $N_c$ QCD. The coefficients $\lambda_{N^n}$ defined in \eqref{lambdaNn} are therefore identified with the nucleon decay constants.

\subsection{Spectrum of nucleons}\label{spectrumnucleons}

In this subsection we obtain the spectrum of nucleons solving the eigenvalue equation \eqref{fnlr} for the normalisable modes. Before doing that it is interesting to rewrite \eqref{fnlr} as Schr\"odinger equations and investigate the corresponding Schr\"odinger potentials. 

Using the Bogoliubov transformation
\begin{eqnarray}
    f_{R/L}^n(z) = e^{- 2A_s (z)} \xi_{R/L}^{n} (z)\label{Bogoliubovnucleons} \, ,
\end{eqnarray}
 in \eqref{fnlr} we obtain
\begin{equation}
\left [ - \partial_z^2  + V_{R/L} \right ] \xi_{R/L}^{n} =  m_{N^n}^{2} \xi_{R/L}^{n} \label{fnlrSchrod}
\end{equation}
where the Schr\"odinger potential $V_{R/L}$ are given by
\begin{align}
V_{R/L} &= \pm \partial_z \Big ( e^{A_s} \tilde m \Big ) + ( e^{A_s} \tilde m )^2  \,. \label{fLRSchrodPot}
\end{align}
Motivated by the Schr\"odinger potential potential behavior of vector mesons \eqref{potentialvec}, which is a combination of the warp factor and dilation derivatives, we postulate the following mass coupling for our model
\begin{eqnarray}
    \tilde{m} = e^{-A_{s}}\left(\dfrac{1}{2}\Phi' - m A_{s}^{\prime}\right)\, .\label{couplingnucleon}
\end{eqnarray}
The coefficients were fixed in order to recover on the one hand the 5d constant mass $m$ in the AdS limit and on the other hand to guarantee a quadratic behaviour for the Schr\"odinger potential at large $z$ compatible with the soft wall model. The latter is a necessary requirement for obtaining asymptotically linear Regge trajectories for the nucleons, i.e. $m_{n}^{2}\sim n$ at large $n$ \footnote{This can be easily checked using a WKB approximation.}.

\begin{figure}[htp!]%
    \centering
    {{\includegraphics[width=7cm]{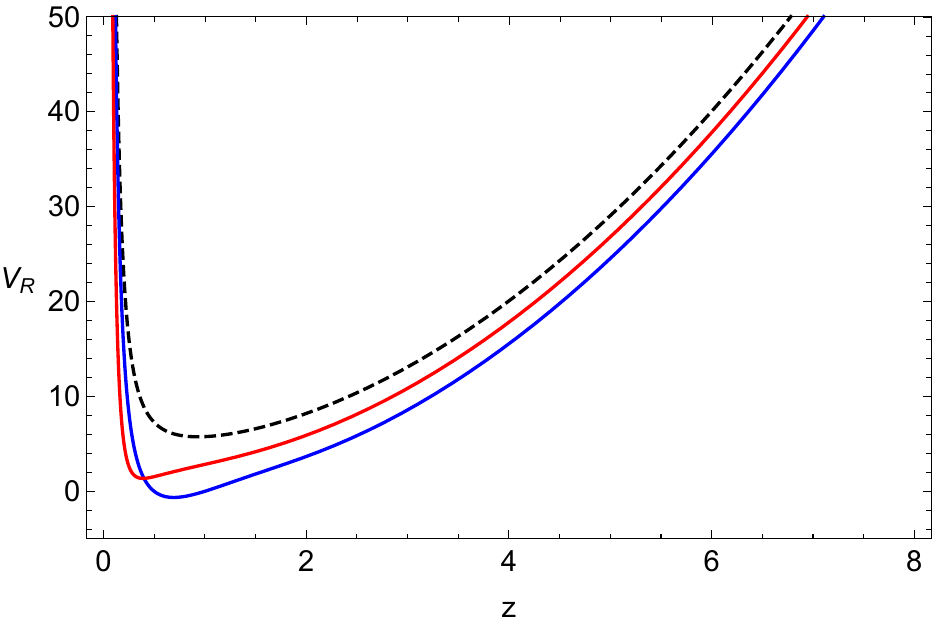} }}%
    \qquad
    {{\includegraphics[width=7cm]{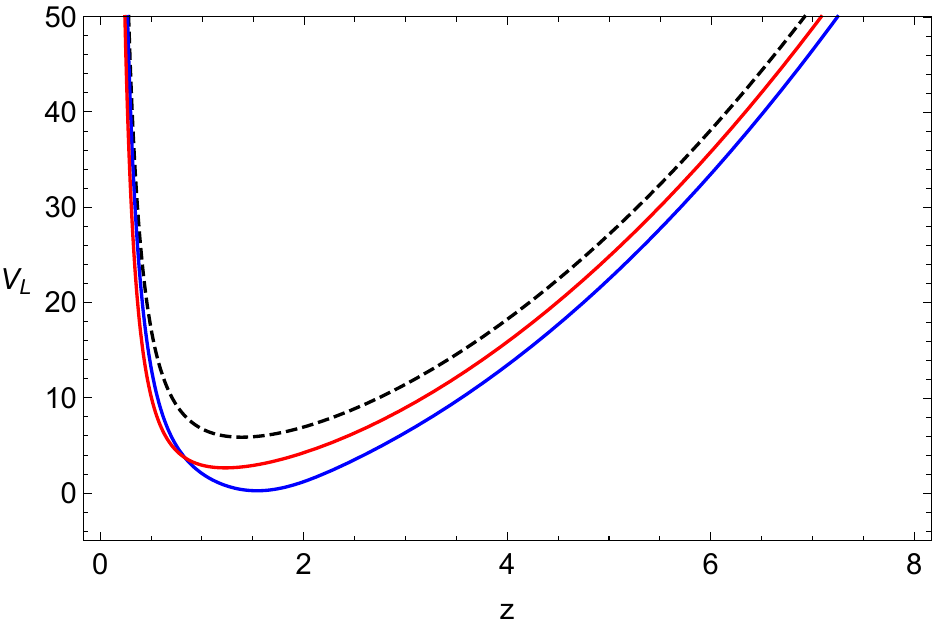} }}%
    \caption{Schr\"odinger potentials $V_R$ (left panel) and $V_L$ (right panel) for Einstein-dilaton models I and II (blue and red lines) and the soft wall model (black dashed lines) in the case $m=3/2$.}%
    \label{Fig:VRVLmeq3ov2}%
\end{figure}

\begin{figure}[htp!]%
    \centering
    {{\includegraphics[width=7cm]{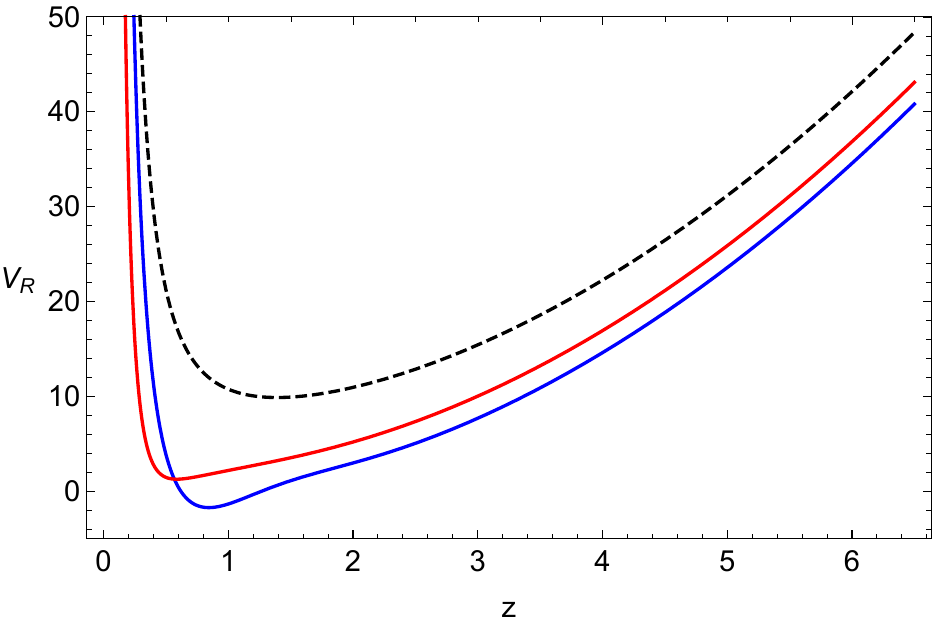} }}%
    \qquad
    {{\includegraphics[width=7cm]{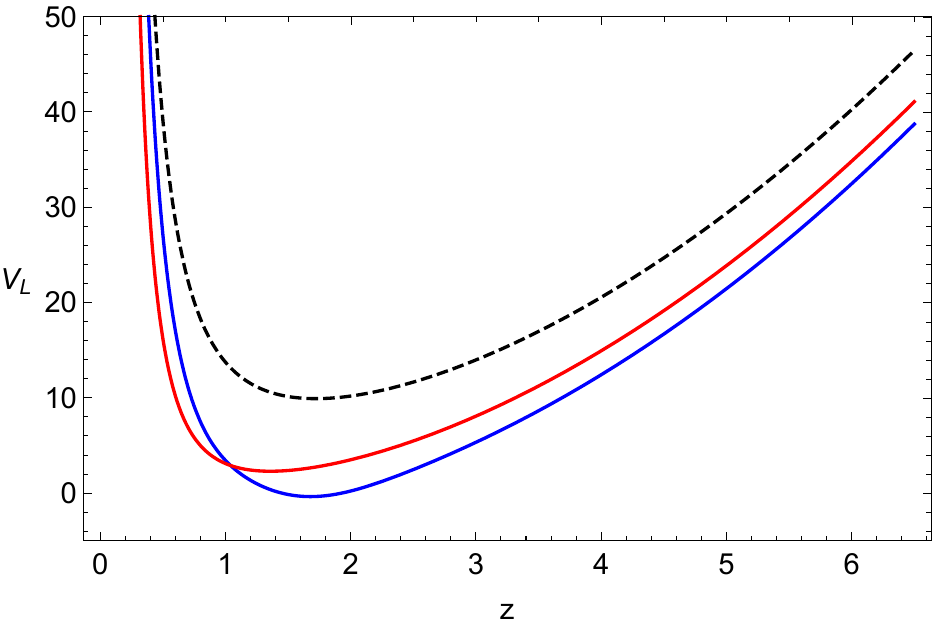} }}%
    \caption{Schr\"odinger potentials $V_R$ (left panel) and $V_L$ (right panel) for Einstein-dilaton models I and II (blue and red lines) and the soft wall model (black dashed lines) in the case $m=5/2$.}%
    \label{Fig:VRVLmeq5ov2}%
\end{figure}

Figure \ref{Fig:VRVLmeq3ov2} shows the results for the Schr\"odinger potentials $V_R$ (left panel) and $V_L$ (right panel) for the nucleons in the case $m=3/2$.  The blue and red lines represent the results for the Einstein-dilaton models I and II, respectively, whereas the black dashed lines represent the results for the soft wall model. Figure \ref{Fig:VRVLmeq5ov2} shows the results for the Schr\"odinger potentials in the case $m=5/2$. Note that the Schr\"odinger potentials for the soft wall model present a minimum at a higher value compared with the minima for the Einstein-dilaton models I and II. Note that this effect is enhanced as we go from the case $m=3/2$ to the case $m=5/2$.

\medskip 

{\bf Asymptotic solution and numerical integration}

To find the spectrum of nucleons we need to solve the eigenvalue equations \eqref{fnlr} or equivalently the Schr\"odinger equations \eqref{fnlrSchrod}. As expected, the eigenvalues for the left and right sector are the same since these two sectors are coupled. 

The eigenvalues and eigenfunctions are found numerically. 
For the numerical integration, we use for the initial conditions the asymptotic solution  at small $z$
\begin{eqnarray}
    f_{R}^n = N_{R}^n  z^{2+m} \quad , \quad 
   f_{L}^n = N_{L}^n  z^{3+m} \,. 
\end{eqnarray}
The normalisation constants $N_R^n$ and $N_L^n$ can be obtained imposing the condition that the eigenfunctions $\xi_R^{n}$ and $\xi_L^{n}$, defined in \eqref{Bogoliubovnucleons}, are normalised to $1$. The numerical integration is carried from small $z$ to large $z$ where we impose the asymptotic behaviour
\begin{equation}
\lim_{z \to \infty} \sqrt{z} \,  \xi_{R/L}^{n} = 0 \,.     
\end{equation}
Using this shooting method we find the set of eigenvalues $m_{N^n}$ corresponding to the 4d nucleon masses. 

\medskip 

{\bf Spectrum}

In tables \ref{Table:NucleonsMasses} and \ref{Table:NucleonsMasses2} we present our results for the nucleon masses for the Einstein-dilaton models I and II and compare them against the soft wall model, the hard wall model as well as the experimental results.  Table \ref{Table:NucleonsMasses} displays the results when the conformal dimension is fixed as $\Delta =7/2$ whereas table \ref{Table:NucleonsMasses2} corresponds to the case $\Delta = 9/2$.  The former takes into account the possible contribution from the anomalous dimension whilst the latter sets the anomalous dimension to zero.  The results for the soft wall model and the hard wall model were obtained following  \cite{Abidin:2009hr} and \cite{Hong:2006ta} respectively. We briefly review those works in appendices \ref{App:reviewsoftwall} and \ref{App:reviewhardwall}.

\begin{figure}[htp!]%
    \centering
    {{\includegraphics[width=7cm]{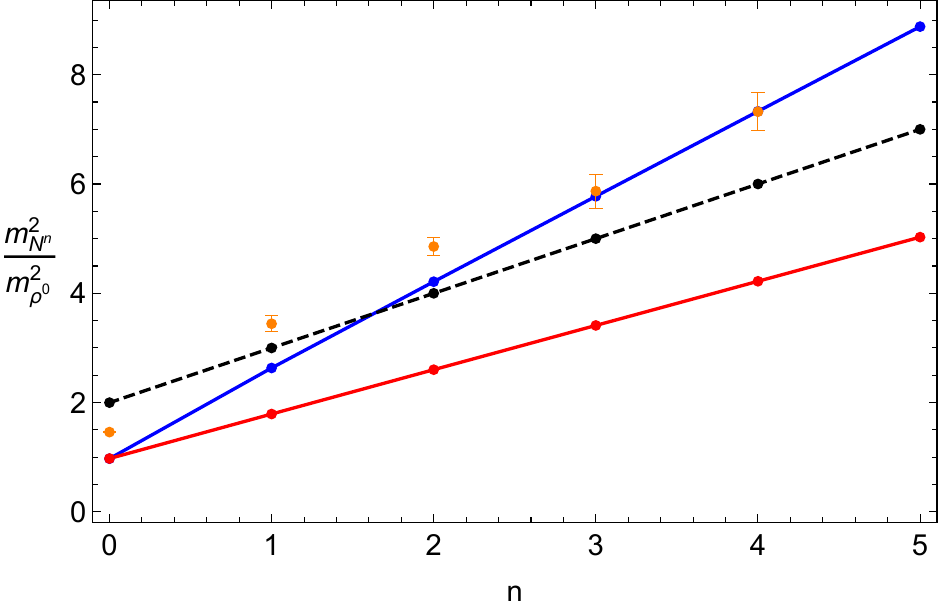} }}%
    \qquad
    {{\includegraphics[width=7cm]{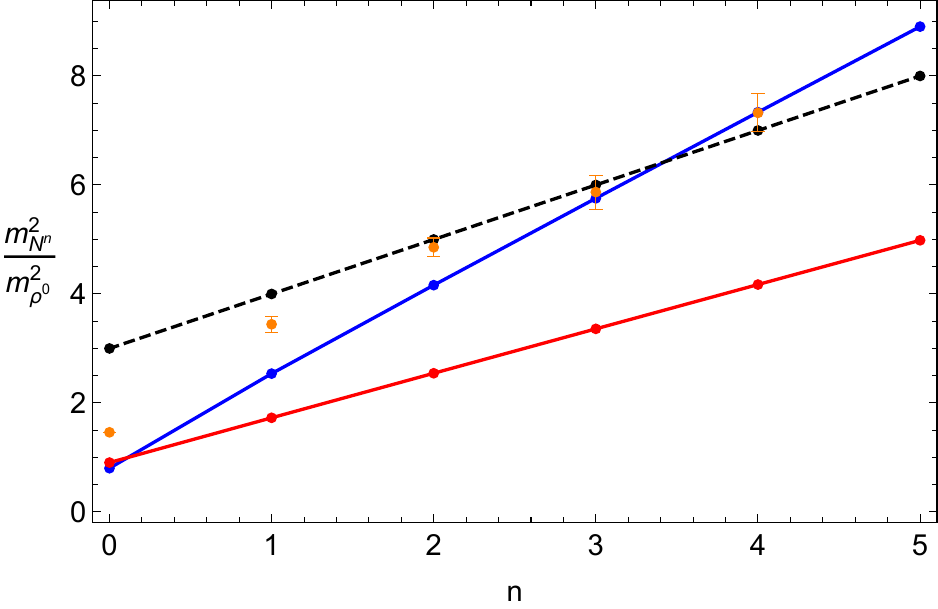} }}%
    \caption{{\bf Left:} Dimensionless squared mass ratios $m_{N^n}^2/m_{\rho^0}^2$ in the case $\Delta = 7/2$ ($m=3/2$) for nucleons in the Einstein-dilaton models I and II (blue and red solid lines with dots) and the soft wall model (black dashed line with dots), compared against experimental data (orange dots and error bars). {\bf Right:} Same as Left but this time $\Delta = 9/2$ ($m = 5/2$).}
    \label{Fig:Reggenucleons}%
\end{figure}

From our analysis we conclude that the Einstein-dilaton model I provide the results that are closest to the experimental data in both cases $\Delta = 7/2$ and $\Delta =9/2$. This can also be seen in figure \ref{Fig:Reggenucleons} where we plot the squared masses of the first six nucleon states as a function of the radial excitation number for the Einstein-dilaton models I and II (blue and red solid lines with dots) and the soft wall model (black dashed line with dots). As expected, the Regge trajectories in the Einstein-dilaton models I and II are approximately linear whilst the Regge trajectory in the soft wall model is exactly linear.

\begin{table}[ht]
\centering
\begin{tabular}{l |c|c|c|c|c}
\hline 
\hline
Ratio & Model I & Model II & Soft wall  & Hard wall & Experimental \cite{Workman:2022ynf} \\
\hline 
 $m_{N^0}/m_{\rho^0} $ & 0.987   & 0.988    & 1.414   & 1.593 & $1.209 \pm 0.002$  \\
 $m_{N^1}/m_{\rho^0} $ & 1.623   & 1.339   & 1.732   & 2.917  & $1.856 \pm 0.039$  \\
$m_{N^2}/m_{\rho^0} $ & 2.053  & 1.613  & 2        & 4.23   & $2.204 \pm 0.039$   \\
$m_{N^3}/m_{\rho^0} $ & 2.403    & 1.847  & 2.236  & 5.54  & $2.423 \pm 0.065$ \\
$m_{N^4}/m_{\rho^0} $ & 2.707  & 2.054  & 2.449  &   6.849   & $2.706 \pm 0.065$   \\
\hline\hline
\end{tabular}
\caption{
Nucleon masses divided by the mass of the $\rho_0$ meson in the case $\Delta =7/2$ ($m=3/2$) in the Einstein-dilaton models, the soft wall model and the hard wall model compared against the experimental results from PDG  \cite{Workman:2022ynf}.  The numerical error in our computations of mass ratios in Einstein-dilaton models I and II was of the order of $10^{-6}$.
}
\label{Table:NucleonsMasses}
\end{table}

\begin{table}[ht]
\centering
\begin{tabular}{l |c|c|c|c|c}
\hline 
\hline
Ratio & Model I & Model II & Soft wall  & Hard wall & Experimental \cite{Workman:2022ynf} \\
\hline 
 $m_{N^0}/m_{\rho^0} $ & 0.896    & 0.952     & 1.732   & 2.136  & $1.209 \pm 0.002$  \\
 $m_{N^1}/m_{\rho^0} $ & 1.593   & 1.314    &  2  & 3.5  & $1.856 \pm 0.039$  \\
$m_{N^2}/m_{\rho^0} $ & 2.04   &  1.595  &  2.236       &  4.832   & $2.204 \pm 0.039$   \\
$m_{N^3}/m_{\rho^0} $ & 2.399    & 1.833  & 2.449  & 6.153   & $2.423 \pm 0.065$ \\
$m_{N^4}/m_{\rho^0} $ &2.708   & 2.043   & 2.646  & 7.468     & $2.706 \pm 0.065$   \\
\hline\hline
\end{tabular}
\caption{
Nucleon masses divided by the mass of the $\rho_0$ meson in the case $\Delta =9/2$ ($m=5/2$) in the Einstein-dilaton models, the soft wall model and the hard wall model compared against the experimental results from PDG  \cite{Workman:2022ynf}. The numerical error in our computations of mass ratios in Einstein-dilaton models I and II was of the order of $10^{-6}$.
}
\label{Table:NucleonsMasses2}
\end{table}

\subsection{Wave functions and nucleon decay constants}

\begin{figure}[htp!]%
    \centering
    {{\includegraphics[width=7cm]{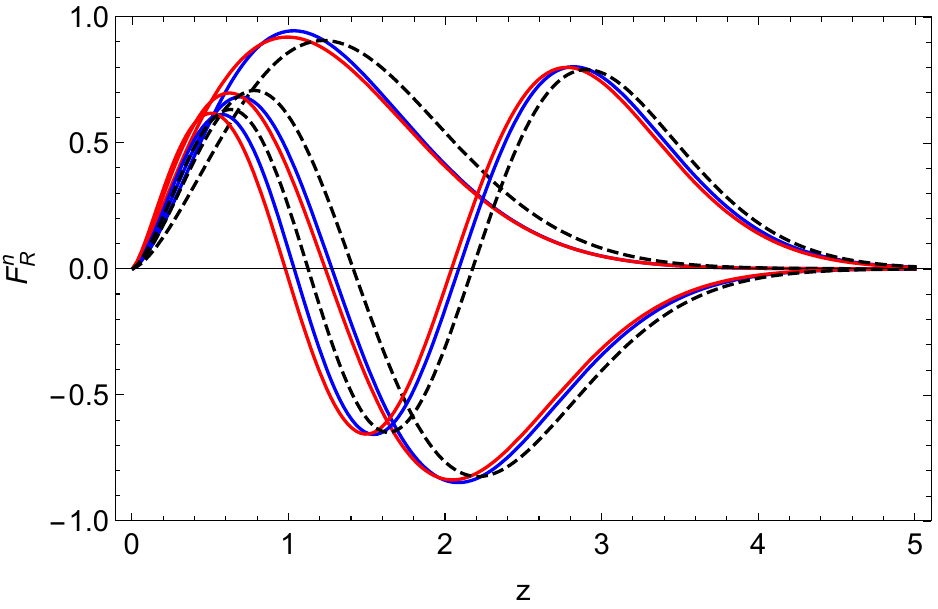} }}%
    \qquad
    {{\includegraphics[width=7cm]{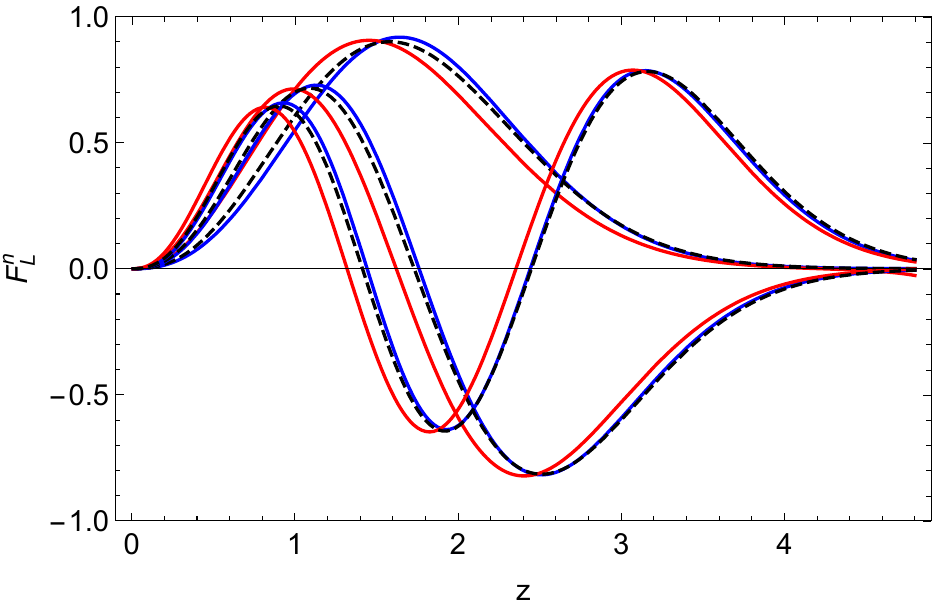} }}%
    \caption{ Normalised wave functions $F_R^n(z)$ (left panel) and $F_L^n(z)$ (right panel) for the Einstein-dilaton models I (blue), II (red) and the soft wall model (black dashed) for the case $\Delta = 7/2$ ($m = 3/2$).}%
    \label{NFRFLmeq3ov2}%
\end{figure}

\begin{figure}[htp!]%
    \centering
    {{\includegraphics[width=7cm]{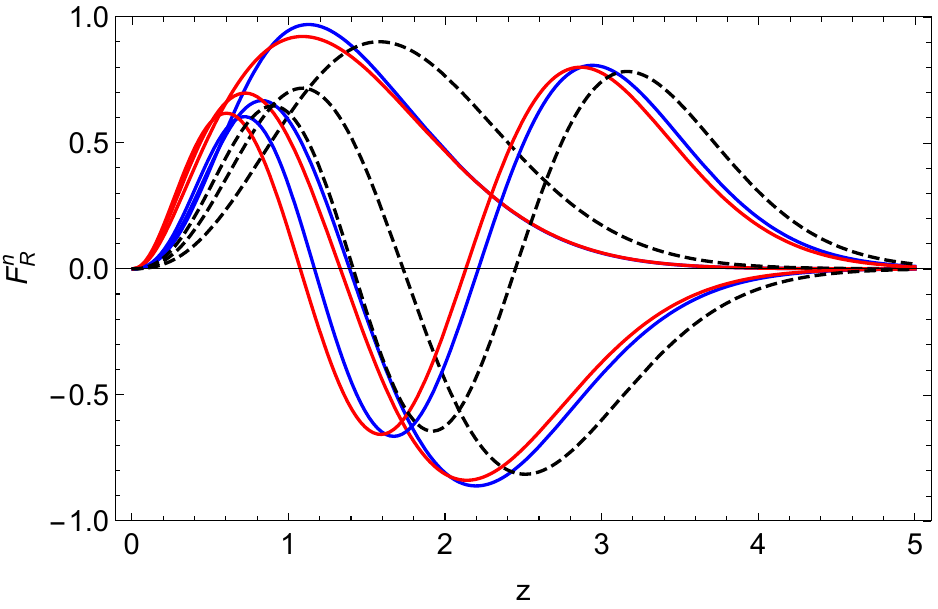} }}%
    \qquad
    {{\includegraphics[width=7cm]{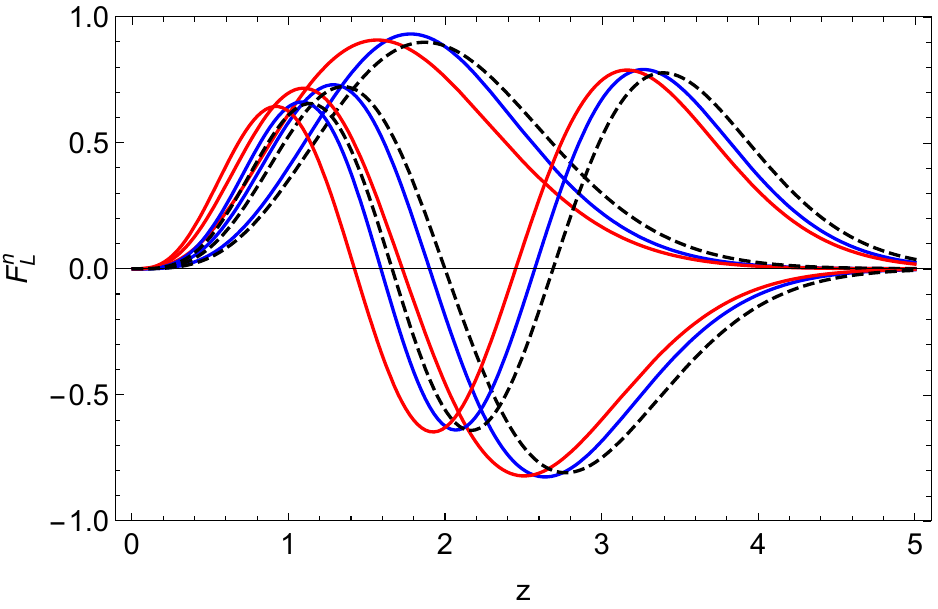} }}%
    \caption{ Normalised wave functions $F_R^n(z)$ (left panel) and $F_L^n(z)$ (right panel) for the Einstein-dilaton models I (blue), II (red) and the soft wall model (black dashed) for the case $\Delta = 9/2$ ($m = 5/2$).}%
    \label{NFRFLmeq5ov2}%
\end{figure}

Figures \ref{NFRFLmeq3ov2} and \ref{NFRFLmeq5ov2} display the normalised eigenfunctions $f_R^n(z)$ and $f_L^n(z)$, representing the nucleon states, for the cases $\Delta = 7/2$ ($m = 3/2$) and  $\Delta = 9/2$ ($m = 5/2$) respectively.  The blue and red solid lines correspond to Einstein-dilaton models I and II, respectively, whilst the black dashed line represents the soft wall model. These results confirm that the first nucleon masses obtained in the previous subsection correspond to the fundamental state and the first excited states. 

Previously in this section we obtained the holographic dictionary for the nucleon decay constants \eqref{lambdaNn}.  We can finally evaluate this formula using the normalised eigenfunctions and obtain
\begin{equation}
\lambda_{N^n} = \sqrt{G_F} \Big [ z^{-2-m}  f_{R,n}(z) \Big ]_{z= \epsilon} 
= \sqrt{G_F} N_R^n \, ,    \label{lambdaNnv2}
\end{equation}
where $N_R^n$ is the normalisation constant  in the right sector. The coupling constant $G_F$ in the fermionic sector was already fixed in \eqref{GF} in order to reproduce the perturbative QCD result for the correlation function. 

We display in tables \ref{Table:NucleonDecayConstants} and \ref{Table:NucleonDecayConstants2} our results for the nucleon decay constants in the cases $\Delta = 7/2$ ($m=3/2$) and $\Delta = 9/2$ ($m=5/2$) respectively. In the latter case we also present the result for the fundamental state in lattice QCD obtained in \cite{Leinweber:1994nm}  using a nucleon operator similar the one presented in \eqref{nucleonop}. Note that the Einstein-dilaton model I and the soft wall model provide results that are closer to lattice QCD.  It is interesting to note that holographic QCD models provide results for the excited nucleon states which, as far as we are concerned, are not available in other non-perturbative approaches. In particular, all the holographic QCD models predic that the nucleon decay constants grow with the radial excitation number. 

\begin{table}[ht]
\centering
\begin{tabular}{l |c|c|c|c}
\hline 
\hline
Ratio & Model I & Model II & Soft wall  & Hard wall    \\
\hline 
 $\lambda_{N^0}/m_{\rho^0}^{\alpha} $  & 0.1108  & 0.09835   & 0.0507    & 0.1667     \\
 $\lambda_{N^1}/m_{\rho^0}^{\alpha} $  & 0.1302  & 0.1158    & 0.0716    & 0.4096     \\
$\lambda_{N^2}/m_{\rho^0}^{\alpha}  $  & 0.1519  & 0.1284    & 0.0877    & 0.7138     \\
$\lambda_{N^3}/m_{\rho^0}^{\alpha}  $  & 0.1708  & 0.1388    & 0.1013    & 1.069      \\
$\lambda_{N^4}/m_{\rho^0}^{\alpha}  $  & 0.1878  & 0.1478    & 0.1133    & 1.469       \\
\hline\hline
\end{tabular}
\caption{
Nucleon decay constants $\lambda_{N^n}$ divided by $m_{\rho_0}^{\alpha}$ with $\alpha = \Delta - 3/2 = m + 1/2 $ in the case $\Delta =7/2$ ($m=3/2$) in the Einstein-dilaton models, the soft wall model and the hard wall model.   The numerical error in our computations of $\lambda_{N^n}/m_{\rho^0}^{\alpha}$ in Einstein-dilaton models I and II was of the order of $10^{-3}$. 
}
\label{Table:NucleonDecayConstants}
\end{table}

\begin{table}[ht]
\centering
\begin{tabular}{l |c|c|c|c|c}
\hline 
\hline
Ratio & Model I & Model II & Soft wall  & Hard wall & Lattice QCD \cite{Leinweber:1994nm}  \\
\hline 
 $\lambda_{N^0}/m_{\rho^0}^{\alpha} $ & 0.1055  & 0.158    & 0.01791   & 0.1414   & $ 0.05778 \pm 0.0107$  \\
 $\lambda_{N^1}/m_{\rho^0}^{\alpha} $ & 0.1201  & 0.1906   & 0.03102   & 0.4755   & -  \\
$\lambda_{N^2}/m_{\rho^0}^{\alpha} $  & 0.1462  & 0.2172   & 0.04387   & 1.058    & -   \\
$\lambda_{N^3}/m_{\rho^0}^{\alpha} $  & 0.172   & 0.2409   & 0.05664   & 1.931    & -   \\
$\lambda_{N^4}/m_{\rho^0}^{\alpha} $  & 0.1973  & 0.2627   & 0.06937   & 3.129    & -   \\
\hline\hline
\end{tabular}
\caption{
Nucleon decay constants $\lambda_{N^n}$  divided by $m_{\rho_0}^{\alpha}$ with $\alpha = \Delta - 3/2 = m + 1/2 $  in the case $\Delta =9/2$ ($m=5/2$) in the Einstein-dilaton models, the soft wall model and the hard wall model. In this case we also compare against the lattice QCD result. The numerical error in our computations of $\lambda_{N^n}/m_{\rho^0}^{\alpha}$ in Einstein-dilaton models I and II was of the order of $10^{-3}$. 
}
\label{Table:NucleonDecayConstants2}
\end{table}

\section{Conclusions}
\label{Sec:Conclusions}

We have built in this paper a bottom-up holographic QCD model that provides a unified description of vector mesons and nucleons in a confining background based on Einstein-dilaton gravity. Our model has three parameters: $g_5^2$ associated with the two-point correlation function in the vector meson sector, $G_{F}$ associated with the two-point correlation function in the nucleon sector and $k$ the infrared mass scale associated with the spectrum. We fixed $g_5^2$ and $G_F$ using the perturbative QCD results for the hadronic correlators in the high energy regime. Since we worked with dimensionless ratios for the hadronic masses and decay constants we did not need to fix the constant $k$. 

To investigate the spectral decomposition of the hadronic correlators and the associated decay constants, we applied the Sturm-Liouville theory inspired by previous works \cite{Erlich:2005qh, Grigoryan:2007vg,Grigoryan:2007my}. The Sturm-Liouville theory allowed us to find  spectral decompositions for the 5d bulk to boundary propagators and the 4d correlation functions. We showed that the latter are compatible  with QCD in the large $N_{c}$ limit. We obtained a holographic dictionary for the hadronic decay constants that is valid for a general class of holographic models based on Einstein-dilaton gravity. The Sturm-Liouville also led naturally to the completeness relation for the normalisable modes, identified as the Sturm-Liouville modes.  

We would like to remark that our model allows for a clean description of the spectrum and decay constants of nucleons and vector mesons in a confining background based on Einstein-dilaton gravity. This improves previous bottom-up approaches in many aspects. The results for the vector meson and nucleon masses lie on approximate linear Regge trajectories and are very close to experimental data. The results for the vector meson and nucleon decay constants provide a guide for future phenomenological studies, in particular for the higher excited states. Below we  discuss some of the main results that are relevant to hadron phenomenology.

In table \ref{Table:VectorMesonMasses} we presented our results for the spectrum of vector mesons in terms of the ratio between the masses of excited states, i.e. $m_{\rho^n}$ with $n \geq 1$, and the mass of the ground state $m_{\rho^0}$ for the Einstein-dilaton models I and II, discussed in section \ref{Sec:models}. We compared our results with the soft wall model, the  hard wall model and experimental data. We concluded that the Einstein-dilaton model I and the soft wall model provide results that are the closest to the experimental results. We presented our results for the nucleon spectrum in Eintein-dilaton models I and II for the cases $\Delta=7/2$  and $\Delta=9/2$ in tables \ref{Table:NucleonsMasses}  and \ref{Table:NucleonsMasses2} respectively. Our results for the nucleon spectrum were presented in terms of the ratios between the masses of the nucleon states $m_{N^n}$ with $n=0,1,\dots$ (ground state and excited states) relative to the mass of the vector meson ground state $m_{\rho^0}$.  We compared our results with the soft wall model, the hard wall model and experimental data and concluded that the Einstein-dilaton model I provide the best results compared to experimental data.

In table \ref{Table:VectorMesonDecayConstants} we presented our results for the vector meson decay constants $F_{\rho^n}$ in terms of the dimensionless ratios $\sqrt{F_{\rho^n}}/m_{\rho^0}$ for $n=0,1,\dots$ (ground state and excited states). We compared our results for the Einstein-dilaton models I and II with the soft wall and hard wall model. For the ground state case ($n=0$) we also compared against the only available experimental data and concluded that the hard wall model still provides the closest result.  We presented our results for the nucleon decay constants $\lambda_{N^n}$ for the cases $\Delta = 7/2$ and $\Delta = 9/2$ in tables \ref{Table:NucleonDecayConstants}  and \ref{Table:NucleonDecayConstants2} respectively. The results were presented in terms of the dimensionless ratios $\lambda_{N^n}/m_{\rho}^\alpha$ with $\alpha = \Delta - 3/2$. For the case $\Delta =9/2$ (canonical dimension) and $n=0$ (nucleon ground stated)  we also compared against the lattice QCD result and concluded that the Einstein-dilaton model I and the soft wall model provide results that are closer to lattice QCD. It is important to remark that holographic QCD models are capable of predicting the decay constants of excited states. In particular, we noted that for both light vector mesons and nucleons the decay constants increase with the radial excitation number. We hope that in the future, as more experimental results on hadronic decay constants become available, these findings can be further tested.

A natural continuation of this work would be to investigate the spectrum and decay constants of the Delta baryons that have spin and isospin $3/2$ in the context of holographic QCD models based on Einstein-dilaton gravity. Some works have already been developed using the hard wall model and the soft wall model \cite{Ahn:2009px, deTeramond:2005su,Huseynova:2020gqn}. We also want to apply the Sturm-Liouville theory in that case to obtain the  spectral decomposition for the correlators of Delta baryon operators. We are also interested in studying the strong couplings between vector mesons and baryons, the electromagnetic and the gravitational form factors. Last but not least, we want to investigate the effects of chiral symmetry breaking on the mass generation of nucleons and vector mesons. We intend to develop these works in the near future.

\section*{Acknowledgments}

The work of the author A.B-B is partially funded by Conselho Nacional de Desenvolvimento Cient\'\i fico e Tecnol\'ogico (CNPq, Brazil), Grant No. 314000/2021-6, and Coordena\c{c}\~ao de Aperfei\c{c}oamento do Pessoal de N\'ivel Superior (CAPES, Brazil), Finance Code 001. The work of the author A.S.S. Jr has financial support from Conselho Nacional de Desenvolvimento Cientifico e Tecnologico (CNPq).

\appendix

\section{Sturm-Liouville theory and the spectral decomposition}
\label{App:sturmlitheory}

The Sturm-Liouville theory can be described, for example, by a non-homogeneous one-dimensional second-order differential equation \cite{arfken2013mathematical, butkov1968mathematical}
\begin{eqnarray}
    \dfrac{d}{dz}\left(p(z)\dfrac{dy}{dz}\right) - s(z)y + \lambda r(z)y = f(z)\label{sturm}
\end{eqnarray}
where $y(z)$, $p(z)$, $s(z)$, $r(z)$, and $f(z)$ are functions of $z$ and $\lambda$ is a constant parameter. In the homogeneous case we takes $f(z) = 0$ in \eqref{sturm} equation. From the two first terms in \eqref{sturm}, we can define the Sturm-Liouville operator,
\begin{eqnarray}
    \mathcal{L}\equiv \dfrac{d}{dz}\left(p(z)\dfrac{d}{dz}\right) - s(z).
\end{eqnarray}
This is a second-order self-adjoint operator with eigenvalue $\lambda$. Rewriting the equation \eqref{sturm} in terms of $\mathcal{L}$, we have
\begin{eqnarray}
    \left[\mathcal{L} + \lambda \, r(z)\right]y = f(z).\label{sturmoperator}
\end{eqnarray}
We will be particularly interested in the solution of the homogeneous case $f(z)=0$, we will call this solution $y_0(z)$. We can obtain the Green's functions that satisfy \eqref{sturmoperator} starting from
\begin{eqnarray}
    \left[\mathcal{L} + \lambda \, r(z)\right]G(z; z^{\prime}) =\delta(z - z^{\prime})\label{liouvillegreen}
\end{eqnarray}
where $G(z; z^{\prime})$ is the Green function that must obey some boundary condition. We can expand the Green's functions into a series of eigenfunctions expressed as
\begin{eqnarray}
    G(z; z^{\prime}) =\sum_{n} a_{n}(z^{\prime})\varphi_{n}(z)\label{greenexpand}
\end{eqnarray}
Plugging \eqref{greenexpand} into \eqref{liouvillegreen} and imposing the orthonormality condition
\begin{eqnarray}
    \int dz\, r(z)\,\varphi_{m}(z)\varphi_{n}(z) =\delta_{mn} \label{normal}
\end{eqnarray}
we have
\begin{eqnarray}
    a_{n}(z^{\prime}) =\dfrac{\varphi_{n}(z^{\prime})}{\lambda - \lambda_{n}}\label{coefficient}
\end{eqnarray}
Substituting \eqref{coefficient} in \eqref{greenexpand}, we obtain the spectral decomposition for the Green's function 
\begin{eqnarray}
    G(z ; z^{\prime}) = \sum_{n}\dfrac{\varphi_{n}(z)\varphi_{n}(z^{\prime})}{\lambda - \lambda_{n}}. \label{GreenDecomp}
\end{eqnarray}
The eigenfunctions obey the equation
\begin{eqnarray}
    \left[\mathcal{L} + \lambda_{n} r(z)\right]\varphi_{n}(z) = 0.
\end{eqnarray}
We can relate the homogeneous solution $y_0(z)$ to the Green's function $G(z; z^{\prime})$ as follows. Multiplying  both sides of \eqref{liouvillegreen}  by $y_0(z)$, integrating by parts twice over $z$ and using the homogeneous equation for $y_0(z)$ we obtain
\begin{eqnarray}
    y_0(z^{\prime}) = \big[r(z)\big(y_0(z)\partial_{z}G(z; z^{\prime}) - G(z; z^{\prime})\partial_{z}y_0(z)\big)\big]_{z = z_{i}}^{z = z_{f}}\label{wronskian}
\end{eqnarray}
Note that this result has the form of a Wronskian in the $z$ variable for the functions $y_0(z)$ and $G(z,z')$. The limits of integration $z_{i}$ and $z_{f}$ depend on the boundary conditions of the problem.  

Plugging the spectral decomposition \eqref{GreenDecomp} into \eqref{wronskian} we find
the expansion
\begin{align}
    y_0(z') &=  \sum_n \frac{\varphi_n (z')}{\lambda - \lambda_n} 
    \Big [ r(z) \big (y_0(z) \partial_{z} \varphi_n (z) - \varphi_n (z) \partial_{z}y_0(z) \big ) \Big ]_{z=z_i}^{z=z_f} \equiv \sum_n \alpha_n \varphi_n (z') \,. \label{ydecomp}
\end{align}
Using the orthonormality condition, we have
\begin{eqnarray}
    \int dz\, r(z) \varphi_{m}(z)y_0(z) =\alpha_{m}\, .\label{alpham}
\end{eqnarray}
From \eqref{ydecomp} and  \eqref{alpham} we obtain
\begin{eqnarray}
    y_0(z^{\prime}) &=&\sum_{n}\left[\int dz\, r(z)\varphi_{n}(z)y_0(z)\varphi_{n}(z')\right]\\
                  &=&\int dz\, y_0(z)\left[\sum_{n}r(z)\varphi_{n}(z)\varphi_{n}(z^{\prime})\right]\\
                  &=&\int dz\, \delta(z - z^{\prime})y_0(z)
\end{eqnarray}
In this way, we find the completeness relation
\begin{eqnarray}
    \sum_{n}r(z)\varphi_{n}(z)\varphi_{n}(z^{\prime}) =\delta(z - z^{\prime}).
\end{eqnarray}
\section{The Proca Field Propagator}\label{App:procapropagator}

In this appendix, we briefly discuss the Proca field with an additional term that acts as a Lagrange multiplier; see, for example \cite{Itzykson:1980rh,Peskin:1995ev}. 

Consider the Proca Lagrangian for a massive spin $1$ particle in four dimensions
\begin{align}
    \mathcal{L} = - \dfrac{1}{4}(F_{\mu\nu})^{2} + \dfrac{1}{2}m^{2}A_{\mu}^{2} - \dfrac{1}{2}\left(1 - \dfrac{1}{\chi}\right)\left(\partial_{\mu}A^{\mu}\right)^{2}\label{procaLagrangian}
\end{align}
where
\begin{align}
    F_{\mu\nu} =\partial_{\mu}A_{\nu} - \partial_{\nu}A_{\mu}
\end{align}
being $F_{\mu\nu}$ the field strength usual, $A^{\mu}$ is the gauge field, $m$ is the mass and $\chi$ the Lagrange multiplier. The equations of motions obtained from \eqref{procaLagrangian} written in momentum space takes the form
\begin{eqnarray}
    \left[\eta^{\mu\nu}\left(k^{2} + m^{2}\right) - \left(1 - \dfrac{1}{\chi}\right)k^{\mu}k^{\nu}\right]A_{\mu} = 0\label{procaeqmmoment}
\end{eqnarray}
To obtain the two-point correlation function, let us write the above operator in \eqref{procaeqmmoment} as
\begin{align}
  \eta^{\mu\nu} (k^{2} + m^{2}) - k^{\mu}k^{\nu} + \dfrac{1}{\chi}k^{\mu}k^{\nu} 
    &= \left(\eta^{\mu\nu} - \dfrac{k^{\mu}k^{\nu}}{k^{2}}\right)\left(k^{2} + m^{2}\right) + \left(\dfrac{k^{\mu}k^{\nu}}{k^{2}}\right)\dfrac{1}{\chi}\left(k^{2} + \chi m^{2}\right)\label{procaeqmoment1}
\end{align}
The Proca propagator is obtained by inverting the terms that multiply the projectors in \eqref{procaeqmoment1}
\begin{align}
    \langle A^{\mu}(k)A^{\nu}(-k)\rangle &= \dfrac{-i}{k^{2} + m^{2}}\left(\eta^{\mu\nu} - \dfrac{k^{\mu\nu}}{k^{2}}\right) + \dfrac{-i\chi}{k^{2} + \chi m^{2}}\left(\dfrac{k^{\mu}k^{\nu}}{k^{2}}\right)\\
    &=\dfrac{-i}{k^{2} + m^{2}}\left[\eta^{\mu\nu} - \dfrac{k^{\mu}k^{\nu}}{k^{2} + \chi m^{2}}\left(1 + \chi\right)\right]\label{procacorrelator}
\end{align}
From the result \eqref{procacorrelator} above, we can take some limits. For $\chi = 0$, we have
\begin{align}
     \langle A^{\mu}(k)A^{\nu}(-k)\rangle\Big|_{\chi = 0} = \dfrac{-i}{k^{2} + m^{2}}\left[\eta^{\mu\nu} - \dfrac{k^{\mu}k^{\nu}}{k^{2}}\right]\label{procachizero}
\end{align}
For $\chi\to \infty$,
\begin{align}
     \langle A^{\mu}(k)A^{\nu}(-k)\rangle\Big|_{\chi = 0} = \dfrac{-i}{k^{2} + m^{2}}\left[\eta^{\mu\nu} - \dfrac{k^{\mu}k^{\nu}}{m^{2}}\right]\label{procachiinfty}
\end{align}
Lastly, for $\chi = -1$,
\begin{align}
     \langle A^{\mu}(k)A^{\nu}(-k)\rangle\Big|_{\chi = 0} = \dfrac{-i \eta^{\mu\nu}}{k^{2} + m^{2}}\, .\label{procachiminusone}
\end{align}
\section{Vector mesons and nucleons in the soft wall model}\label{App:reviewsoftwall}

The soft wall model, initially proposed in the seminal paper by Karch, Katz, Son, and Stephanov in \cite{Karch:2006pv}, has been demonstrated to effectively capture the Regge trajectories of various particles, including vector mesons and nucleons. In order to break the conformal symmetry, the soft wall model incorporates a dilaton field $\Phi(z)$ into its action. In this appendix we provide a succinct overview of the field equations, normalisable solutions, spectrum and decay constants of vector mesons and nucleons within the framework of the soft wall model.

\subsection{Vector mesons}
In section \ref{Sec:VectorMesons} we study the vector mesons using the Einstein-dilaton model. We can get the vector meson results for the soft wall model from the equations of the Einstein-dilaton model by considering the AdS limit. We start with the equation \eqref{ELeqsfluctsv5}
\begin{eqnarray}
    \Big [ \partial_z + A_s' - \Phi' \Big ] \partial_z V^{\hat \mu,a}_{\perp} -  q^2   V^{\hat \mu,a}_{\perp}  = 0\, .
\end{eqnarray}
In the soft wall, the warp factor is $A_{s} = - \ln z$ and the dilaton remains quadratic, $\Phi = k z^{2}$. For simplicity, let's consider $k = 1$ in our equations. It is convenient to write the vector field as $V_{\perp}^{\hat{\mu}, a} =\eta_{\perp}^{\hat{\mu}}v(q^{2}, z)$ where $\eta_{\perp}^{\hat{\mu}}$ is a transverse polarisation vector, i.e.  $q_{\hat{\mu}}\eta_{\perp}^{\hat{\mu}} = 0$. Writing the vector field this way, the above equation reduces to
\begin{eqnarray}
    \left[z^{2}\partial_{z}^{2} - (1 + 2z^{2})z\partial_{z}\right]v - q^{2}z^{2}v = 0
\end{eqnarray}
This differential equation has an analytical solution that can be written as
\begin{eqnarray}
    v(Q, z) = c_{1}\, z^{2}U\left(1 - \dfrac{Q^{2}}{4}, 2, z^{2}\right) + c_{2}\, z^{2} M\left(1 - \dfrac{Q^{2}}{4}, 2, z^{2}\right)
\end{eqnarray}
where $Q = \sqrt{-q^2}$,  $c_{1}$ and $c_{2}$ are constant coefficients whereas $U(a, b, x)$ and $M(a, b, x)$ are the Tricomi and Kummer confluent hypergeometric functions. To guarantee the regularity of the solution far from the boundary, we must take $c_{2} = 0$. In this way, the solution reduces to the Tricomi function
\begin{eqnarray}
     v(Q, z) = c_{1}\, z^{2}U\left(1 - \dfrac{Q^{2}}{4}, 2, z^{2}\right)\, . \label{tricomi}
\end{eqnarray}
In order to obtain the normalisable solution the first argument of the Tricomi function in \eqref{tricomi} has to be $-n$ with $n$ a non-negative integer. This leads to the spectrum of the vector mesons
\begin{eqnarray}
    m^{2}_{v_{n}} = 4(n + 1)\quad \textnormal{with}\quad n = 0, 1, 2, 3, ...
\end{eqnarray}
The normalisable solutions take the form 
\begin{eqnarray}
    v_{n}(z) = N_{v^{n}} z^{2}L_{n}^{1}(z^{2})\quad \textnormal{with}\quad n = 0, 1, 2, 3, ...
\end{eqnarray}
where $L_n^k(x)$ are the associated Laguerre polynomials and 
\begin{eqnarray}
    N_{v^{n}} = \sqrt{\dfrac{2}{n + 1}}
\end{eqnarray}
are the normalisation constants that can be obtained using the condition
\begin{eqnarray}
    \int dz\, e^{A_{s} - \Phi}v_{n}^{2} = 1\, .
\end{eqnarray}
For small $z$ (near the boundary) the normalisable solution takes the form
\begin{align}
    v_{n}(z) = c_{2, n}\, z^{2} + c_{4, n}\, z^{4} + \, ...\, ,
\end{align}
where
\begin{align}
    c_{2, n} = (n + 1)N_{v^{n}}\quad \textnormal{and}\quad c_{4, n} = -\dfrac{1}{2}n(n + 1)N_{v^{n}} = -\dfrac{n}{2}c_{2, n}\, .
\end{align}
The decay constants can be obtained from \eqref{vecdecayconstants} considering $A_{s} = -\ln z$, reproducing the results of  \cite{Grigoryan:2007my,Ballon-Bayona:2021ibm}
\begin{eqnarray}
    f_{v^{n}} = \dfrac{1}{g_{5}}\left[e^{A_{s} - \Phi}\partial_{z}v_{n}(z)\right]_{z\to 0} =\dfrac{2 c_{2, n}}{g_{5}} =\dfrac{2(n + 1)N_{v^{n}}}{g_{5}} =\dfrac{1}{g_{5}}\sqrt{8(n + 1)}\, .
\end{eqnarray}

\subsection{Nucleons}

In this appendix, we obtain the results for the soft wall model as a particular case of the results of section \ref{nucleonseinsteindilaton} for the Einstein-dilaton models. The starting point is the equation \eqref{fLReq2ndv2} given by
\begin{align}
&\Big [ \partial_z^2 + 4 A_s' \partial_z + 2 A_s'' + 4 A_s'^2 \mp \partial_z \Big ( e^{A_s}  \tilde m \Big ) - e^{2A_s} \tilde m^2 + Q^{2} \Big ] F_{R/L} = 0 \,.
\end{align}
Considering the warp factor as $A_{s} = - \ln z$ and $\tilde{m} = m + \Phi$ with $\Phi = k z^{2}$, we obtain the equation for the soft wall model
\begin{eqnarray}
    \left[(z\partial_{z})^{2} - 5z\partial_{z} + 6 + z^{2}(Q^{2} \mp k) \pm - (m + kz^{2})^{2}\right] F_{R/L} = 0
\end{eqnarray}
Again, for simplicity we will take $k=1$. The general solution of this equation that is regular at large $z$ (far from the boundary) takes the form
\begin{align}
F_{R}(z) &=  e^{- z^{2}/2}  (z^{2})^{1 + \frac{m}{2}} \Big [ d_1 U ( m + \frac12 - \frac{Q^2}{4 } , m  + \frac12 , z^{2}) \Big ] \, ,\nonumber \\
F_{L} (z) &=  e^{-  z^{2}/2} (z^{2})^{\frac32 + \frac{m}{2}} \Big [ c_1 U ( m + \frac12 - \frac{Q^2}{4 k} , m + \frac32 , z^{2})  
  \Big ] \, ,
\end{align}
where $d_1$ and $c_1$ are constant coefficients. 
By arguments similar to the previous section, the spectrum of nucleons is given by
\begin{eqnarray}
    m_{N^{n}}^{2} =4n + 4m + 2 \, ,\quad n = 0, 1, 2, 3, ...
\end{eqnarray}
where $m$ is usually chosen as $m = 3/2$ ($\Delta = 7/2$) or $m = 5/2$ ($\Delta = 9/2$). The normalisable solutions are expressed by
\begin{equation}
f_{R/L}^{n}(z) = N_{R/L} e^{- z^2/2} z^{5/2+ (m \mp 1/2) } L_n^{m \mp 1/2}(z^2) \,  ,
\end{equation}
with the normalisation constants
\begin{equation}
N_R = \sqrt{ \frac{2 \, \Gamma(n+1)}{\Gamma (n + m + 1/2)}} \quad , \quad N_L = \sqrt{ \frac{2 \, \Gamma(n+1)}{\Gamma (n + m + 3/2)}} \, ,
\end{equation}
obtained from the normalisation condition
\begin{eqnarray}
    \int dz\, e^{4A(z)} f_{R/L}^{m}(z)f_{R/L}^{n}(z) = \delta^{mn}
\end{eqnarray}
Using the holographic dictionary \eqref{lambdaNn}, the nucleon decay constants in the soft wall model take the form 
\begin{align}
\lambda_{N^n} &= \sqrt{G_F} f_n = \sqrt{G_F} \Big [ z^{-2-m}  f_{R,n}(z) \Big ]_{z= \epsilon} \nonumber \\
&= \sqrt{G_{F}}  N_R  L_n^{m-1/2}(0) =  \frac{2 \sqrt{G_F} }{N_R \Gamma(m+1/2)} \,. 
\end{align}
which is compatible with \cite{Abidin:2009hr}. 
\section{Vector mesons and nucleons in the hard wall model}
\label{App:reviewhardwall}

In the context of holographic QCD models in the bottom-up approach, the hard wall model is the pioneer. This model was proposed by Polchinski and Strassler in the study of glueball scattering in the fixed angle regime \cite{Polchinski:2001tt}. Further investigations of glueballs \cite{Boschi-Filho:2002xih}, mesons and chiral symmetry breaking \cite{Erlich:2005qh} showed that the hard wall model constitutes a very good toy model for investigating hadronic physics.  The model consists of cutting the AdS space limiting the holographic coordinate  to the region $0 < z \leq z_0$ and imposing boundary conditions for the 5d fields. By slicing the AdS space, the conformal symmetry is broken and this allows a mass gap to be generated. In this appendix, we briefly review the field equations and solutions, the spectrum and the decay constants for the case of vector mesons and nucleons.

\subsection{Vector mesons}

In order to describe vector mesons in the hard wall model, we consider $A_{s} = -\ln z$ and $\Phi = 0$ in equation \eqref{ELeqsfluctsv5} and take the vector field as $V_{\perp}^{\hat{\mu}, a} =\eta_{\perp}^{\hat{\mu}}v(q^{2}, z)$. This reduces the equation to
\begin{align}
\left[z^{2}\partial_{z}^{2} - z\partial_{z}\right]v - q^{2}z^{2}v = 0\, .\label{vechw1}
\end{align}
This differential equation has analytical solutions, that can be written as
\begin{align}
    v(Q, z) = z \left[c_{1}\, J_{1}(Qz) + c_{2}\, Y_{1}(Qz)\right]\label{vechw2} \, ,
\end{align}
where $J_m(x)$ and $Y_m(x)$ are Bessel functions of the first and second kind, respectively, and $Q = \sqrt{-q^2}$. The normalisable solution is given by 
\begin{align}
    v_{n}(z) = N_{v^{n}}zJ_{1}(Q_{n} z)
\end{align}
The spectrum of the vector mesons can be obtained by imposing a Neumann boundary condition at the hard wall $z=z_0$. For simplicity, we will work in units where $z_0=1$. The condition at the hard wall becomes 
\begin{align}
    J_{0}(Q_{n}) = 0
\end{align}
and the spectrum is given by
\begin{align}
   m_{V^n} = Q_{n} =j_{0, n} \quad n=1,2,\dots \, ,
\end{align}
where $j_{0, n}$ are zeros of the Bessel function $J_{0}(x)$. The behavior of the normalisable solution for small $z$ is
\begin{align}
    v_{n}(z) = c_{2, n}z^{2} + c_{4, n}z^{4} + ...
\end{align}
where we can identify the coefficients as
\begin{align}
    c_{2, n} =\dfrac{1}{2}N_{v^{n}}Q_{n},\quad c_{4, n} = -\dfrac{Q_{n}^{2}}{8}c_{2, n}\, .
\end{align}
Using the orthonormality condition
\begin{align}
    \int_{0}^{z_{0}} dz\, e^{A_{s}}v_{m}v_{n} =\delta_{mn}
\end{align}
we obtain  the normalisation constant
\begin{align}
    N_{v^{n}} =\dfrac{\sqrt{2}}{J_{1}(j_{0,n})}\, .
\end{align}
The holographic dictionary for the decay constants can be written as, see for instance \cite{Grigoryan:2007vg,Ballon-Bayona:2021ibm}, 
\begin{equation}
f_{v^n} = \frac{1}{g_5} \Big [ e^{A_s} \partial_z v_n(z)  \Big ]_{z \to 0} = \frac{2 c_{2,n}}{g_5} = \frac{N_{v^n} Q_n}{g_5} \,. 
\end{equation}

\subsection{Nucleons}

The nucleons can be described in the hard wall model from \eqref{fLReq2ndv2} considering $A_{s} = -\ln z$ and $\Phi = 0$, thus $\tilde{m} = m$. In this way, the equation \eqref{fLReq2ndv2} reduces to
\begin{eqnarray}
    \Big\{(z\partial_{z})^{2} - 5z\partial_{z} + 6\pm m - m^{2} + Q^{2}z^{2}\Big\}F_{R/L} = 0 \, , 
\end{eqnarray}
whose solution is
\begin{eqnarray}
     F_{R/L}(z) = c_{1} z^{5/2}J_{m\mp\frac{1}{2}}(Qz) + c_{2}z^{5/2}Y_{m\mp \frac{1}{2}}(Qz)
\end{eqnarray}
The normalisable solutions are given by
\begin{eqnarray}
    f_{R/L}^{n}(z) = N_{R/L}^{n} z^{5/2}J_{m \mp \frac{1}{2}}(Q_{n}z)\, .
\end{eqnarray}
Again we will work fix the hard wall position as $z_0=1$ for simplicity. In the hard wall model there are two possible boundary conditions for the normalisable solutions in the nucleon sector, fixing either  $f_{R}^{n}$ (model I) or  $f_{L}^{n}$ (model I) at the hard wall. We will be interested only in model I because the model II allows for a zero mode not present in the nucleon spectrum \cite{Hong:2006ta,Abidin:2009hr}. Fixing $f_{R}^{n}$ at the hard wall $z = z_{0}$ we obtain
\begin{eqnarray}
    J_{m - \frac{1}{2}}(Q_{n}) = 0
\end{eqnarray}
The spectrum of nucleons becomes
\begin{align}
     Q_{n} &= j_{1,n}\quad \textnormal{when}\quad m = 3/2 \, , \\
     Q_{n} &= j_{2,n}\quad \textnormal{when}\quad  m = 5/2 \, ,
\end{align}
where $j_{1,n}$ and $j_{2,n}$ are zeros of the Bessel functions $J_{1}(x)$ and $J_{2}(x)$, respectively. The normalisation condition  is given by
\begin{eqnarray}
    \int dz\, e^{4A_{s}}f_{R/L}^{m}(z)f_{R/L}^{n}(z) = \delta^{mn}
\end{eqnarray}
Using this condition we find the normalisation constants \begin{eqnarray}
    N_{R/L}  &=&\dfrac{\sqrt{2}}{J_{m + \frac{1}{2}}(Q_{n})} \, .
\end{eqnarray}
Using the holographic dictionary \eqref{lambdaNn} we find for the hard wall model that
\begin{eqnarray}
    \lambda_{N^n}   = \sqrt{G_F} \Big [ z^{-2-m}  f_{R,n}(z) \Big ]_{z= \epsilon}  = \sqrt{G_{F}} N_{R/L} \frac{ 2^{1/2 - m} Q_{n}^{m-1/2}}{\Gamma(m+1/2)} \,, 
\end{eqnarray}
which is compatible with \cite{Abidin:2009hr}.

\bibliographystyle{utphys}

\bibliography{Baryons}

\providecommand{\href}[2]{#2}\begingroup\raggedright\begin{thebibliography}{10}

\bibitem{Miransky:1994vk}
V.~A. Miransky, {\em {Dynamical symmetry breaking in quantum field theories}}.
\newblock World Scientific, 1994.

\bibitem{Shuryak:2004pry}
E.~V. Shuryak, \href{http://dx.doi.org/10.1142/5367}{{\em {The QCD vacuum,
  hadrons and the superdense matter}}}, vol.~71.
\newblock 10, 2004.

\bibitem{Nambu:1961tp}
Y.~Nambu and G.~Jona-Lasinio, ``{Dynamical Model of Elementary Particles Based
  on an Analogy with Superconductivity. 1.},''
  \href{http://dx.doi.org/10.1103/PhysRev.122.345}{{\em Phys. Rev.} {\bfseries
  122} (1961) 345--358}.

\bibitem{Nambu:1961fr}
Y.~Nambu and G.~Jona-Lasinio, ``{Dynamical model of elementary particles based
  on an analogy with superconductivity. II.},''
  \href{http://dx.doi.org/10.1103/PhysRev.124.246}{{\em Phys. Rev.} {\bfseries
  124} (1961) 246--254}.

\bibitem{Gell-Mann:1960mvl}
M.~Gell-Mann and M.~Levy, ``{The axial vector current in beta decay},''
  \href{http://dx.doi.org/10.1007/BF02859738}{{\em Nuovo Cim.} {\bfseries 16}
  (1960) 705}.

\bibitem{Scherer:2002tk}
S.~Scherer, ``{Introduction to chiral perturbation theory},'' {\em Adv. Nucl.
  Phys.} {\bfseries 27} (2003) 277,
  \href{http://arxiv.org/abs/hep-ph/0210398}{{\ttfamily arXiv:hep-ph/0210398}}.

\bibitem{Wilson:1974sk}
K.~G. Wilson, ``{Confinement of Quarks},''
  \href{http://dx.doi.org/10.1103/PhysRevD.10.2445}{{\em Phys. Rev. D}
  {\bfseries 10} (1974) 2445--2459}.

\bibitem{Shifman:1978bx}
M.~A. Shifman, A.~I. Vainshtein, and V.~I. Zakharov, ``{QCD and Resonance
  Physics. Theoretical Foundations},''
  \href{http://dx.doi.org/10.1016/0550-3213(79)90022-1}{{\em Nucl. Phys. B}
  {\bfseries 147} (1979) 385--447}.

\bibitem{Colangelo:2000dp}
P.~Colangelo and A.~Khodjamirian, ``{QCD sum rules, a modern perspective},''
  \href{http://arxiv.org/abs/hep-ph/0010175}{{\ttfamily arXiv:hep-ph/0010175}}.

\bibitem{Maldacena:1997re}
J.~M. Maldacena, ``{The Large N limit of superconformal field theories and
  supergravity},'' \href{http://dx.doi.org/10.4310/ATMP.1998.v2.n2.a1}{{\em
  Adv. Theor. Math. Phys.} {\bfseries 2} (1998) 231--252},
  \href{http://arxiv.org/abs/hep-th/9711200}{{\ttfamily arXiv:hep-th/9711200}}.

\bibitem{Witten:1998qj}
E.~Witten, ``{Anti-de Sitter space and holography},''
  \href{http://dx.doi.org/10.4310/ATMP.1998.v2.n2.a2}{{\em Adv. Theor. Math.
  Phys.} {\bfseries 2} (1998) 253--291},
  \href{http://arxiv.org/abs/hep-th/9802150}{{\ttfamily arXiv:hep-th/9802150}}.

\bibitem{Gubser:1998bc}
S.~S. Gubser, I.~R. Klebanov, and A.~M. Polyakov, ``{Gauge theory correlators
  from noncritical string theory},''
  \href{http://dx.doi.org/10.1016/S0370-2693(98)00377-3}{{\em Phys. Lett. B}
  {\bfseries 428} (1998) 105--114},
  \href{http://arxiv.org/abs/hep-th/9802109}{{\ttfamily arXiv:hep-th/9802109}}.

\bibitem{Polchinski:2001tt}
J.~Polchinski and M.~J. Strassler, ``{Hard scattering and gauge / string
  duality},'' \href{http://dx.doi.org/10.1103/PhysRevLett.88.031601}{{\em Phys.
  Rev. Lett.} {\bfseries 88} (2002) 031601},
  \href{http://arxiv.org/abs/hep-th/0109174}{{\ttfamily arXiv:hep-th/0109174}}.

\bibitem{Boschi-Filho:2002xih}
H.~Boschi-Filho and N.~R.~F. Braga, ``{Gauge / string duality and scalar
  glueball mass ratios},''
  \href{http://dx.doi.org/10.1088/1126-6708/2003/05/009}{{\em JHEP} {\bfseries
  05} (2003) 009}, \href{http://arxiv.org/abs/hep-th/0212207}{{\ttfamily
  arXiv:hep-th/0212207}}.

\bibitem{Erlich:2005qh}
J.~Erlich, E.~Katz, D.~T. Son, and M.~A. Stephanov, ``{QCD and a holographic
  model of hadrons},''
  \href{http://dx.doi.org/10.1103/PhysRevLett.95.261602}{{\em Phys. Rev. Lett.}
  {\bfseries 95} (2005) 261602},
  \href{http://arxiv.org/abs/hep-ph/0501128}{{\ttfamily arXiv:hep-ph/0501128}}.

\bibitem{DaRold:2005mxj}
L.~Da~Rold and A.~Pomarol, ``{Chiral symmetry breaking from five dimensional
  spaces},'' \href{http://dx.doi.org/10.1016/j.nuclphysb.2005.05.009}{{\em
  Nucl. Phys. B} {\bfseries 721} (2005) 79--97},
  \href{http://arxiv.org/abs/hep-ph/0501218}{{\ttfamily arXiv:hep-ph/0501218}}.

\bibitem{Karch:2006pv}
A.~Karch, E.~Katz, D.~T. Son, and M.~A. Stephanov, ``{Linear confinement and
  AdS/QCD},'' \href{http://dx.doi.org/10.1103/PhysRevD.74.015005}{{\em Phys.
  Rev. D} {\bfseries 74} (2006) 015005},
  \href{http://arxiv.org/abs/hep-ph/0602229}{{\ttfamily arXiv:hep-ph/0602229}}.

\bibitem{Gursoy:2007cb}
U.~Gursoy and E.~Kiritsis, ``{Exploring improved holographic theories for QCD:
  Part I},'' \href{http://dx.doi.org/10.1088/1126-6708/2008/02/032}{{\em JHEP}
  {\bfseries 02} (2008) 032}, \href{http://arxiv.org/abs/0707.1324}{{\ttfamily
  arXiv:0707.1324 [hep-th]}}.

\bibitem{Gursoy:2007er}
U.~Gursoy, E.~Kiritsis, and F.~Nitti, ``{Exploring improved holographic
  theories for QCD: Part II},''
  \href{http://dx.doi.org/10.1088/1126-6708/2008/02/019}{{\em JHEP} {\bfseries
  02} (2008) 019}, \href{http://arxiv.org/abs/0707.1349}{{\ttfamily
  arXiv:0707.1349 [hep-th]}}.

\bibitem{Gubser:2008ny}
S.~S. Gubser and A.~Nellore, ``{Mimicking the QCD equation of state with a dual
  black hole},'' \href{http://dx.doi.org/10.1103/PhysRevD.78.086007}{{\em Phys.
  Rev. D} {\bfseries 78} (2008) 086007},
  \href{http://arxiv.org/abs/0804.0434}{{\ttfamily arXiv:0804.0434 [hep-th]}}.

\bibitem{Cai:2012xh}
R.-G. Cai, S.~He, and D.~Li, ``{A hQCD model and its phase diagram in
  Einstein-Maxwell-Dilaton system},''
  \href{http://dx.doi.org/10.1007/JHEP03(2012)033}{{\em JHEP} {\bfseries 03}
  (2012) 033}, \href{http://arxiv.org/abs/1201.0820}{{\ttfamily arXiv:1201.0820
  [hep-th]}}.

\bibitem{Li:2013oda}
D.~Li and M.~Huang, ``{Dynamical holographic QCD model for glueball and light
  meson spectra},'' \href{http://dx.doi.org/10.1007/JHEP11(2013)088}{{\em JHEP}
  {\bfseries 11} (2013) 088}, \href{http://arxiv.org/abs/1303.6929}{{\ttfamily
  arXiv:1303.6929 [hep-ph]}}.

\bibitem{Ballon-Bayona:2017sxa}
A.~Ballon-Bayona, H.~Boschi-Filho, L.~A.~H. Mamani, A.~S. Miranda, and V.~T.
  Zanchin, ``{Effective holographic models for QCD: glueball spectrum and trace
  anomaly},'' \href{http://dx.doi.org/10.1103/PhysRevD.97.046001}{{\em Phys.
  Rev. D} {\bfseries 97} no.~4, (2018) 046001},
  \href{http://arxiv.org/abs/1708.08968}{{\ttfamily arXiv:1708.08968
  [hep-th]}}.

\bibitem{Karch:2002sh}
A.~Karch and E.~Katz, ``{Adding flavor to AdS / CFT},''
  \href{http://dx.doi.org/10.1088/1126-6708/2002/06/043}{{\em JHEP} {\bfseries
  06} (2002) 043}, \href{http://arxiv.org/abs/hep-th/0205236}{{\ttfamily
  arXiv:hep-th/0205236}}.

\bibitem{Sakai:2004cn}
T.~Sakai and S.~Sugimoto, ``{Low energy hadron physics in holographic QCD},''
  \href{http://dx.doi.org/10.1143/PTP.113.843}{{\em Prog. Theor. Phys.}
  {\bfseries 113} (2005) 843--882},
  \href{http://arxiv.org/abs/hep-th/0412141}{{\ttfamily arXiv:hep-th/0412141}}.

\bibitem{Grigoryan:2007vg}
H.~R. Grigoryan and A.~V. Radyushkin, ``{Form Factors and Wave Functions of
  Vector Mesons in Holographic QCD},''
  \href{http://dx.doi.org/10.1016/j.physletb.2007.05.044}{{\em Phys. Lett. B}
  {\bfseries 650} (2007) 421--427},
  \href{http://arxiv.org/abs/hep-ph/0703069}{{\ttfamily arXiv:hep-ph/0703069}}.

\bibitem{Grigoryan:2007my}
H.~R. Grigoryan and A.~V. Radyushkin, ``{Structure of vector mesons in
  holographic model with linear confinement},''
  \href{http://dx.doi.org/10.1103/PhysRevD.76.095007}{{\em Phys. Rev. D}
  {\bfseries 76} (2007) 095007},
  \href{http://arxiv.org/abs/0706.1543}{{\ttfamily arXiv:0706.1543 [hep-ph]}}.

\bibitem{Forkel:2007cm}
H.~Forkel, M.~Beyer, and T.~Frederico, ``{Linear square-mass trajectories of
  radially and orbitally excited hadrons in holographic QCD},''
  \href{http://dx.doi.org/10.1088/1126-6708/2007/07/077}{{\em JHEP} {\bfseries
  07} (2007) 077}, \href{http://arxiv.org/abs/0705.1857}{{\ttfamily
  arXiv:0705.1857 [hep-ph]}}.

\bibitem{dePaula:2008fp}
W.~de~Paula, T.~Frederico, H.~Forkel, and M.~Beyer, ``{Dynamical AdS/QCD with
  area-law confinement and linear Regge trajectories},''
  \href{http://dx.doi.org/10.1103/PhysRevD.79.075019}{{\em Phys. Rev. D}
  {\bfseries 79} (2009) 075019},
  \href{http://arxiv.org/abs/0806.3830}{{\ttfamily arXiv:0806.3830 [hep-ph]}}.

\bibitem{FolcoCapossoli:2019imm}
E.~Folco~Capossoli, M.~A. Mart\'\i{}n~Contreras, D.~Li, A.~Vega, and
  H.~Boschi-Filho, ``{Hadronic spectra from deformed AdS backgrounds},''
  \href{http://dx.doi.org/10.1088/1674-1137/44/6/064104}{{\em Chin. Phys. C}
  {\bfseries 44} no.~6, (2020) 064104},
  \href{http://arxiv.org/abs/1903.06269}{{\ttfamily arXiv:1903.06269
  [hep-ph]}}.

\bibitem{Iatrakis:2010jb}
I.~Iatrakis, E.~Kiritsis, and A.~Paredes, ``{An AdS/QCD model from tachyon
  condensation: II},'' \href{http://dx.doi.org/10.1007/JHEP11(2010)123}{{\em
  JHEP} {\bfseries 11} (2010) 123},
  \href{http://arxiv.org/abs/1010.1364}{{\ttfamily arXiv:1010.1364 [hep-ph]}}.

\bibitem{Arean:2013tja}
D.~Are\'an, I.~Iatrakis, M.~J\"arvinen, and E.~Kiritsis, ``{The discontinuities
  of conformal transitions and mass spectra of V-QCD},''
  \href{http://dx.doi.org/10.1007/JHEP11(2013)068}{{\em JHEP} {\bfseries 11}
  (2013) 068}, \href{http://arxiv.org/abs/1309.2286}{{\ttfamily arXiv:1309.2286
  [hep-ph]}}.

\bibitem{He:2013qq}
S.~He, S.-Y. Wu, Y.~Yang, and P.-H. Yuan, ``{Phase Structure in a Dynamical
  Soft-Wall Holographic QCD Model},''
  \href{http://dx.doi.org/10.1007/JHEP04(2013)093}{{\em JHEP} {\bfseries 04}
  (2013) 093}, \href{http://arxiv.org/abs/1301.0385}{{\ttfamily arXiv:1301.0385
  [hep-th]}}.

\bibitem{Ballon-Bayona:2023zal}
A.~Ballon-Bayona, T.~Frederico, L.~A.~H. Mamani, and W.~de~Paula, ``{Dynamical
  holographic QCD model for spontaneous chiral symmetry breaking and
  confinement},'' \href{http://dx.doi.org/10.1103/PhysRevD.108.106016}{{\em
  Phys. Rev. D} {\bfseries 108} no.~10, (2023) 106016},
  \href{http://arxiv.org/abs/2308.07503}{{\ttfamily arXiv:2308.07503
  [hep-ph]}}.

\bibitem{deTeramond:2005su}
G.~F. de~Teramond and S.~J. Brodsky, ``{Hadronic spectrum of a holographic dual
  of QCD},'' \href{http://dx.doi.org/10.1103/PhysRevLett.94.201601}{{\em Phys.
  Rev. Lett.} {\bfseries 94} (2005) 201601},
  \href{http://arxiv.org/abs/hep-th/0501022}{{\ttfamily arXiv:hep-th/0501022}}.

\bibitem{Hong:2006ta}
D.~K. Hong, T.~Inami, and H.-U. Yee, ``{Baryons in AdS/QCD},''
  \href{http://dx.doi.org/10.1016/j.physletb.2007.01.030}{{\em Phys. Lett. B}
  {\bfseries 646} (2007) 165--171},
  \href{http://arxiv.org/abs/hep-ph/0609270}{{\ttfamily arXiv:hep-ph/0609270}}.

\bibitem{Brodsky:2008pg}
S.~J. Brodsky and G.~F. de~Teramond, ``{AdS/CFT and Light-Front QCD},''
  \href{http://dx.doi.org/10.1142/9789814293242_0008}{{\em Subnucl. Ser.}
  {\bfseries 45} (2009) 139--183},
  \href{http://arxiv.org/abs/0802.0514}{{\ttfamily arXiv:0802.0514 [hep-ph]}}.

\bibitem{Abidin:2009hr}
Z.~Abidin and C.~E. Carlson, ``{Nucleon electromagnetic and gravitational form
  factors from holography},''
  \href{http://dx.doi.org/10.1103/PhysRevD.79.115003}{{\em Phys. Rev. D}
  {\bfseries 79} (2009) 115003},
  \href{http://arxiv.org/abs/0903.4818}{{\ttfamily arXiv:0903.4818 [hep-ph]}}.

\bibitem{Gutsche:2011vb}
T.~Gutsche, V.~E. Lyubovitskij, I.~Schmidt, and A.~Vega, ``{Dilaton in a
  soft-wall holographic approach to mesons and baryons},''
  \href{http://dx.doi.org/10.1103/PhysRevD.85.076003}{{\em Phys. Rev. D}
  {\bfseries 85} (2012) 076003},
  \href{http://arxiv.org/abs/1108.0346}{{\ttfamily arXiv:1108.0346 [hep-ph]}}.

\bibitem{Nascimento:2023dzx}
A.~d. C. P.~d. Nascimento and H.~Boschi-Filho, ``{Comparison between
  holographic deformed AdS and soft wall models for fermions},''
  \href{http://dx.doi.org/10.1103/PhysRevD.108.106008}{{\em Phys. Rev. D}
  {\bfseries 108} no.~10, (2023) 106008},
  \href{http://arxiv.org/abs/2306.17315}{{\ttfamily arXiv:2306.17315
  [hep-th]}}.

\bibitem{Hata:2007mb}
H.~Hata, T.~Sakai, S.~Sugimoto, and S.~Yamato, ``{Baryons from instantons in
  holographic QCD},'' \href{http://dx.doi.org/10.1143/PTP.117.1157}{{\em Prog.
  Theor. Phys.} {\bfseries 117} (2007) 1157},
  \href{http://arxiv.org/abs/hep-th/0701280}{{\ttfamily arXiv:hep-th/0701280}}.

\bibitem{Hashimoto:2008zw}
K.~Hashimoto, T.~Sakai, and S.~Sugimoto, ``{Holographic Baryons: Static
  Properties and Form Factors from Gauge/String Duality},''
  \href{http://dx.doi.org/10.1143/PTP.120.1093}{{\em Prog. Theor. Phys.}
  {\bfseries 120} (2008) 1093--1137},
  \href{http://arxiv.org/abs/0806.3122}{{\ttfamily arXiv:0806.3122 [hep-th]}}.

\bibitem{Pomarol:2008aa}
A.~Pomarol and A.~Wulzer, ``{Baryon Physics in Holographic QCD},''
  \href{http://dx.doi.org/10.1016/j.nuclphysb.2008.10.004}{{\em Nucl. Phys. B}
  {\bfseries 809} (2009) 347--361},
  \href{http://arxiv.org/abs/0807.0316}{{\ttfamily arXiv:0807.0316 [hep-ph]}}.

\bibitem{Cherman:2009gb}
A.~Cherman, T.~D. Cohen, and M.~Nielsen, ``{Model Independent Tests of
  Skyrmions and Their Holographic Cousins},''
  \href{http://dx.doi.org/10.1103/PhysRevLett.103.022001}{{\em Phys. Rev.
  Lett.} {\bfseries 103} (2009) 022001},
  \href{http://arxiv.org/abs/0903.2662}{{\ttfamily arXiv:0903.2662 [hep-ph]}}.

\bibitem{Sutcliffe:2015sta}
P.~Sutcliffe, ``{Holographic Skyrmions},''
  \href{http://dx.doi.org/10.1142/S0217984915400515}{{\em Mod. Phys. Lett. B}
  {\bfseries 29} no.~16, (2015) 1540051}.

\bibitem{Hayashi:2020ipd}
Y.~Hayashi, T.~Ogino, T.~Sakai, and S.~Sugimoto, ``{Stringy excited baryons in
  holographic quantum chromodynamics},''
  \href{http://dx.doi.org/10.1093/ptep/ptaa045}{{\em PTEP} {\bfseries 2020}
  no.~5, (2020) 053B04}, \href{http://arxiv.org/abs/2001.01461}{{\ttfamily
  arXiv:2001.01461 [hep-th]}}.

\bibitem{Jarvinen:2022gcc}
M.~J\"arvinen, E.~Kiritsis, F.~Nitti, and E.~Pr\'eau, ``{The V-QCD baryon:
  numerical solution and baryon spectrum},''
  \href{http://dx.doi.org/10.1007/JHEP05(2023)081}{{\em JHEP} {\bfseries 05}
  (2023) 081}, \href{http://arxiv.org/abs/2212.06747}{{\ttfamily
  arXiv:2212.06747 [hep-th]}}.

\bibitem{Abt:2019tas}
R.~Abt, J.~Erdmenger, N.~Evans, and K.~S. Rigatos, ``{Light composite fermions
  from holography},'' \href{http://dx.doi.org/10.1007/JHEP11(2019)160}{{\em
  JHEP} {\bfseries 11} (2019) 160},
  \href{http://arxiv.org/abs/1907.09489}{{\ttfamily arXiv:1907.09489
  [hep-th]}}.

\bibitem{Nakas:2020hyo}
T.~Nakas and K.~S. Rigatos, ``{Fermions and baryons as open-string states from
  brane junctions},'' \href{http://dx.doi.org/10.1007/JHEP12(2020)157}{{\em
  JHEP} {\bfseries 12} (2020) 157},
  \href{http://arxiv.org/abs/2010.00025}{{\ttfamily arXiv:2010.00025
  [hep-th]}}.

\bibitem{Kinar:1998vq}
Y.~Kinar, E.~Schreiber, and J.~Sonnenschein, ``{Q anti-Q potential from strings
  in curved space-time: Classical results},''
  \href{http://dx.doi.org/10.1016/S0550-3213(99)00652-5}{{\em Nucl. Phys. B}
  {\bfseries 566} (2000) 103--125},
  \href{http://arxiv.org/abs/hep-th/9811192}{{\ttfamily arXiv:hep-th/9811192}}.

\bibitem{Andreev:2006ct}
O.~Andreev and V.~I. Zakharov, ``{Heavy-quark potentials and AdS/QCD},''
  \href{http://dx.doi.org/10.1103/PhysRevD.74.025023}{{\em Phys. Rev. D}
  {\bfseries 74} (2006) 025023},
  \href{http://arxiv.org/abs/hep-ph/0604204}{{\ttfamily arXiv:hep-ph/0604204}}.

\bibitem{Ramallo:2013bua}
A.~V. Ramallo, ``{Introduction to the AdS/CFT correspondence},''
  \href{http://dx.doi.org/10.1007/978-3-319-12238-0_10}{{\em Springer Proc.
  Phys.} {\bfseries 161} (2015) 411--474},
  \href{http://arxiv.org/abs/1310.4319}{{\ttfamily arXiv:1310.4319 [hep-th]}}.

\bibitem{Witten:1979kh}
E.~Witten, ``{Baryons in the 1/n Expansion},''
  \href{http://dx.doi.org/10.1016/0550-3213(79)90232-3}{{\em Nucl. Phys. B}
  {\bfseries 160} (1979) 57--115}.

\bibitem{Son:2003et}
D.~T. Son and M.~A. Stephanov, ``{QCD and dimensional deconstruction},''
  \href{http://dx.doi.org/10.1103/PhysRevD.69.065020}{{\em Phys. Rev. D}
  {\bfseries 69} (2004) 065020},
  \href{http://arxiv.org/abs/hep-ph/0304182}{{\ttfamily arXiv:hep-ph/0304182}}.

\bibitem{Ballon-Bayona:2020qpq}
A.~Ballon-Bayona and L.~A.~H. Mamani, ``{Nonlinear realization of chiral
  symmetry breaking in holographic soft wall models},''
  \href{http://dx.doi.org/10.1103/PhysRevD.102.026013}{{\em Phys. Rev. D}
  {\bfseries 102} no.~2, (2020) 026013},
  \href{http://arxiv.org/abs/2002.00075}{{\ttfamily arXiv:2002.00075
  [hep-ph]}}.

\bibitem{Ballon-Bayona:2021ibm}
A.~Ballon-Bayona, L.~A.~H. Mamani, and D.~M. Rodrigues, ``{Spontaneous chiral
  symmetry breaking in holographic soft wall models},''
  \href{http://dx.doi.org/10.1103/PhysRevD.104.126029}{{\em Phys. Rev. D}
  {\bfseries 104} no.~12, (2021) 126029},
  \href{http://arxiv.org/abs/2107.10983}{{\ttfamily arXiv:2107.10983
  [hep-ph]}}.

\bibitem{OBELIX:1997zla}
{\bfseries OBELIX} Collaboration, A.~Bertin {\em et~al.}, ``{Study of anti-p p
  --\ensuremath{>} 2pi+ 2pi- annihilation from S states},''
  \href{http://dx.doi.org/10.1016/S0370-2693(97)01189-1}{{\em Phys. Lett. B}
  {\bfseries 414} (1997) 220--228}.

\bibitem{Workman:2022ynf}
{\bfseries Particle Data Group} Collaboration, R.~L. Workman {\em et~al.},
  ``{Review of Particle Physics},''
  \href{http://dx.doi.org/10.1093/ptep/ptac097}{{\em PTEP} {\bfseries 2022}
  (2022) 083C01}.

\bibitem{Donoghue:1992dd}
J.~F. Donoghue, E.~Golowich, and B.~R. Holstein,
  \href{http://dx.doi.org/10.1017/CBO9780511524370}{{\em {Dynamics of the
  standard model}}}, vol.~2.
\newblock CUP, 2014.

\bibitem{Ioffe:1981kw}
B.~L. Ioffe, ``{Calculation of Baryon Masses in Quantum Chromodynamics},''
  \href{http://dx.doi.org/10.1016/0550-3213(81)90259-5}{{\em Nucl. Phys. B}
  {\bfseries 188} (1981) 317--341}. [Erratum: Nucl.Phys.B 191, 591--592
  (1981)].

\bibitem{Cohen:1994wm}
T.~D. Cohen, R.~J. Furnstahl, D.~K. Griegel, and X.-m. Jin, ``{QCD sum rules
  and applications to nuclear physics},''
  \href{http://dx.doi.org/10.1016/0146-6410(95)00043-I}{{\em Prog. Part. Nucl.
  Phys.} {\bfseries 35} (1995) 221--298},
  \href{http://arxiv.org/abs/hep-ph/9503315}{{\ttfamily arXiv:hep-ph/9503315}}.

\bibitem{Leinweber:1995fn}
D.~B. Leinweber, ``{QCD sum rules for skeptics},''
  \href{http://dx.doi.org/10.1006/aphy.1996.5641}{{\em Annals Phys.} {\bfseries
  254} (1997) 328--396}, \href{http://arxiv.org/abs/nucl-th/9510051}{{\ttfamily
  arXiv:nucl-th/9510051}}.

\bibitem{Leinweber:1994nm}
D.~B. Leinweber, ``{Nucleon properties from unconventional interpolating
  fields},'' \href{http://dx.doi.org/10.1103/PhysRevD.51.6383}{{\em Phys. Rev.
  D} {\bfseries 51} (1995) 6383--6393},
  \href{http://arxiv.org/abs/nucl-th/9406001}{{\ttfamily
  arXiv:nucl-th/9406001}}.

\bibitem{Ahn:2009px}
H.~C. Ahn, D.~K. Hong, C.~Park, and S.~Siwach, ``{Spin 3/2 Baryons and Form
  Factors in AdS/QCD},''
  \href{http://dx.doi.org/10.1103/PhysRevD.80.054001}{{\em Phys. Rev. D}
  {\bfseries 80} (2009) 054001},
  \href{http://arxiv.org/abs/0904.3731}{{\ttfamily arXiv:0904.3731 [hep-ph]}}.

\bibitem{Huseynova:2020gqn}
N.~J. Huseynova, ``{Coupling Constant of a Vector Meson with Delta Baryons in
  the Soft-Wall Ads/QCD Model},''
  \href{http://dx.doi.org/10.1007/s11182-020-02109-0}{{\em Russ. Phys. J.}
  {\bfseries 63} no.~5, (2020) 860--866}.

\bibitem{arfken2013mathematical}
G.~B. Arfken, H.~J. Weber, and F.~E. Harris, {\em Mathematical methods for
  physicists}.
\newblock Elsevier, Waltham, MA, 2013.

\bibitem{butkov1968mathematical}
E.~Butkov, {\em Mathematical physics}.
\newblock Addison-Wesley, Reading, MA, 1968.

\bibitem{Itzykson:1980rh}
C.~Itzykson and J.~B. Zuber, {\em {Quantum Field Theory}}.
\newblock International Series In Pure and Applied Physics. McGraw-Hill, New
  York, 1980.

\bibitem{Peskin:1995ev}
M.~E. Peskin and D.~V. Schroeder, {\em {An Introduction to quantum field
  theory}}.
\newblock Addison-Wesley, Reading, USA, 1995.

\end{thebibliography}\endgroup

\end{document}